\documentclass{emulateapj}
\usepackage{amssymb,amsmath}
\usepackage{verbatim}
\usepackage{color,hyperref}
\definecolor{linkcolor}{rgb}{0,0,0.25}
\hypersetup{
  colorlinks=true,        
  linkcolor=linkcolor,    
  citecolor=linkcolor,    
  filecolor=linkcolor,    
  urlcolor=linkcolor      
}
\newcommand{\ie}{i.e.}
\newcommand{\etal}{et al.}
\newcommand{\dd}{\mathrm{d}}
\newcommand{\eg}{e.g.}
\newcommand{\eqnname}{equation}
\newcommand{\Eqnname}{Equation}
\newcommand{\equationname}{\eqnname}

\renewcommand{\figurename}{Figure}

\newcommand{\sectionname}{$\mathsection$}

\newcommand{\feh}{\ensuremath{[\mathrm{Fe/H}]}}
\newcommand{\afe}{\ensuremath{[\alpha\mathrm{/Fe}]}}

\renewcommand{\vec}[1]{\ensuremath{\mathbf{#1}}}

\newcommand{\dex}{\ensuremath{\,\mathrm{dex}}}
\newcommand{\yr}{\ensuremath{\,\mathrm{yr}}}
\newcommand{\Myr}{\ensuremath{\,\mathrm{Myr}}}
\newcommand{\Gyr}{\ensuremath{\,\mathrm{Gyr}}}
\newcommand{\kpc}{\ensuremath{\,\mathrm{kpc}}}
\newcommand{\pc}{\ensuremath{\,\mathrm{pc}}}
\newcommand{\kms}{\ensuremath{\,\mathrm{km\ s}^{-1}}}
\newcommand{\msun}{\ensuremath{\,\mathrm{M}_{\odot}}}
\newcommand{\mas}{\ensuremath{\,\mathrm{mas}}}
\newcommand{\muas}{\ensuremath{\,\mu\mathrm{as}}}
\newcommand{\inv}{\ensuremath{^{-1}}}

\newcommand{\magunit}{\,\mbox{mag}}
\newcommand{\ks}{\ensuremath{K_s}}

\newcommand{\apogee}{APOGEE}
\newcommand{\rc}{RC}
\newcommand{\logg}{\ensuremath{\log g}}
\newcommand{\teff}{\ensuremath{T_{\mathrm{eff}}}}

\newcommand{\ncatalog}{10,341}

\submitted{}

\usepackage{atbegshi}
\def\hackaltaffiltext#1#2{\AtBeginShipoutNext{\footnotetext[#1]{#2}\stepcounter{footnote}}}

\begin{document}

\title{The APOGEE red-clump catalog: Precise distances, velocities,
  and high-resolution elemental abundances over a large area of the
  Milky Way's disk}

\author{Jo~Bovy\altaffilmark{1,2},
  David~L.~Nidever\altaffilmark{3}, 
  Hans-Walter~Rix\altaffilmark{4}, 
  L\'{e}o~Girardi\altaffilmark{5,6}, 
  Gail~Zasowski\altaffilmark{7}, 
  S.~Drew~Chojnowski\altaffilmark{8}, 
  Jon~Holtzman\altaffilmark{9}, 
  Courtney~Epstein\altaffilmark{10}, 
  Peter~M.~Frinchaboy\altaffilmark{11}, 
  Michael~R.~Hayden\altaffilmark{9}, 
  Tha\'{i}se~S.~Rodrigues\altaffilmark{5,6}, 
  Steven~R.~Majewski\altaffilmark{8}, 
  Jennifer~A.~Johnson\altaffilmark{10,12}, 
  Marc~H.~Pinsonneault\altaffilmark{10}, 
  Dennis~Stello\altaffilmark{13,14}, 
  Carlos~Allende~Prieto\altaffilmark{15,16}, 
  Brett~Andrews\altaffilmark{10}, 
  Sarbani~Basu\altaffilmark{17}, 
  Timothy~C.~Beers\altaffilmark{18,19}, 
  Dmitry~Bizyaev\altaffilmark{20}, 
  Adam~Burton\altaffilmark{8}, 
  William~J.~Chaplin\altaffilmark{14,21}, 
  Katia~Cunha\altaffilmark{22,23}, 
  Yvonne~Elsworth\altaffilmark{14,21}, 
  Rafael~A.~Garc\'{\i}a\altaffilmark{24}, 
  Domingo~A.~Garc\'{\i}a-Her\'{n}andez\altaffilmark{15,16}, 
  Ana~E.~Garc\'{\i}a~P\'{e}rez\altaffilmark{8}, 
  Fred~R.~Hearty\altaffilmark{25}, 
  Saskia~Hekker\altaffilmark{26}, 
  Thomas~Kallinger\altaffilmark{27}, 
  Karen~Kinemuchi\altaffilmark{20}, 
  Lars~Koesterke\altaffilmark{28}, 
  Szabolcs~M{\'e}sz{\'a}ros\altaffilmark{29}, 
  Beno\^\i t~Mosser\altaffilmark{30}, 
  Robert~W.~O'Connell\altaffilmark{8}, 
  Daniel~Oravetz\altaffilmark{20}, 
  Kaike~Pan\altaffilmark{20}, 
  Annie~C.~Robin\altaffilmark{31}, 
  Ricardo~P.~Schiavon\altaffilmark{32}, 
  Donald~P.~Schneider\altaffilmark{25,33}, 
  Mathias~Schultheis\altaffilmark{34}, 
  Aldo~Serenelli\altaffilmark{35}, 
  Matthew~Shetrone\altaffilmark{36}, 
  Victor~Silva~Aguirre\altaffilmark{14}, 
  Audrey~Simmons\altaffilmark{20}, 
  Michael~Skrutskie\altaffilmark{8}, 
  Verne~V.~Smith\altaffilmark{22,37}, 
  Keivan~Stassun\altaffilmark{38,39}, 
  David~H.~Weinberg\altaffilmark{10,12}, 
  John~C.~Wilson\altaffilmark{8}, 
  and Olga Zamora\altaffilmark{15,16} 
  }
\altaffiltext{1}{
  Institute for Advanced Study, Einstein Drive, Princeton, NJ 08540, USA; bovy@ias.edu~}
\altaffiltext{2}{
  Hubble fellow}
\altaffiltext{3}{
  Department of Astronomy, University of Michigan, Ann Arbor, MI 48109, USA}
\altaffiltext{4}{
  Max-Planck-Institut f\"ur Astronomie, K\"onigstuhl 17, D-69117 Heidelberg, Germany}
\altaffiltext{5}{
  Osservatorio Astronomico di Padova - INAF, Vicolo dell'Osservatorio 5, I-35122 Padova, Italy}
\altaffiltext{6}{
  Laborat\'{o}rio Interinstitucional de e-Astronomia - LIneA, Rua Gal. Jos\'e Cristino 77, Rio de Janeiro, RJ - 20921-400, Brazil}
\altaffiltext{7}{
  Department of Physics and Astronomy, Johns Hopkins University, Baltimore, MD 21218, USA}
\altaffiltext{8}{
  Department of Astronomy, University of Virginia, Charlottesville, VA, 22904, USA}
\altaffiltext{9}{
  New Mexico State University, Las Cruces, NM 88003, USA}
\altaffiltext{10}{
  Department of Astronomy, The Ohio State University, Columbus, OH 43210, USA}
\altaffiltext{11}{
  Department of Physics and Astronomy, Texas Christian University, Fort Worth, TX 76129, USA}
\altaffiltext{12}{
  Center for Cosmology and Astro-Particle Physics, The Ohio State University, Columbus, OH 43210, USA}
\altaffiltext{13}{
  Sydney Institute for Astronomy (SIfA), School of Physics, University of Sydney, NSW 2006, Australia}
\altaffiltext{14}{
  Stellar Astrophysics Centre, Department of Physics and Astronomy, Aarhus University, DK-8000 Aarhus C, Denmark\phantom{filler filler filler filler filler filler filler}}
\hackaltaffiltext{15}{
  Instituto de Astrof{\'{\i}}sica de Canarias (IAC), E-38200 La Laguna, Tenerife, Spain} 
\hackaltaffiltext{16}{
  Departamento de Astrof{\'{\i}}sica, Universidad de La Laguna (ULL), E-38206 La Laguna, Tenerife, Spain}
\hackaltaffiltext{17}{
  Department of Astronomy, Yale University, P.O. Box 208101, New Haven, CT 06520-8101, USA}
\hackaltaffiltext{18}{
  National Optical Astronomy Observatory, Tucson, AZ 85719, USA}
\hackaltaffiltext{19}{
  JINA: Joint Institute for Nuclear Astrophysics}
\hackaltaffiltext{20}{
  Apache Point Observatory and New Mexico State University, P.O. Box 59, Sunspot, NM, 88349-0059, USA}
\hackaltaffiltext{21}{
  University of Birmingham, School of Physics and Astronomy, Edgbaston, Birmingham B15 2TT, UK}
\hackaltaffiltext{22}{
  Observat\'{o}rio Nacional, Rio de Janeiro, RJ 20921-400, Brazil}
\hackaltaffiltext{23}{
  Steward Observatory, U. Arizona, Tucson, AZ 85719, USA}
\hackaltaffiltext{24}{
  Laboratoire AIM, CEA/DSM-CNRS-Universite Paris Diderot, IRFU/SAp, Centre de Saclay, F-91191 Gif-sur-Yvette Cedex, France}
\hackaltaffiltext{25}{
  Department of Astronomy \& Astrophysics, The Pennsylvania State University, 525 Davey Laboratory, University Park PA  16802, USA}
\hackaltaffiltext{26}{
  Max-Planck-Institut f\"ur Sonnensystemforschung, Justus-von-Liebig-Weg 3, 37077 G\"ottingen, Germany}
\hackaltaffiltext{27}{
  Institute for Astronomy, University of Vienna, Turkenschanzstrasse 17, A-1180 Vienna, Austria}
\hackaltaffiltext{28}{
  Texas Advanced Computing Center, University of Texas, Austin, TX 78759, USA}
\hackaltaffiltext{29}{
  Department of Astronomy, Indiana University, Bloomington, IN 47405-7105, USA}
\hackaltaffiltext{30}{
  LESIA, CNRS, Universit\'e Pierre et Marie Curie, Universit\'e Denis Diderot, Observatoire de Paris, 92195 Meudon, France}
\hackaltaffiltext{31}{
  Institute Utinam, CNRS UMR6213, Universit\'{e} de Franche-Comt\'{e}, OSU THETA de Franche-Comt\'{e}-Bourgogne, Besancon, France}
\hackaltaffiltext{32}{
  Astrophysics Research Institute, IC2, Liverpool Science Park, Liverpool John Moores University, 146 Brownlow Hill, Liverpool, L3 5RF, UK}
\hackaltaffiltext{33}{
  Institute for Gravitation and the Cosmos, The Pennsylvania State University, University Park, PA 16802}
\hackaltaffiltext{34}{
  Laboratoire Lagrange (UMR7293), Universit\'{e} de Nice Sophia Antipolis, CNRS, Observatoire de la C\^{o}te d'Azur, BP 4229, 06304 Nice Cedex 4, France}
\hackaltaffiltext{35}{
  Institute of Space Sciences (CSIC-IEEC), Campus UAB, Bellaterra, 08193, Spain}
\hackaltaffiltext{36}{
  The University of Texas at Austin, McDonald Observatory, TX 79734, USA}
\hackaltaffiltext{37}{
  National Optical Astronomy Observatory, Tucson, AZ 85719, USA}
\hackaltaffiltext{38}{
  Physics and Astronomy Department, Vanderbilt University, 1807 Station B, Nashville, TN 37235, USA}
\hackaltaffiltext{39}{
  Department of Physics, Fisk University, 1000 17th Avenue North, Nashville, TN 37208, USA}

\begin{abstract}  
  The Sloan Digital Sky Survey III's Apache Point Observatory Galactic
  Evolution Experiment (APOGEE) is a high-resolution near-infrared
  spectroscopic survey covering all of the major components of the
  Galaxy, including the dust-obscured regions of the inner Milky Way
  disk and bulge. Here we present a sample of \ncatalog\ likely
  red-clump stars (\rc) from the first two years of APOGEE operations,
  selected based on their position in
  color--metallicity--surface-gravity--effective-temperature space
  using a new method calibrated using stellar-evolution models and
  high-quality asteroseismology data. The narrowness of the RC locus
  in color--metallicity--luminosity space allows us to assign
  distances to the stars with an accuracy of $5$ to $10\,\%$. The
  sample extends to typical distances of about 3\kpc\ from the Sun,
  with some stars out to 8\kpc, and spans a volume of approximately
  $100\kpc^3$ over $5\kpc \lesssim R \lesssim 14\kpc$, $|Z| \lesssim
  2\kpc$, and $-15^\circ \lesssim \mathrm{Galactocentric\ azimuth}
  \lesssim 30^\circ$. The APOGEE red-clump (APOGEE-RC) catalog
  contains photometry from 2MASS, reddening estimates, distances,
  line-of-sight velocities, stellar parameters and elemental
  abundances determined from the high-resolution APOGEE spectra, and
  matches to major proper motion catalogs. We determine the survey
  selection function for this data set and discuss how the RC
  selection samples the underlying stellar populations. We use this
  sample to limit any azimuthal variations in the median metallicity
  within the $\approx 45^\circ$ azimuthal region covered by the
  current sample to be $\leq0.02\dex$, which is more than an order of
  magnitude smaller than the radial metallicity gradient. This result
  constrains coherent non-axisymmetric flows within a few kpc from the
  Sun.
\end{abstract}


\section{Introduction}

The Milky Way as a galaxy constitutes a unique laboratory for galaxy
formation and evolution studies, because, unlike for external
galaxies, we can determine the high-dimensional stellar distribution
of positions, velocities, ages, elemental abundances, etc. in detail
using observations of individual stars
\citep[\eg,][]{Rix13a}. Unraveling the history of the Milky Way in
large part amounts to understanding the multifarious correlations
among these observables, the question of distinct or smoothly blended
sub-components, and variations in this distribution, and deducing its
co-evolution with the other major components of the Galaxy---gas and
dark matter---over the course of cosmic history.

Elemental abundances and their correlation with the spatial
distribution and kinematics of stars are key to improving our
knowledge of the chemo-dynamical structure of the Milky Way
\citep[\eg,][]{BovyMAPkinematics,BovyMAPstructure}. As detailed
elemental-abundance patterns cannot be derived from broadband
photometry, spectroscopic surveys are and will remain a necessary
complement to large-area photometric and astrometric surveys (\eg,
Pan-STARRS, \citealt{Kaiser02}; SkyMapper, \citealt{Keller07a};
\emph{Gaia}, \citealt{deBruijne2012}). Using current measurements of
the distribution of stellar mass \citep{Bovy13a}, 80\,\% of the stars
in the Milky Way disk lie at $R < R_0$ and $|Z| < 1\kpc$, where $R_0$
is the distance from the Sun to the Galactic center and $R$ and $Z$
are Galactocentric cylindrical coordinates. Much of this volume is
heavily extincted by dust obscuration and therefore difficult to
access using optical spectroscopic surveys or \emph{Gaia}. SDSS-III's
APOGEE (\citealt{Eisenstein11a}; S.~R.~Majewski, \etal\ 2014, in
preparation) is a high-resolution spectroscopic survey that
circumvents this difficulty by operating in the near-infrared
($\approx1.6\mu$m) where extinction by dust is almost an order of
magnitude smaller than at visible wavelengths.

In order to obtain high-resolution spectra in the infrared for stars
out to large distances in the disk, APOGEE is primarily a survey of
bright giants, for which distances are typically quite imprecise
($\gtrsim20\,\%$ uncertainty; \eg, \citealt{Binney13a},
\citealt{Santiago14a}, M.~Hayden, \etal\ 2014, in preparation). This
makes it difficult to study the fine-grained distribution of
velocities and elemental abundances as a function of location in the
disk, because at several kpc (or kiloparsecs), distance uncertainties
are comparable to the scale over which significant gradients in
kinematics (HI streaming motions over $\approx 1\kpc$; \eg,
\citealt{Levine08a}), metallicity ($\approx-0.1\dex\kpc^{-1}$ near the
plane; \eg, \citealt{Anders14a,Hayden14a}; see also
\sectionname~\ref{sec:science}), or level of alpha-enhancement exist
(\eg, \citealt{BovyMAPstructure}). In this paper, we identify a sample
of red-clump star (RC) candidates in the APOGEE data for which the
narrow distribution in the color--magnitude space allows assignment of
distances that are accurate to about $5$ to $10\,\%$, which is smaller
than the scale over which significant Galactic gradients exist even at
the largest distances in our sample.

The \rc\ is a prominent feature in the color--magnitude diagram (CMD)
of stars that corresponds to the core-Helium-burning stage in stellar
evolution of low-mass stars. Because the luminosity distribution of
the \rc\ is very narrow, with a peak magnitude that does not depend
very strongly on age or metallicity (see below), the magnitude of the
\rc\ has been widely used as a distance indicator
\citep{Paczynski98a}. Through isolating the \rc\ in the CMD and
calibrating its luminosity using local data from \emph{Hipparcos}
\citep{ESA97a}, precise distances to the Galactic center
\citep{Paczynski98a}, various parts of the Galactic bulge
\citep{McWilliam10a,Nataf10a}, the Large Magellanic Cloud
\citep{Stanek98b,Udalski98a}, M31 \citep{Stanek98a}, and many other
external galaxies \citep[e.g.,][]{Girardi01a} have been obtained. This
method relies on finding the \rc\ based on its color, and the fact
that it represents a significant overdensity in the CMD of a sample of
stars at approximately a single distance.

Here we identify \emph{individual} \rc\ candidates in order to obtain
distances to individual stars. These stars are \emph{not} at a single
distance, and therefore they cannot be isolated by searching for an
overdensity in the observed CMD. Thus, we require a more detailed
understanding of the \rc\ to select likely \rc\ stars for which we can
obtain precise distances, based on the stars' atmospheric parameters
and colors. In previous studies, the \rc\ method for individual stars
has assumed that the \rc\ has a mean magnitude that does not depend on
color or metallicity, and that the color of the \rc\ is not strongly
affected by metallicity \citep[e.g.,][]{Siebert11a,Williams13a}. It is
clear, however, from theoretical isochrones produced by various
stellar-evolution codes \citep[\eg,][]{Girardi01a} that these
assumptions are not valid, especially the latter, and applying a
metallicity-independent color selection of \rc\ stars will result in
strong contamination ($\gtrsim30\,\%$) from red-giant stars that are
at quite different distances (up to $1\magunit$ in distance modulus;
see \sectionname~\ref{sec:sample}).

The APOGEE-RC sample presented in this paper corresponds to two years
of APOGEE data and contains \ncatalog\ stars over a volume of about
$100\kpc^3$ that approximately spans $5\kpc \lesssim R \lesssim
14\kpc$, $|Z| \lesssim 2\kpc$, and $-15^\circ \lesssim
\mathrm{Galactocentric\ azimuth} \lesssim 30^\circ$. Most stars in the
current sample lie within $5\kpc$ from the Sun. As APOGEE continues
acquiring data, longer integrations designed to reach fainter stars
will extend the reach to $\approx10\kpc$ toward the outer Galaxy,
where extinction is low, and to $\approx8\kpc$ toward the inner Milky
Way. SDSS-IV's APOGEE-2 \citep{Sobeck14a}, a continuation of the
APOGEE survey starting in July 2014, of which a major component is a
similar survey conducted from the southern hemisphere, will extend the
coverage of this red-clump sample to full $360^\circ$ coverage of the
Galactic plane with an additional 10,000+ sample of stars. The
APOGEE-RC catalog will be publicly released with SDSS-III's Data
Release 11/12 in December 2014 (see
\sectionname~\ref{sec:catalog}). We are presenting this description of
the catalog now, because the sample forms the basis of forthcoming
science papers and such that the community can anticipate its
availability.

A brief outline of this paper is as follows: We describe our new
method for selecting RC stars from near-infrared photometry and
high-resolution spectroscopic data in
\sectionname~\ref{sec:sample}. In \sectionname~\ref{sec:distance} we
describe our calibration of the RC distance scale for our sample using
local \emph{Hipparcos} data. We present the sample selection function
for the \apogee\ sample in \sectionname~\ref{sec:ssf} and determine
the manner in which our RC selection samples the underlying Galactic
stellar populations in \sectionname~\ref{sec:astrosf}. We provide a
description of the RC catalog in \sectionname~\ref{sec:catalog}. In
\sectionname~\ref{sec:science} we investigate radial and azimuthal
gradients in the metallicity distribution near the Milky Way's
midplane using the APOGEE-RC sample, allowing us to limit the
ellipticity of the MW disk and constrain significant redistribution of
angular momentum on kpc scales over the last
Gyr. \sectionname~\ref{sec:conclusion} concludes the paper.

In this study, we transform distances and velocities to the
left-handed Galactocentric rest-frame by assuming that the Sun's
displacement from the midplane is 25 pc toward the north Galactic pole
\citep{Chen01a,Juric08a} and that the Sun is located at 8 kpc from the
Galactic center \citep[\eg,][]{Ghez08a,Gillessen09a,Bovy09b}.


\section{Selection of a pure red-clump sample in APOGEE}\label{sec:sample}

\subsection{APOGEE observations and data}\label{sec:apogee}

The \apogee\ is a near-infrared (NIR; $H$-band; 1.51 to 1.70 $\mu$m),
high-resolution ($R\approx 22,500$), spectroscopic survey. The
\apogee\ instrument (\citealt{Wilson10a}, J.~Wilson \etal\ 2014, in
preparation) consists of a spectrograph with 300 $2\arcsec$ fibers
that reaches a signal-to-noise ratio of $100$ per half-resolution
element ($\approx141$ per resolution element) at $H \leq 12.2$ in
three $\approx$one-hour visits during bright time on the 2.5-meter
Sloan telescope, located at the Apache Point Observatory in Sunspot,
NM \citep{Gunn06a}. A detailed description of the target selection and
data reduction pipeline is presented in \citet{Zasowski13a} and
D.~Nidever \etal\ 2014 (in preparation), respectively.

The photometry for all APOGEE targets is corrected for extinction with
the Rayleigh Jeans Color Excess method
\citep[RJCE;][]{Majewski11a}. This technique provides extinction
values $A_{\ks}$ by making use of the near constancy of the
near-to-mid-infrared $(H-[4.5\mu])_0$ color and calculating $A_{\ks}$
as
\begin{equation}
  A_{\ks} = 0.918\,(H -[4.5\mu]-0.08)\,,
\end{equation}
where $H - [4.5\mu]$ is the measured color. NIR photometry comes from
the 2MASS catalog \citep{Skrutskie06a} and mid-IR (MIR) photometry is
obtained from Spitzer-IRAC GLIMPSE-I, -II, and -3D
\citep{Churchwell09a} when available and from WISE \citep{Wright10a}
otherwise. Typical photometric uncertainties for the stars in the RC
catalog defined below are $\sim\!0.02\magunit$ in $(J,H,\ks)$ and
$\sim\!0.05\magunit$ in $[4.5\mu]$, increasing to $\sim\!0.03\magunit$
and $\sim\!0.10\magunit$ in NIR and MIR photometry, respectively, at
the faintest end of the APOGEE catalog ($H \sim\!13.8$). The intrinsic
color spread in $(H-[4.5\mu])_0$ is $\sim\!0.01\magunit$ for evolved
stars with $0.5 \lesssim (J-\ks)_0) \lesssim 0.8$
\citep{Majewski11a}. Therefore, typical random uncertainties in the
$A_{\ks}$ extinction corrections are $\sim\!0.05\magunit$, increasing
to $\sim\!0.10\magunit$ at the faint end. Variations in the adopted
extinction law among different lines of sight can lead to systematic
uncertainties of up to $3.5\,\%$ in $A_{\ks}$ \citep{Zasowski09a},
which are typically $0.005\magunit$ and $<0.023\magunit$ for all but
$1\,\%$ of our RC sample below. The contribution from the uncertainty
in the extinction correction therefore plays only a minor role in the
error budget of the RC distances that we determine below.

For each individual visit, line-of-sight velocities are measured
separately by cross-correlating against a set of $\approx$ 100
synthetic template spectra that span the stellar-parameter range
$3,500$ K $< T_{\mathrm{eff}} < 25,000$ K in effective temperature
$T_{\mathrm{eff}}$, $-2 < \feh < 0.3$ in metallicity \feh, and $2 <
\logg < 5$ in surface gravity \logg\ (see D.~Nidever \etal\ 2014, in
preparation, for further details; the reduction pipeline was also
briefly described in \citealt{Nidever12a}). The rms scatter in the
measured line-of-sight velocity for multiply-observed stars is
typically $0.1\kms$. Field-to-field variations indicate that the zero
point of the velocity scale is stable at the $0.13\kms$ level. A
comparison between the \apogee-measured line-of-sight velocity of 195
stars with literature data in the globular clusters M3, M13, M15, and
M92 shows that the \apogee\ zeropoint accuracy is
$\approx-0.95\pm0.05\kms$.

\begin{figure}[t!]
  \includegraphics[width=0.45\textwidth,clip=]{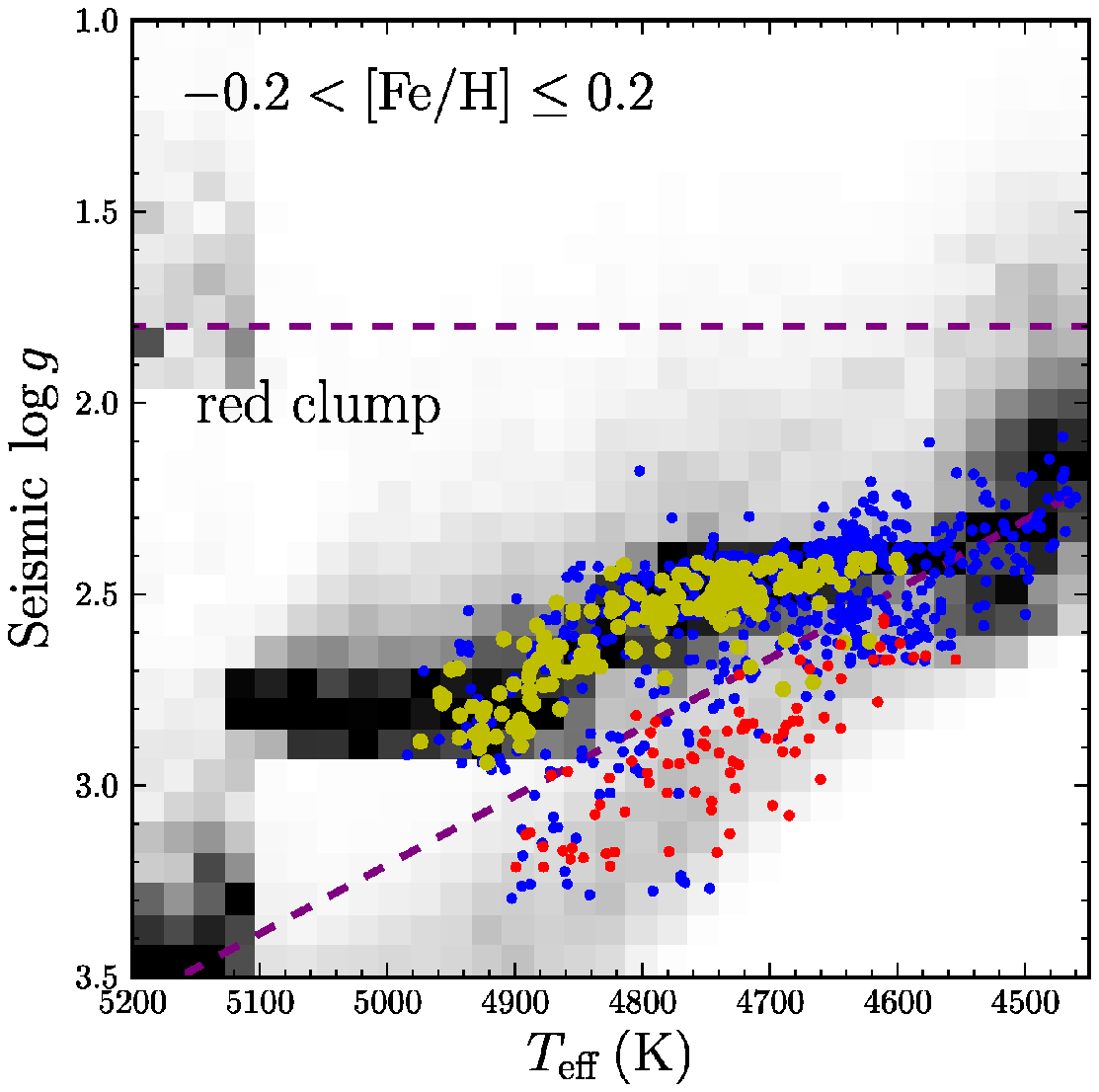}\\
  \includegraphics[width=0.45\textwidth,clip=]{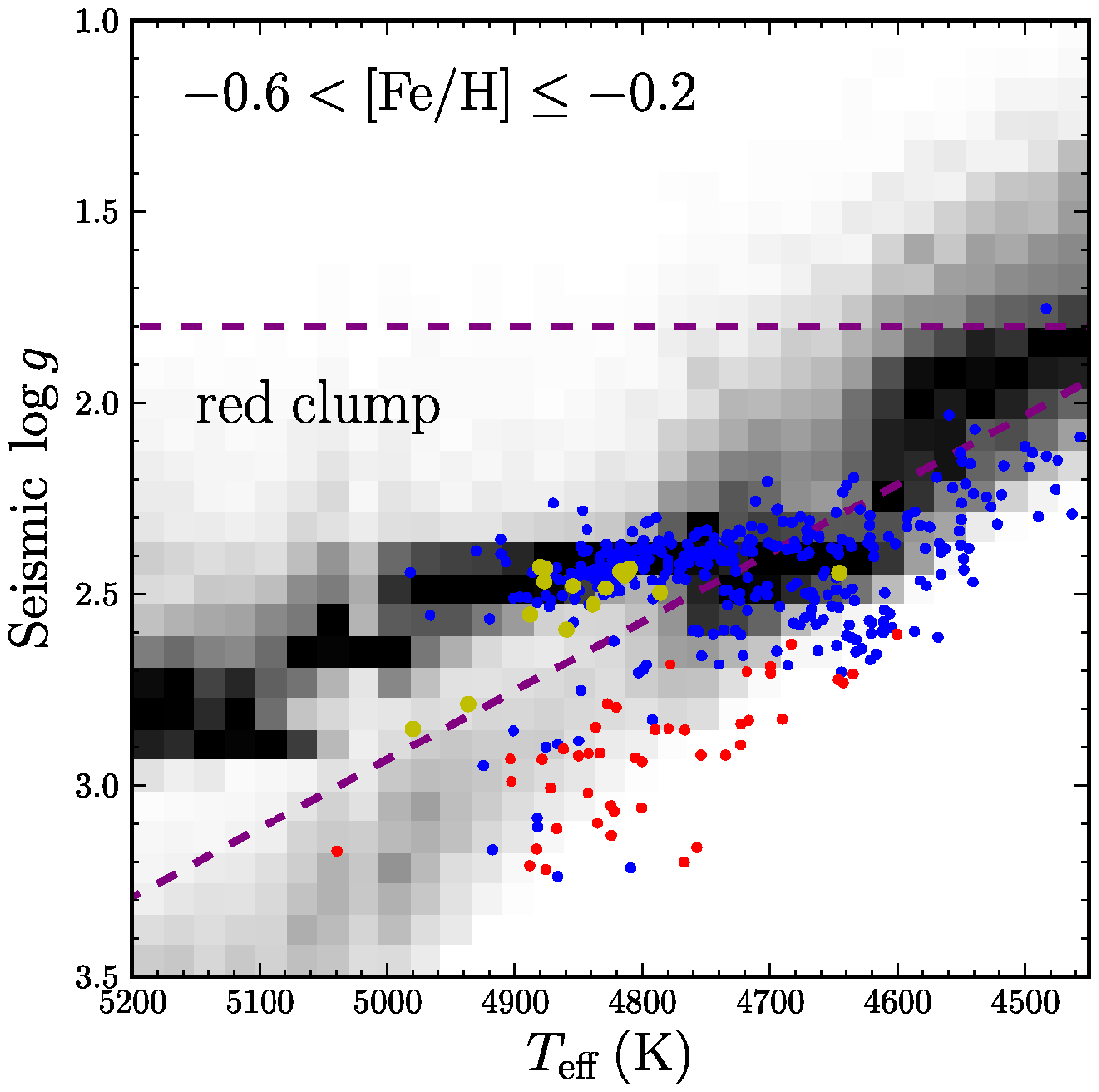}
  \caption{Prediction for the conditional distribution of stars in
    surface gravity given their effective temperature from the PARSEC
    stellar isochrones (see text; density grayscale) and data from the
    APOKASC catalog (colored points), in two metallicity bins. The
    prediction assumes a lognormal \citet{Chabrier01a} model for the
    IMF, a metallicity distribution that matches that of the APOKASC
    data used in each bin, and a constant SFH. The agreement between
    the models and the data is excellent in the region of the red
    clump. Yellow and red points represent stars classified by
    \citet{Stello13a} as RC and RGB stars, respectively, from their
    measured oscillation frequencies; blue points are stars for which
    the evolutionary state is unknown. The purple dashed lines
    represent our cuts in $\logg$ and $\teff$ to separate RC stars
    from less luminous red giant branch stars (see \equationname
    s~[\ref{eq:loggteffcut}] and [\ref{eq:tefffeh}]; evaluated at
    $\feh = 0\dex$ and $\feh = -0.4\dex$ in the top and bottom panel,
    respectively). Additional cuts in $(J-K_s)_0$ and metallicity are
    used to reduce contamination from higher luminosity giants and
    secondary-red-clump stars.}\label{fig:logg-apokasc}

\end{figure}

The APOGEE Stellar Parameter and Chemical Abundances Pipeline (ASPCAP)
extracts stellar parameters and elemental abundances from the
continuum normalized co-added spectra by performing a $\chi^2$
minimization with respect to a pre-computed multi-dimensional grid of
synthetic spectra derived from ATLAS9 model-atmosphere grids
(\citealt{Kurucz79a} and more recent updates) and spectral synthesis
calculations using ASS$\epsilon$T \citep{Koesterke08a,Koesterke09a};
ASPCAP is described in detail in A.~E.~Garc\'{\i}a~P\'{e}rez, 2014, in
preparation. The products of this procedure are best-fitting effective
temperatures $T_{\mathrm{eff}}$, surface gravities $\logg$,
metallicities $\feh$, $\alpha$- $\afe$, carbon- $[\mathrm{C/Fe}]$, and
nitrogen-enhancements $[\mathrm{N/Fe}]$. We find that the differences
with spherical MARCS \citep{Gustafsson08a} / Turbospectrum
\citep{Alvarez98a,Plez12a} spectra are below 5\,\% for the atmospheric
parameters typical of RC stars. Systematic offsets between the ASPCAP
results and literature values for stellar parameters in open and
globular clusters, as well as for seismic \logg\ derived from
asteroseismology of \emph{Kepler} stars observed by APOGEE, were
quantified by \citet{Meszaros13a}. For the purpose of this paper, we
use the corrected stellar parameters and abundances as determined from
this external comparison. Because the ASPCAP fit metallicity was
primarily calibrated against $\feh$ from high-resolution optical
spectroscopy, we refer to the metallicity as \feh. ASPCAP
uncertainties are typically $50$ to $100\,\mathrm{K}$ in \teff,
$0.2\dex$ in \logg, and $0.03$ to $0.08\dex$ in
\feh\ \citep{Meszaros13a}. For identifying a pure sample of RC stars
using the method described below, \logg\ in particular is
crucial. Using the APOKASC asteroseismology data described in more
detail below, we empirically determine the uncertainty in the
spectroscopic \logg\ to be $0.14\dex$ for APOGEE stars with $0.5
\lesssim (J-\ks)_0 \lesssim 0.8$ by comparison with the
highly-accurate seismic \logg. As discussed in more detail in
\sectionname~\ref{sec:contam}, part of this uncertainty is due to a
relative bias in \logg\ between the RC and the red-giant branch (RGB);
when this bias will be removed, we estimate from the seismic
\logg\ that the spectroscopic \logg\ uncertainty will be $0.1\dex$.
In addition to the main atmospheric parameters, the current ASPCAP
pipeline is able to determine elemental abundances for 15 individual
elements. However, the reliability of these still requires more
testing.

\subsection{RC selection and distance determination}\label{sec:kde}

In this section we describe a new method for selecting a sample of
likely red-clump giants from a combination of photometric and
spectroscopic data. This method is developed using isochrone models
calculated by stellar-evolution codes, primarily from the PARSEC
library \citep{Bressan12a}. To test whether these isochrone models are
in agreement with the parameters inferred from \apogee\ spectra, we
use data from the APOKASC catalog (M.~Pinsonneault \etal, 2014, in
preparation). This catalog contains stars in the \emph{Kepler} field
which have stellar parameters (mass, radius, and \logg) determined
from asteroseismology \citep[\eg,][]{Kallinger10a} and \teff\ and
elemental abundances determined by \apogee. The APOKASC asteroseismic
analysis uses spectroscopic \teff\ from APOGEE rather than \teff\ from
the Kepler Input Catalog \citep{Brown11a}. The combination of
\logg\ accurate to a few percent \citep{Hekker13a} and metallicity
measurements allows for a stringent test of the stellar-isochrone
models used here to define a red clump sample.

\begin{figure*}[t!]
\includegraphics[width=0.245\textwidth,clip=]{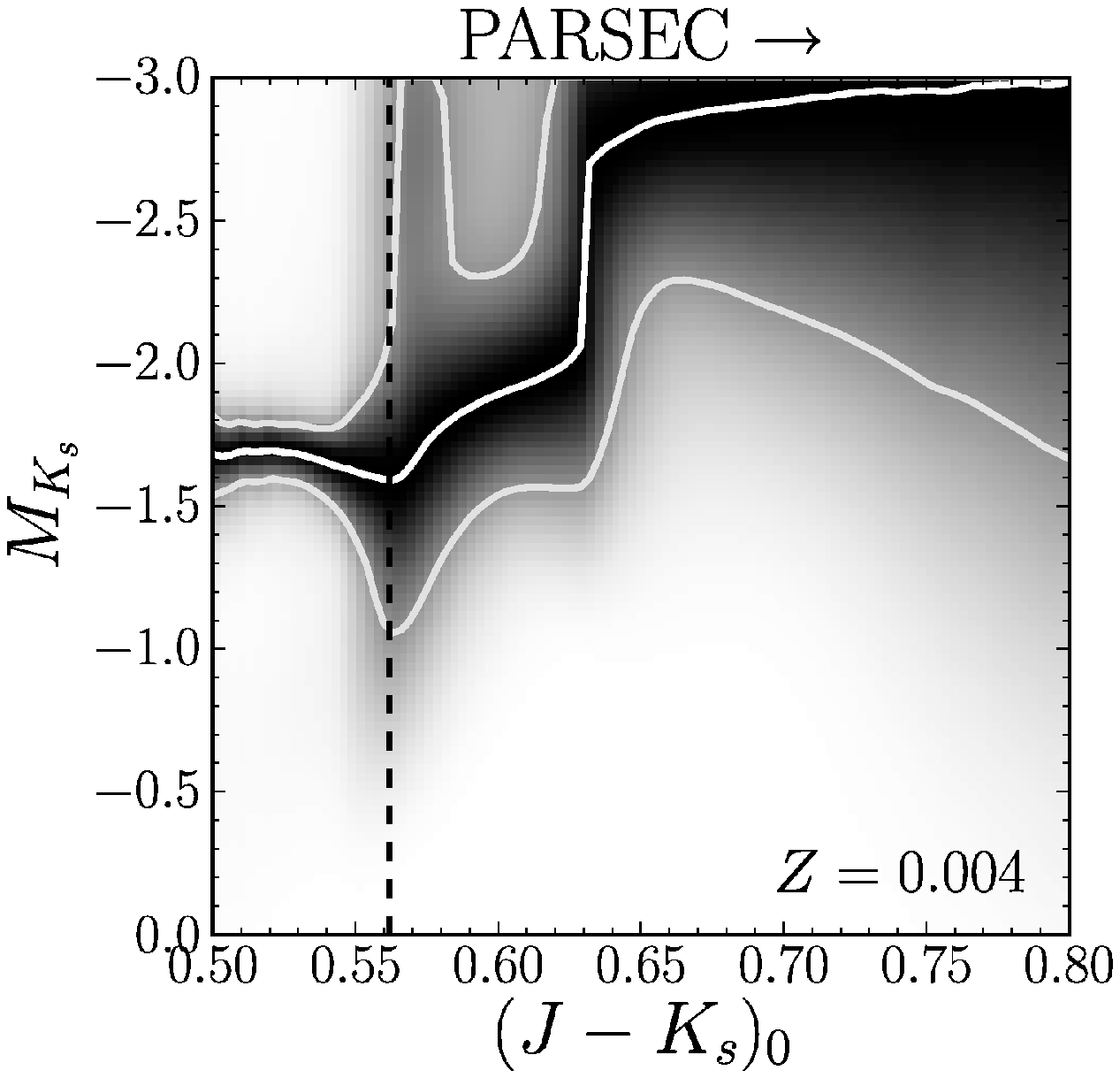}
\includegraphics[width=0.245\textwidth,clip=]{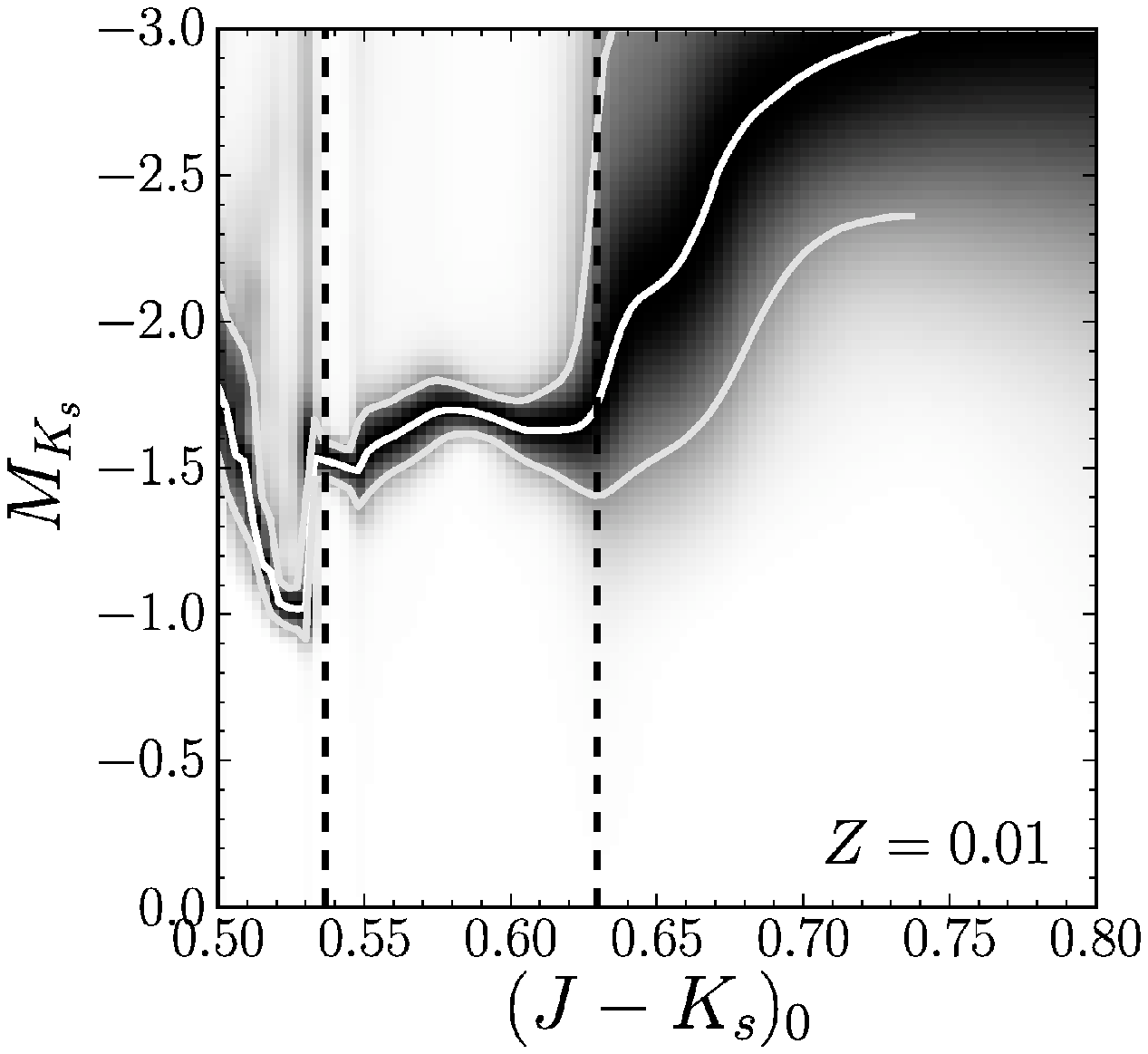}
\includegraphics[width=0.245\textwidth,clip=]{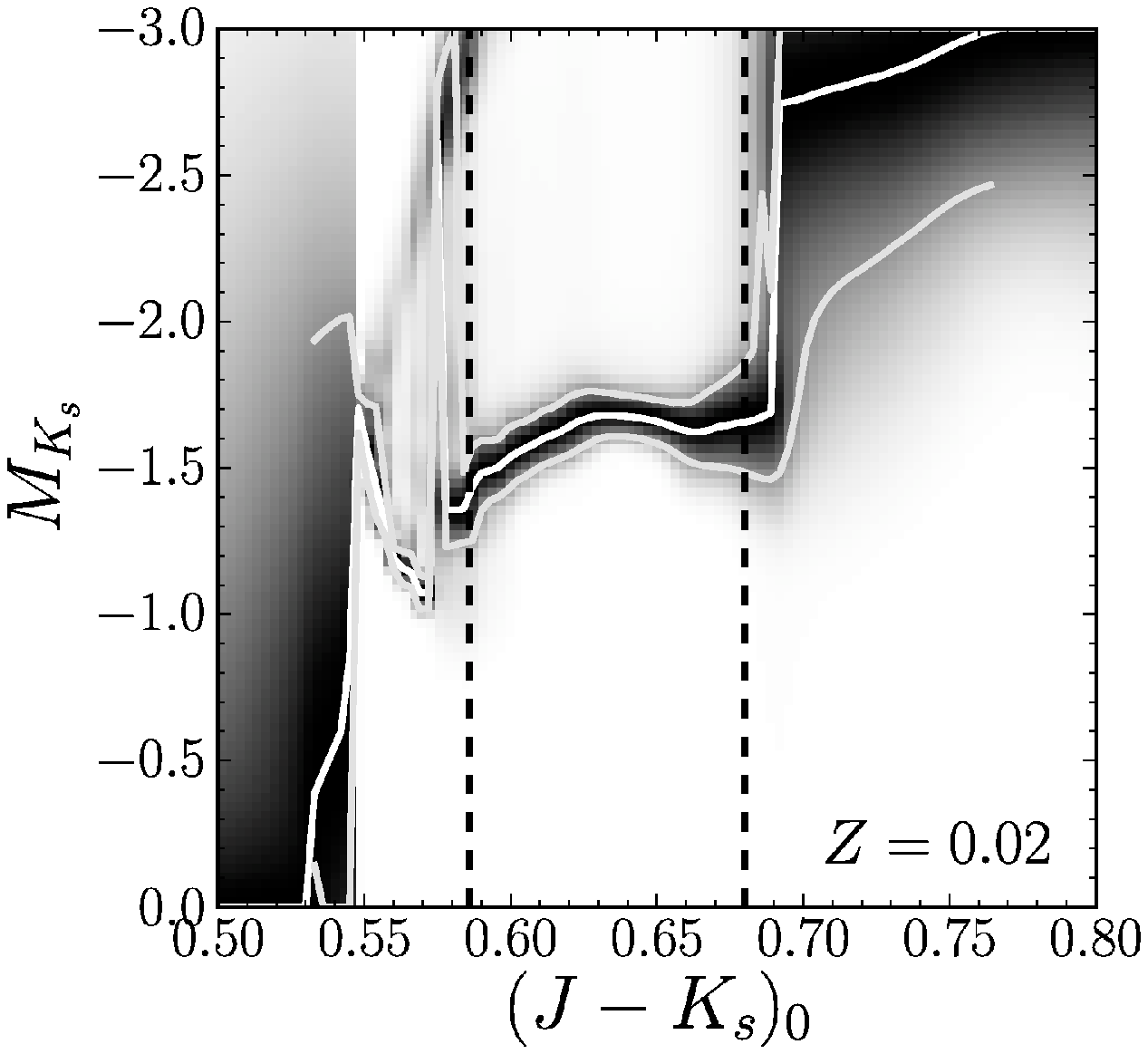}
\includegraphics[width=0.245\textwidth,clip=]{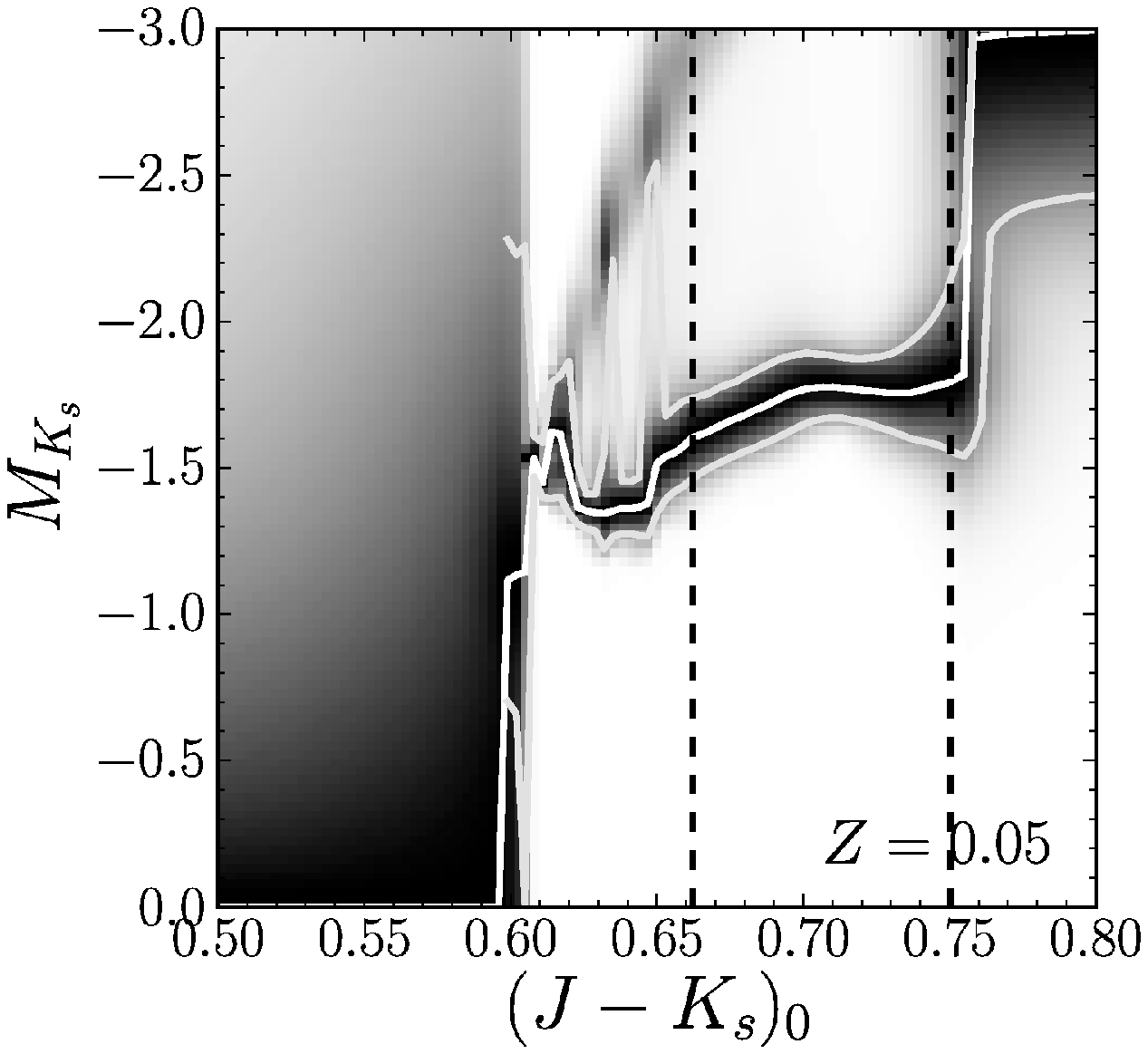}\\
\includegraphics[width=0.245\textwidth,clip=]{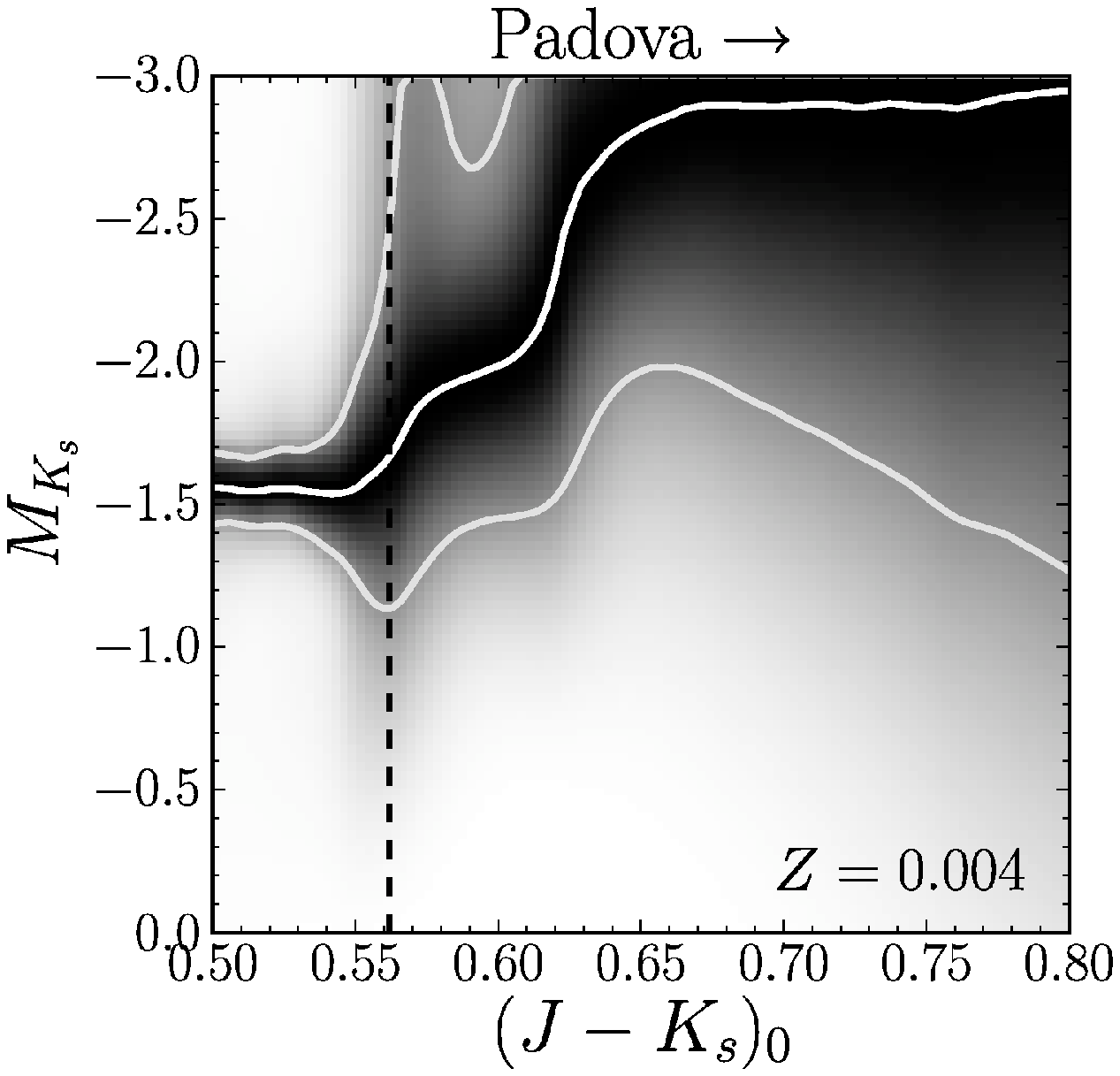}
\includegraphics[width=0.245\textwidth,clip=]{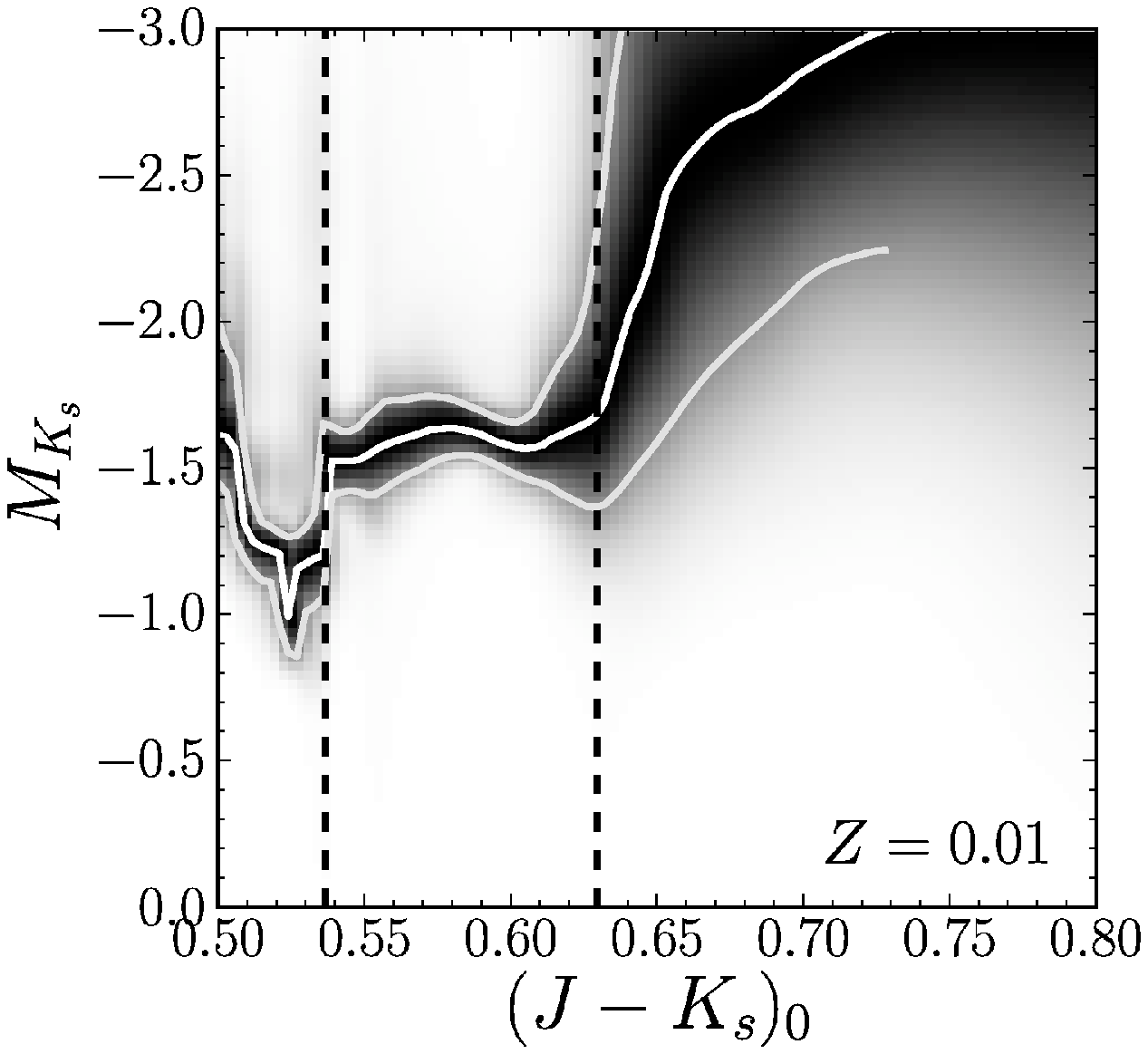}
\includegraphics[width=0.245\textwidth,clip=]{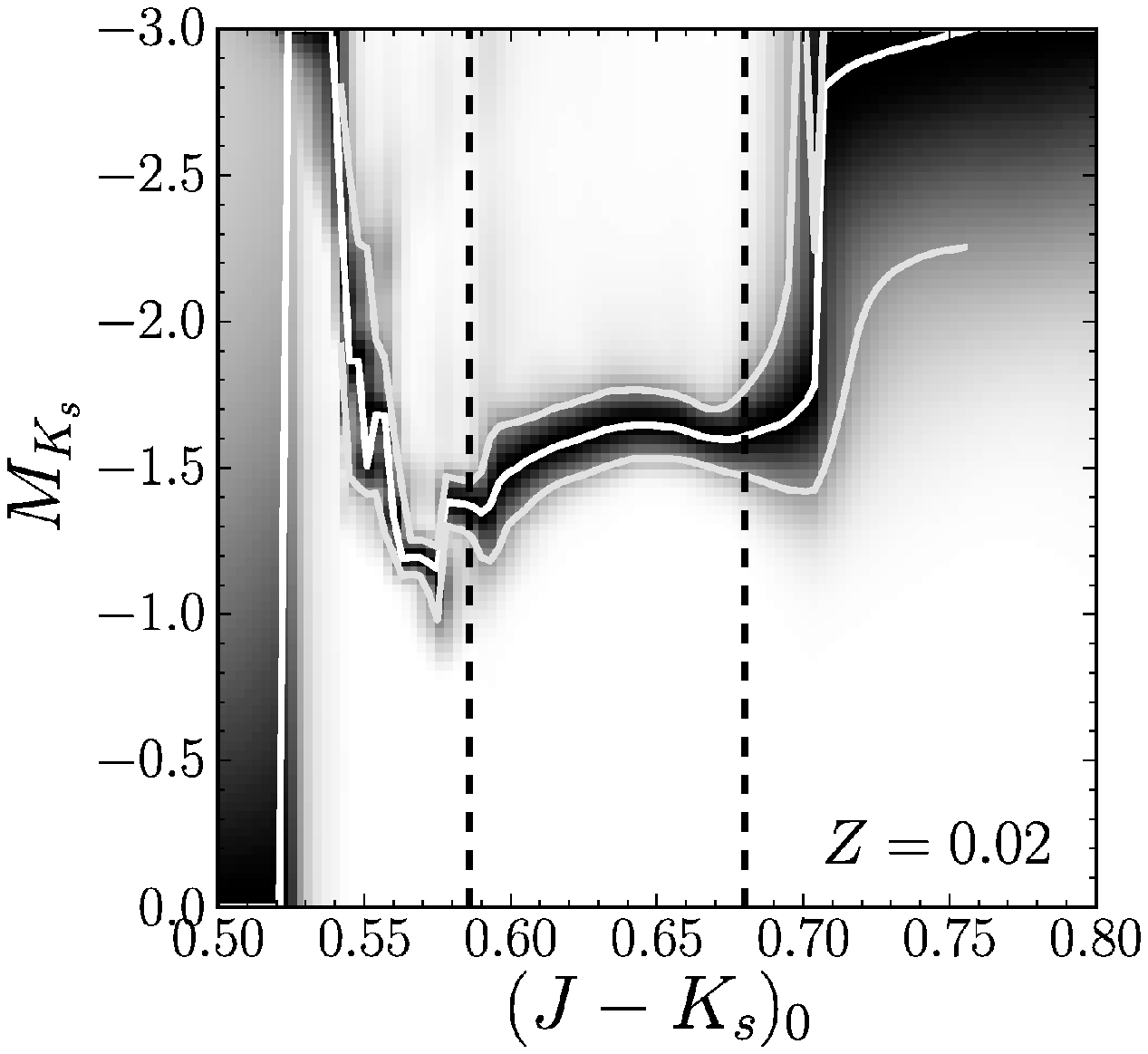}
\includegraphics[width=0.245\textwidth,clip=]{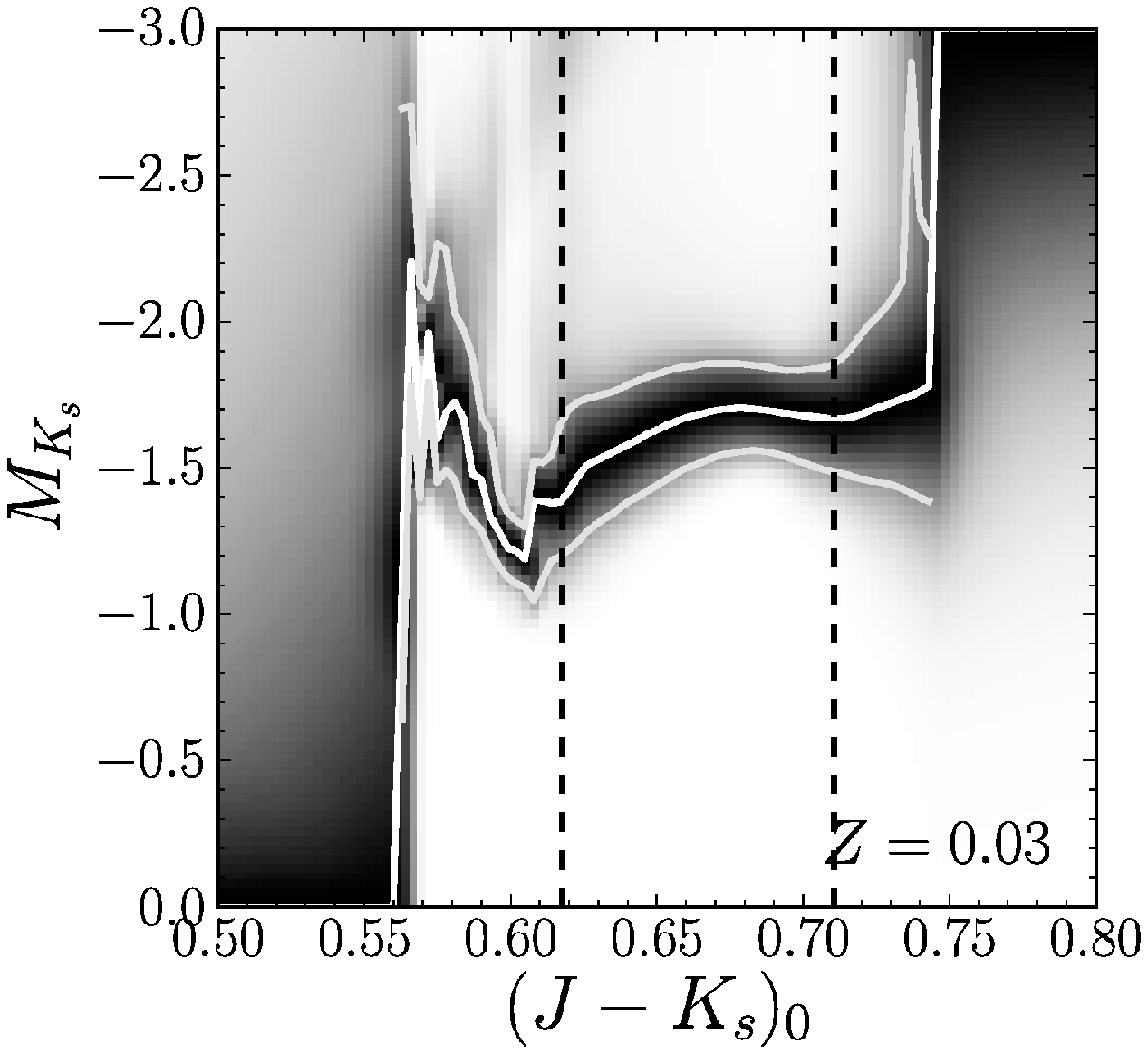}\\
\includegraphics[width=0.245\textwidth,clip=]{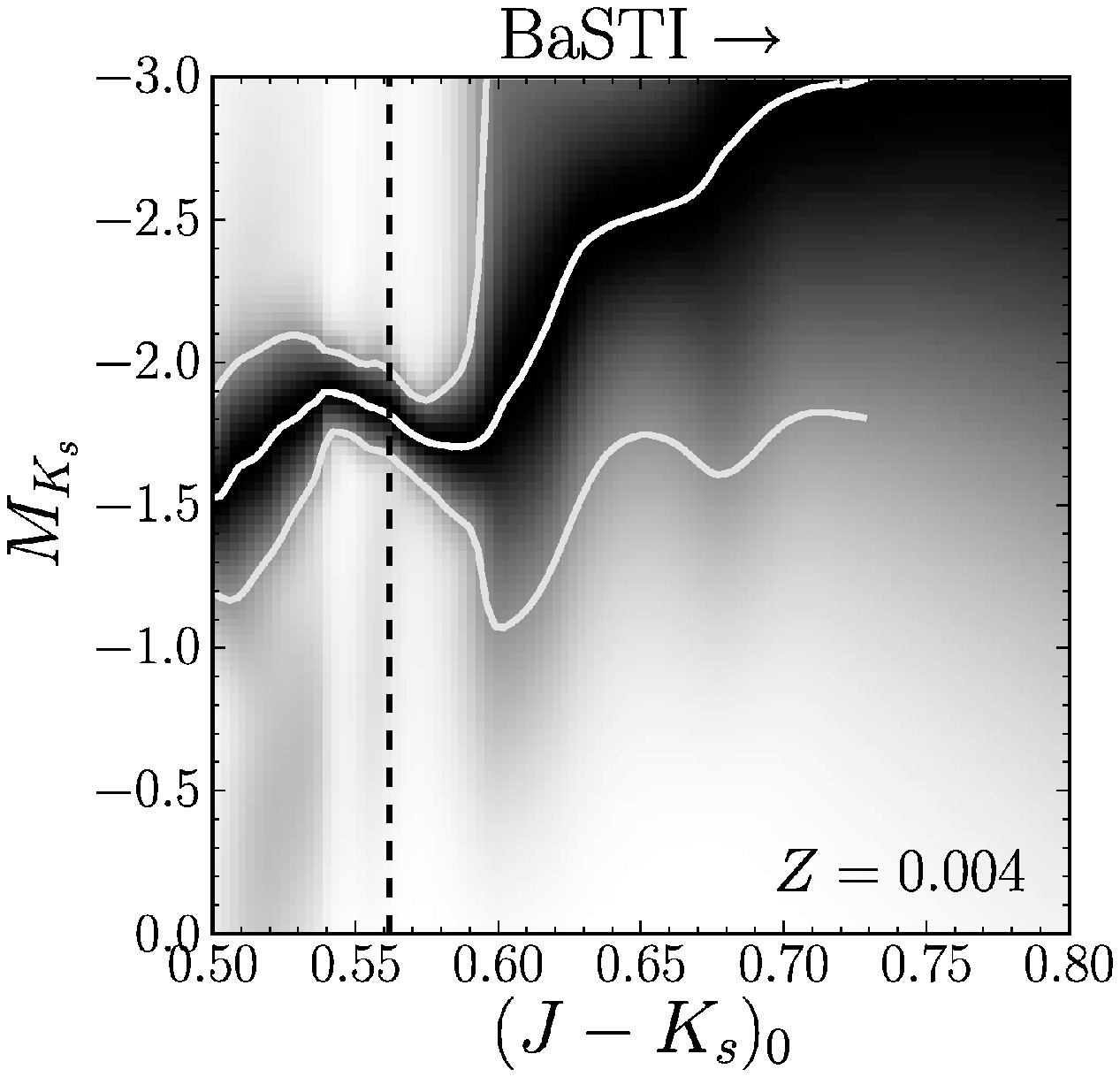}
\includegraphics[width=0.245\textwidth,clip=]{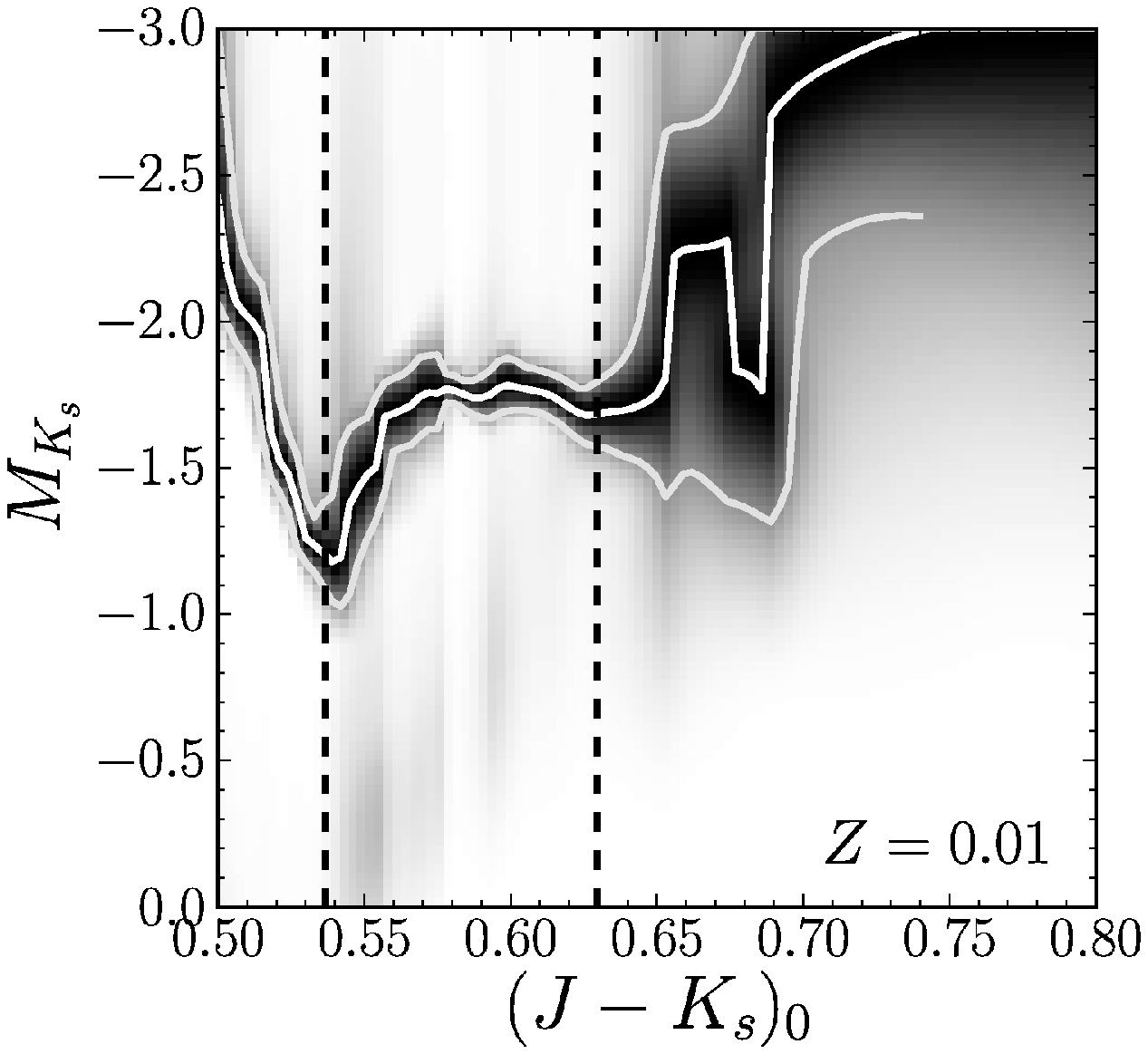}
\includegraphics[width=0.245\textwidth,clip=]{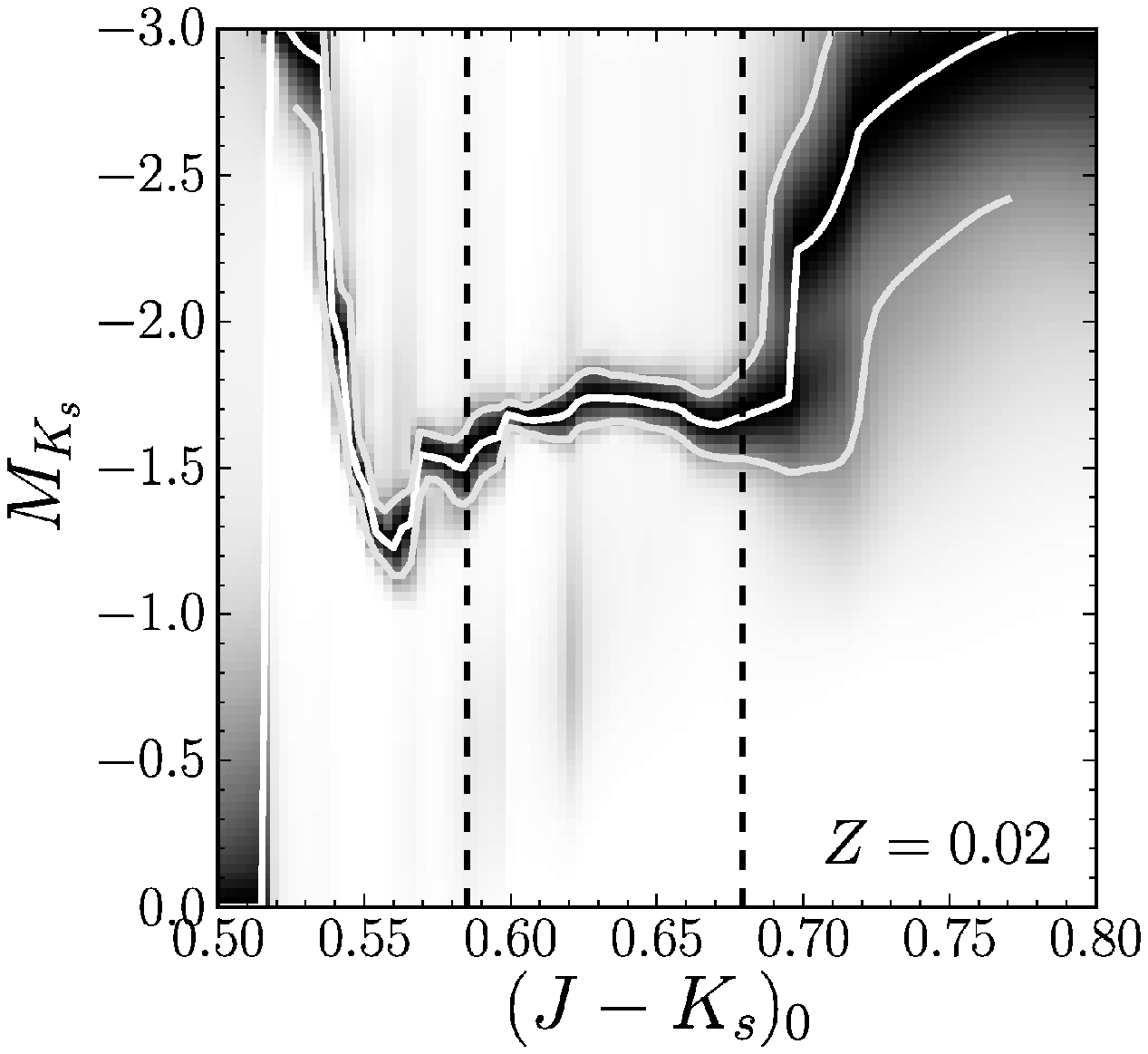}
\includegraphics[width=0.245\textwidth,clip=]{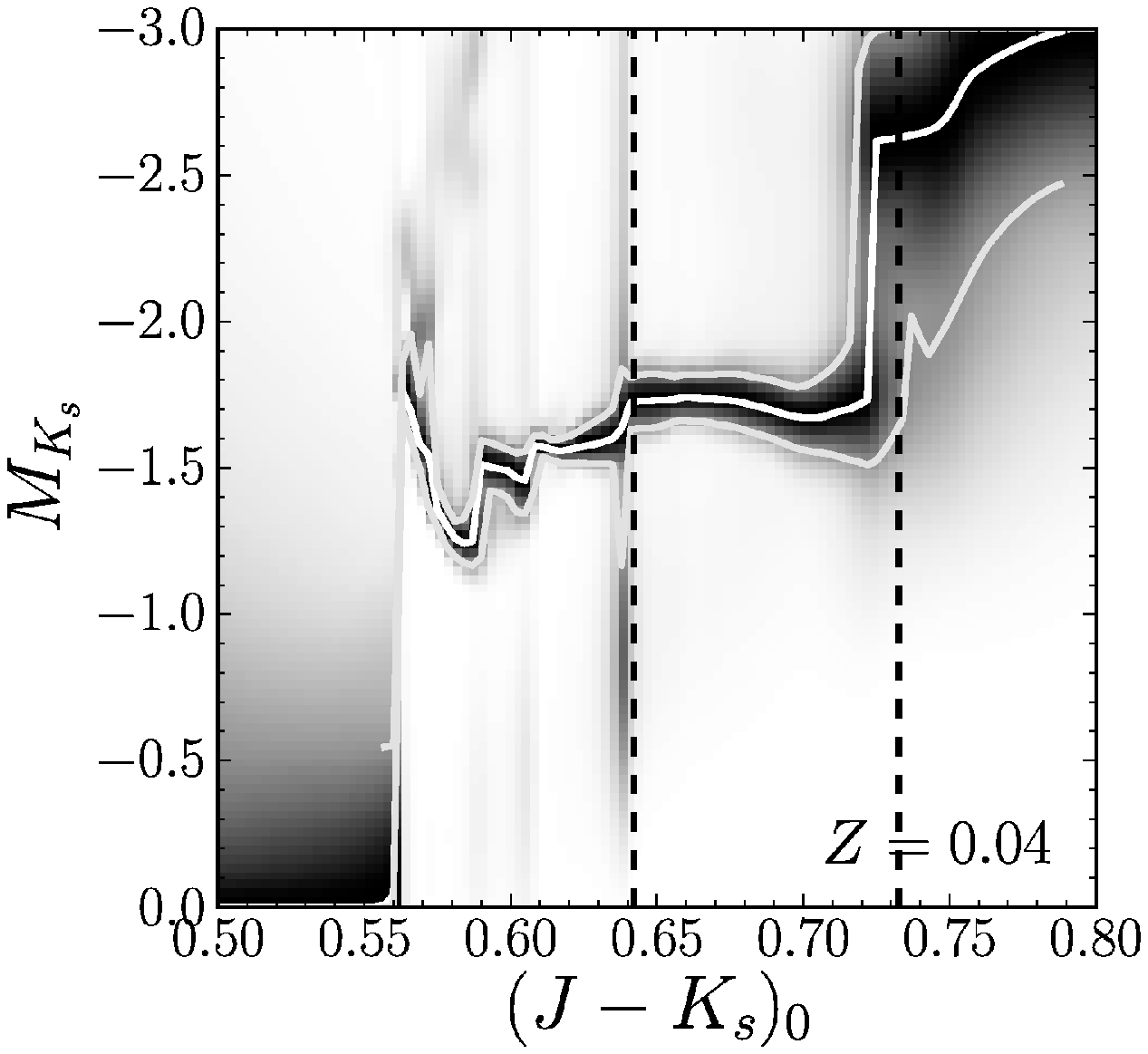}
\caption{Conditional PDF $p(M_{\ks}| [J-\ks]_0)$ for stars satisfying
  the cuts in \equationname s~(\ref{eq:loggteffcut}) and
  (\ref{eq:tefffeh}). This PDF is computed using the KDE method and
  SFH and IMF assumptions described in the text, for different
  metallicities (horizontally; the last column shows the highest
  metallicity Padova and BaSTI models and a high-metallicity PARSEC
  model) and different stellar-evolution codes (PARSEC, upper panels;
  Padova, middle panels; BaSTI, bottom panels). The middle white line
  marks the peak of the PDF and the outer white lines indicate where
  the PDF reaches half of its maximum. At each metallicity there is a
  color range where the PDF is extremely narrow. The vertical dashed
  lines in each panel show the color-cuts of \eqnname
  s~(\ref{eq:jkzcuts}) and (\ref{eq:jkzcuts3}) that isolate the range
  over which the magnitude distribution is narrow. The color range and
  morphology of this locus is similar for the different
  stellar-evolution codes and the dashed lines bound the locus with a
  narrow magnitude range for the three different
  codes.}\label{fig:mx_jkz}
\end{figure*}

\figurename~\ref{fig:logg-apokasc} compares the prediction from the
PARSEC stellar isochrones for the distribution of stars in effective
temperature \teff\ and surface-gravity \logg\ with the data in the
APOKASC catalog, in two different metallicity bins. The predicted
distribution assumes a lognormal \citet{Chabrier01a} model for the
initial mass function (IMF) and a metallicity distribution that
matches that of the APOKASC data used in each bin. For the
distribution of ages we assume a constant star-formation history
(SFH); throughout this paper we only use ages less than 10 Gyr. This
assumption does not strongly affect the locus of the RC and the RGB in
the \logg--\teff\ diagram. \figurename~\ref{fig:logg-apokasc} shows
excellent agreement between the predictions from the stellar-evolution
models and the data in the region of the red clump. Yellow and red
points in the figure are stars in the APOKASC catalog for which the
evolutionary state was measured by \citet{Stello13a} using the period
spacings of gravity-mode stellar oscillations derived with two
different methods \citep{Bedding11a,Mosser11a}. Yellow dots are stars
classified as RC stars, while red dots represent stars identified as
RGB stars. In the \logg--\teff\ plane, the PARSEC isochrone models
match the observed locus of RC and RGB stars, especially around solar
metallicity. We have also compared the stellar-evolution models with
the high-resolution spectroscopic \teff\ and \logg\ of the RC sample
of \citet{Valentini10a} and find similarly good agreement.

Motivated by the good agreement between the stellar isochrones and the
APOKASC data, we use the theoretical models to investigate the
morphology of the \rc, such that we can determine the range of
observed quantities for which the \rc\ is sufficiently narrow in
absolute magnitude to allow for a precise distance measurement for
individual stars. In addition to the PARSEC models, we use various
other stellar-isochrone libraries to determine the theoretical
uncertainty in the models; comparisons with these alternative
libraries are described in detail
below. \figurename~\ref{fig:logg-apokasc} demonstrates that red-clump
stars have $\logg$ between 2.4 and 2.9 and are comfortably separated
in \logg\ and \teff\ from the other main features in the CMD for stars
with precise \logg\ measurements. At high \logg, we adopt a sloping
cut in \logg\ as a function of \teff\ to separate the RC and RGB
branches. From the PARSEC isochrone models, we find that the slope
separating the RC and the RGB does not vary with metallicity, but the
intercept does. We will therefore select red-clump stars starting with
the cuts
\begin{equation}\label{eq:loggteffcut}
1.8 \leq \logg \leq 0.0018\,\mathrm{\dex\,K}^{-1}\,\Big(\teff-\teff^{\mathrm{ref}}(\feh)\Big)+2.5\,,
\end{equation}
where
\begin{equation}\label{eq:tefffeh}
\teff^{\mathrm{ref}}(\feh) = -382.5\,\mathrm{K\,\dex}^{-1}\,\feh+4607\,\mathrm{K}\,.
\end{equation}
It can be seen in \figurename~\ref{fig:logg-apokasc} that most of the
solar-metallicity RC is at $\logg = 2.5$ and $4600\,\mathrm{K}< \teff
< 4850\,\mathrm{K}$, but there is also a significant tail of such
stars extending up to $\teff = 4950\,\mathrm{K}$ and with
\logg\ increasing up to 2.9. These are the secondary-red-clump stars,
\ie, higher-mass helium-burning stars \citep{Girardi99a}, which are
slightly overrepresented in the APOKASC sample (see M.~Pinnsoneault et
al. 2014, in preparation). These stars pass the cuts specified in
\equationname s~(\ref{eq:loggteffcut}) and (\ref{eq:tefffeh}), even
though they are expected to be significantly fainter than the main red
clump (\ie, those RC stars at $\logg=2.5$).

\begin{figure}[t!]
\includegraphics[width=0.48\textwidth,clip=]{rcmodel_mode_jkz_ks_parsec_arxiv.eps}\\
\includegraphics[width=0.48\textwidth,clip=]{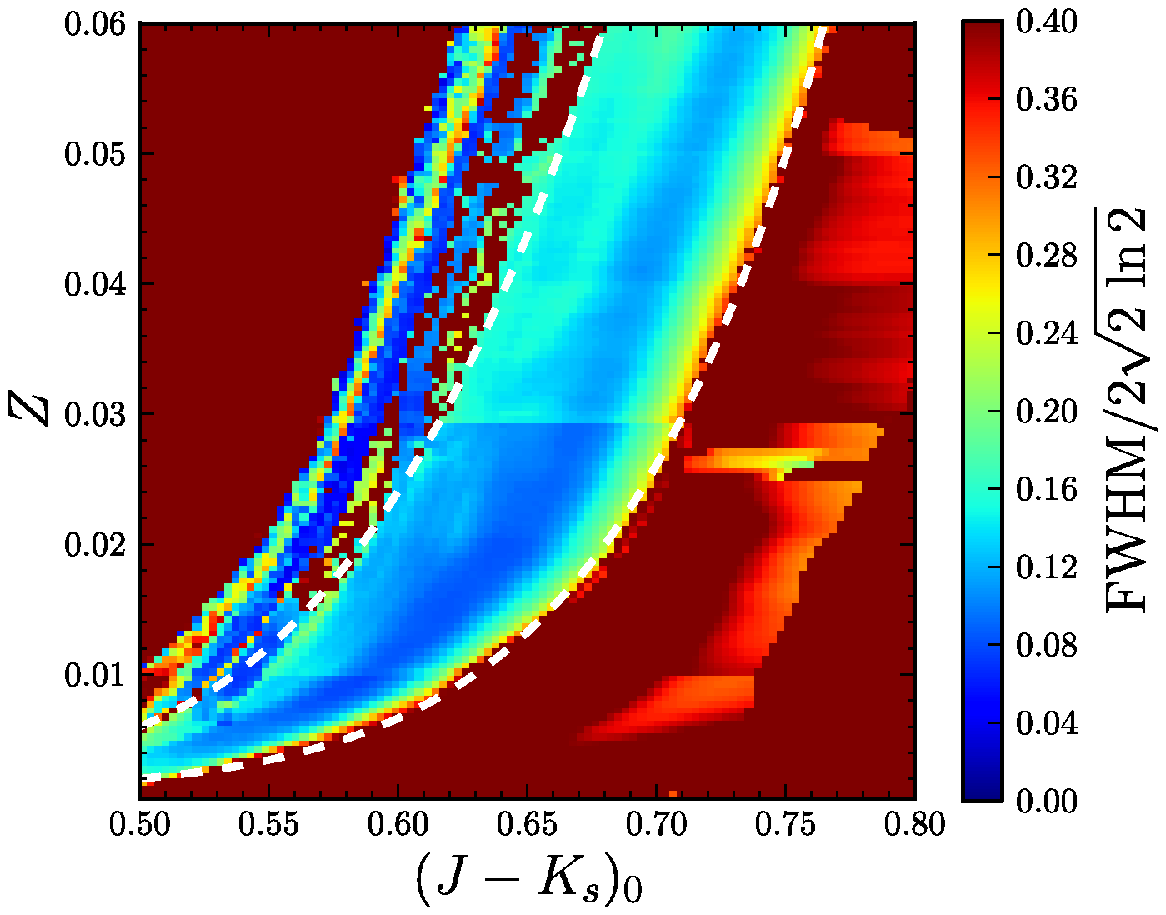}
\caption{Peak of the magnitude PDF $p(M_{\ks}| [J-\ks]_0)$ as a
  function of color and metallicity $Z$ for PARSEC isochrones (top
  panel). The bottom panel shows the FWHM/$2\sqrt{2\log2}$ ($\approx
  \sigma$) of the PDF. The white dashed lines represent the region
  specified by the cuts in \equationname s~(\ref{eq:jkzcuts}) and
  (\ref{eq:jkzcuts3}) over which the distribution of absolute
  magnitudes is narrow. The peak of the magnitude PDF does not
  strongly depend on color or metallicity over the region where the
  PDF is narrow.}\label{fig:sig_jkz}
\end{figure}

\begin{figure*}[t!]
\includegraphics[width=0.48\textwidth,clip=]{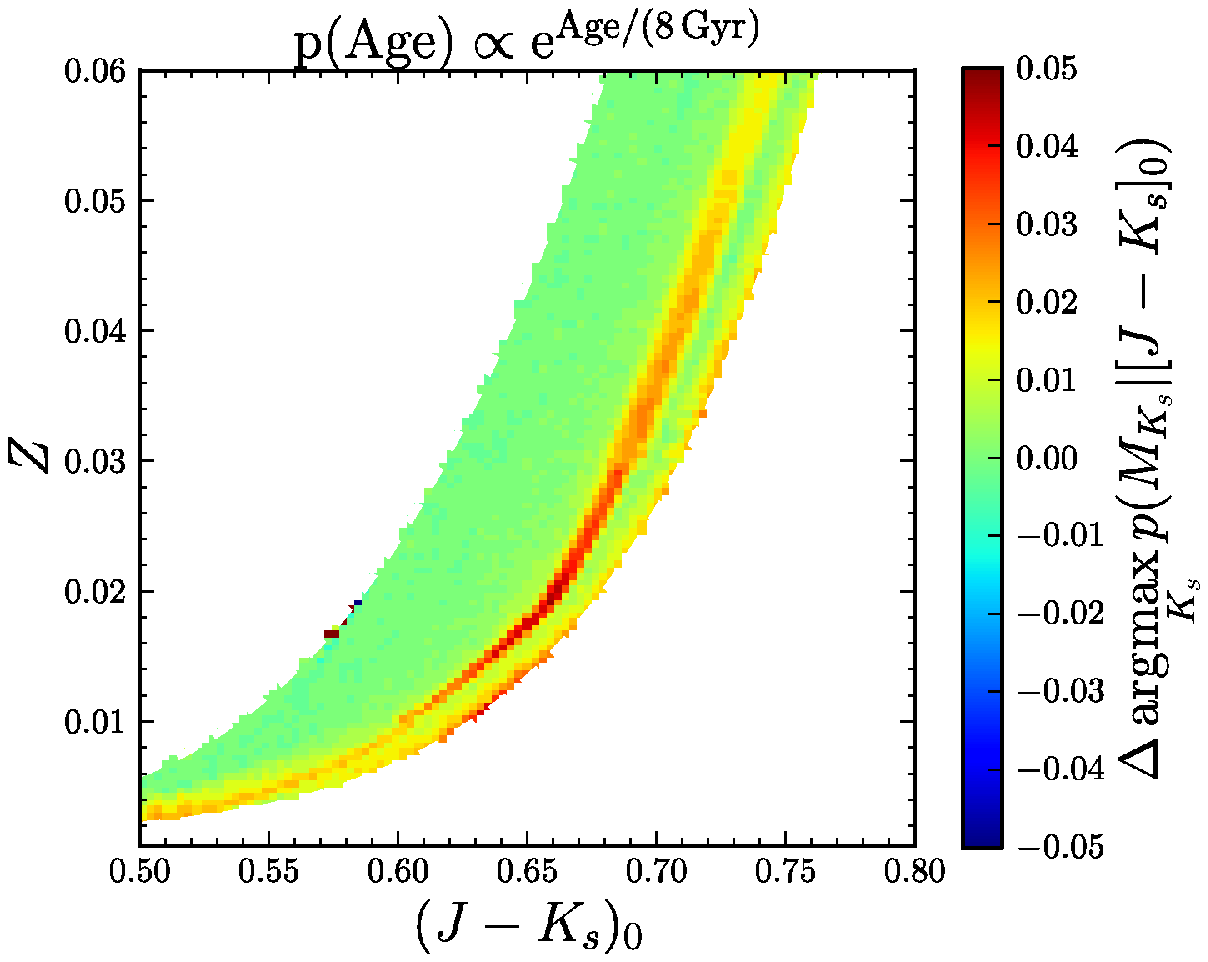}
\includegraphics[width=0.48\textwidth,clip=]{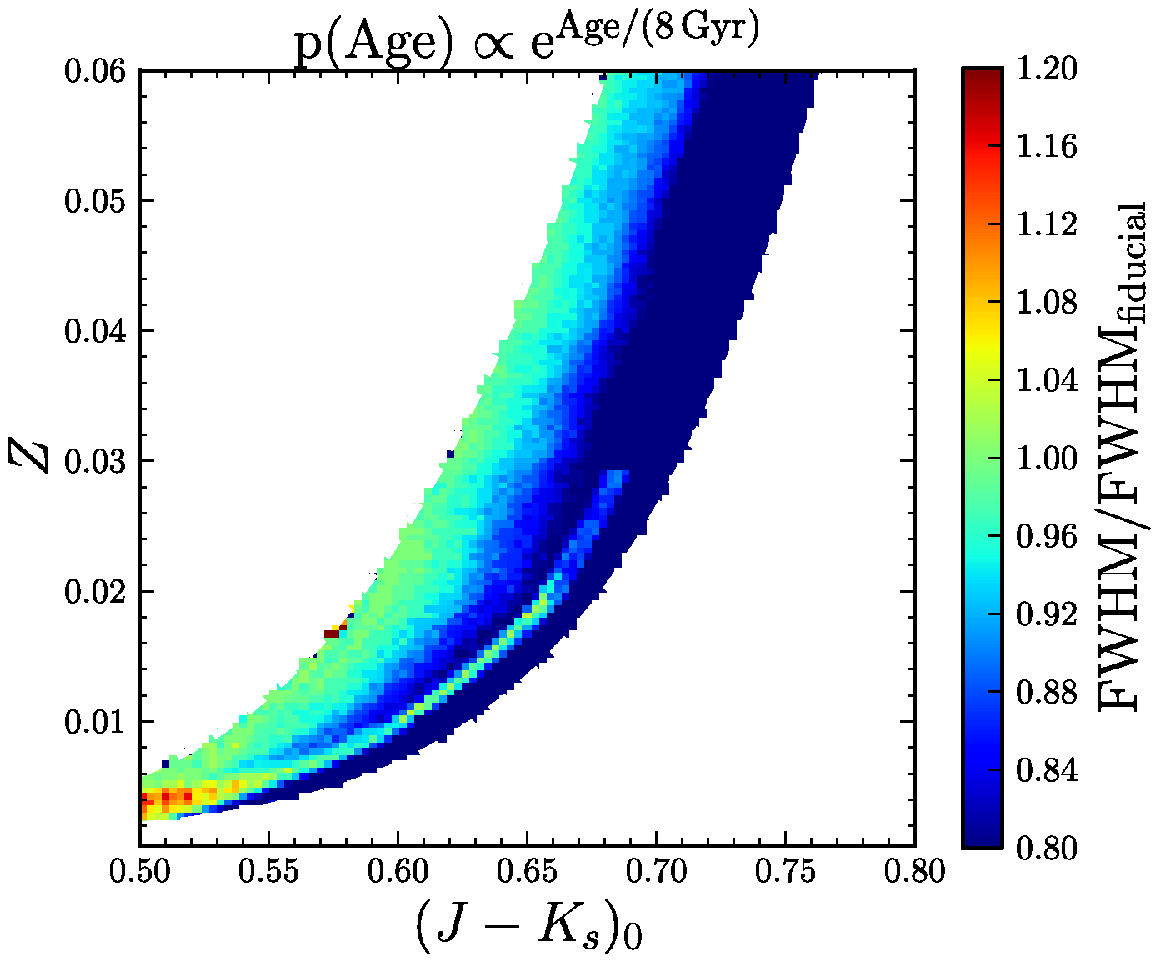}\\
\includegraphics[width=0.48\textwidth,clip=]{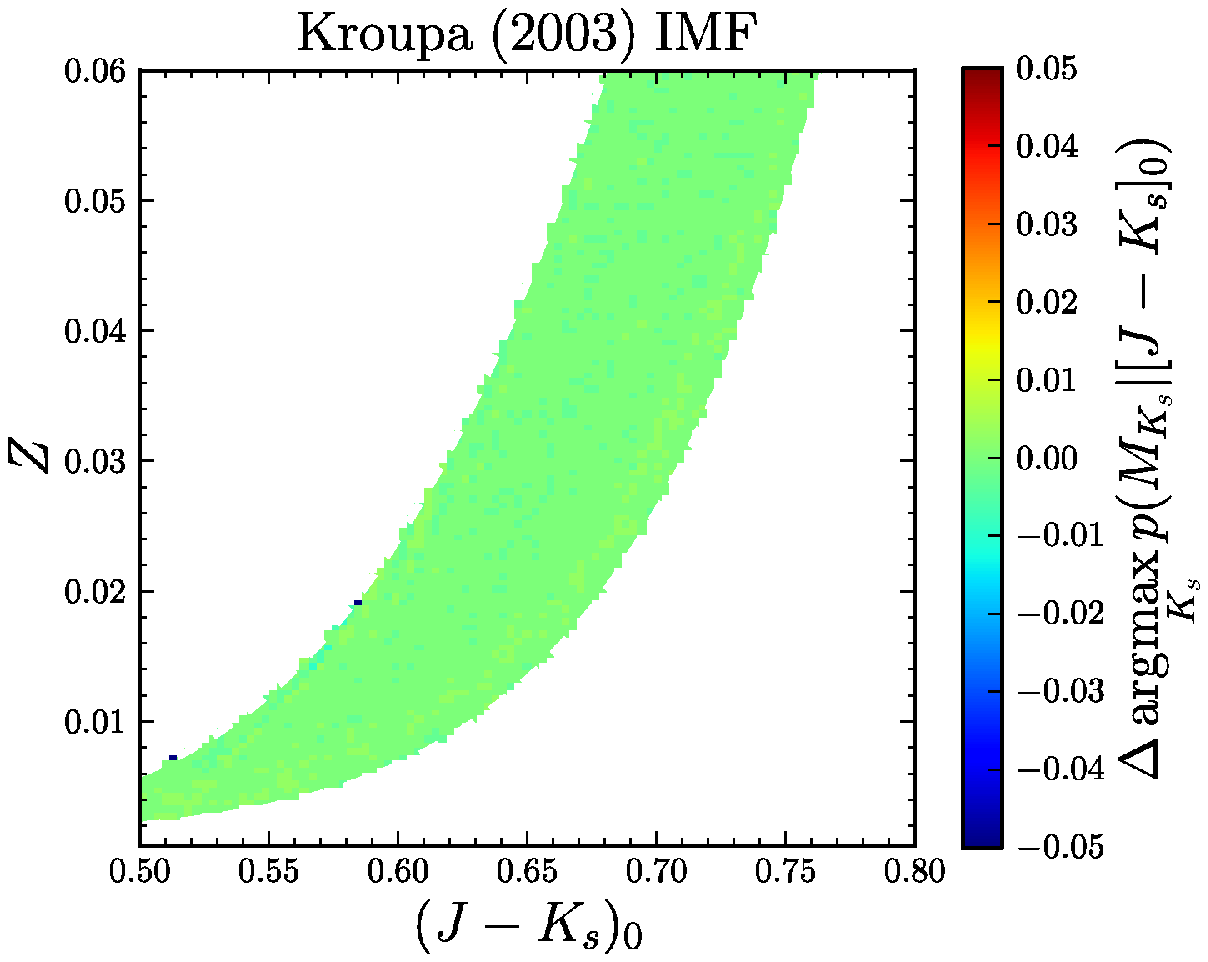}
\includegraphics[width=0.48\textwidth,clip=]{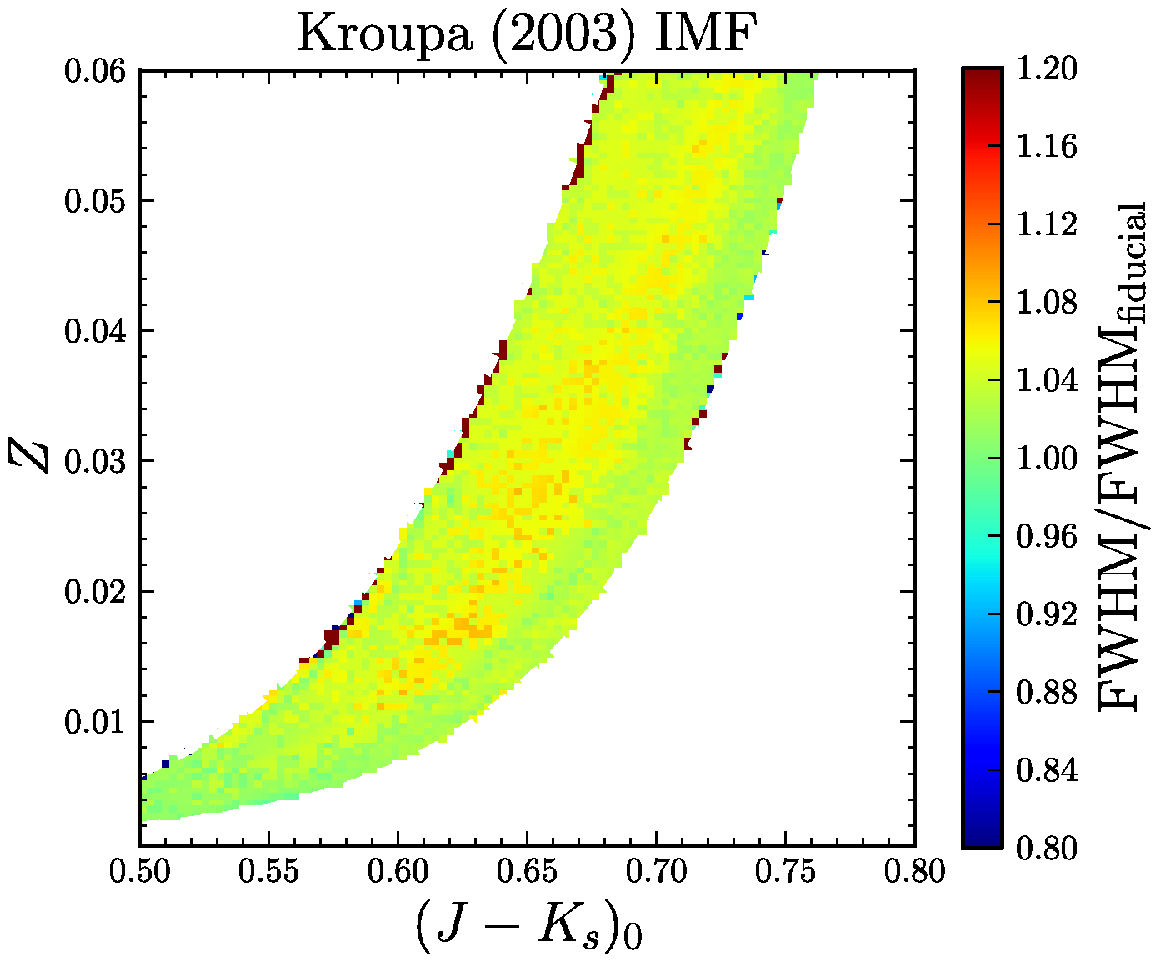}
\caption{Variation of the peak and width of the magnitude PDF
  $p(M_{\ks}| [J-\ks]_0)$ for an exponentially-declining
  star-formation rate or a \citet{Kroupa03a} IMF. This figure shows
  the difference (absolute for the peak magnitude, relative for the
  width) between the results obtained using these alternative
  assumptions and our fiducial model (lognormal \citealt{Chabrier01a}
  IMF and constant SFH). We only display the comparison in the RC
  region defined by the cuts in \equationname~ s(\ref{eq:jkzcuts}) and
  (\ref{eq:jkzcuts3}). The deviations in the mean magnitude are
  typically $\lesssim 0.02\magunit$ or $\lesssim 1\,\%$ in
  distance. The distribution of absolute magnitudes has the same small
  width to within $\approx10\,\%$ regardless of the IMF or SFH
  chosen.}\label{fig:sig_jkz_diff}
\end{figure*}

To further remove contamination from lower \logg\ RGB and
secondary-red-clump stars, and to isolate RC stars for which precise
distances can be derived, we further investigate the
color--magnitude--metallicity distribution. We use the dereddened
$(J-\ks)_0$ color rather than \teff, because of concerns about
systematic biases in the latter in the measurement and in the stellar
models. The $(J-\ks)_0$ color is sensitive to errors in the
dereddening, but from the discussion in \sectionname~\ref{sec:apogee}
it is clear that these are primarily random errors. Because the RC
dominates the star counts at apparent magnitudes probed by APOGEE in
the range $0.5 \lesssim (J-\ks)_0 \lesssim 0.8$ and because the RC
luminosity is only a weak function of color (see below), random
dereddening errors do no lead to significant contamination or distance
errors.

We construct a model for the color--magnitude distribution (CMD)
$(J-\ks)_0$ vs. $M_{\ks}$ for \rc-like stars in the PARSEC isochrones
for a given metallicity $Z$; to increase the theoretical sampling, we
average the isochrones for metallicities $Z, Z-0.0005$, and
$Z+0.0005$. We apply the above cut in \logg, \teff, and \feh\ to the
isochrones, which are sampled at equally-spaced initial masses. We
then create a density model by applying a kernel-density estimation
(KDE) technique with a spherical kernel (after normalizing the color
and magnitude by their standard deviations
$\sigma_{J-\ks}$\footnote{In the discussion of the KDE modeling of the
  CMD we always assume that $J-\ks$ is dereddened, but we do not add
  the superscript `0' to avoid notational clutter.} and
$\sigma_{\ks}$, respectively) that varies with position to account for
the strongly varying density in the color--magnitude plane
\citep{Silverman86a}. Thus, the density at a point $\vec{x} \equiv
(\frac{J-\ks}{\sigma_{J-\ks}},\frac{M_{\ks}}{\sigma_{\ks}})$ is
evaluated as
\begin{equation}
p(\vec{x}) \propto \sum_i{\frac{w_i}{\lambda_i^2}\,K\left(\frac{\vec{x}-\vec{x}_i}{h\,\lambda_i}\right)}\,,
\end{equation}
where $i$ indexes points $\vec{x}_i$ on the isochrones, and $K(\cdot)$
is the biweight kernel $K(\vec{r}) \propto (1-r^2)^2$ for $r \equiv
|\vec{r}| \leq 1$ and zero otherwise. The weights $w_i$ are calculated
using a lognormal \citet{Chabrier01a} model for the IMF and a constant
SFH. The parameter $h$ is the kernel size, set to $N^{-1/5}$, where
$N$ is the number of training points, following ``Scott's rule'' (for
one-dimensional data, as we are primarily interested in the
one-dimensional distribution $p(M_{K_s}|J-\ks)$; \citealt{Scott92a}),
and $\lambda_i$ is a local bandwidth factor, calculated as
\begin{equation}
  \lambda_i = \left(\frac{\hat{p}(\vec{x}_i)}{s}\right)^\alpha\,,
\end{equation}
where $\hat{p}(\vec{x})$ is an estimate of the density distribution,
$s$ is the (straight) geometric mean of the density over $\vec{x}_i$:
$\log s = \sum_i \log \hat{p}(\vec{x}_i) / N$, and $\alpha = 0.5$. We
iterate three times to calculate the weights $\lambda_i$, starting
from uniform $\lambda_i$.

The resulting CMD is shown in the top panels of
\figurename~\ref{fig:mx_jkz} for four values of the overall
metallicity $Z$. The CMD in this figure has been normalized
independently for each color to produce the conditional probability
distribution function (PDF); this approach demonstrates the dependence
of the RC absolute magnitude on color at a given metallicity. The
lines show the peak and half-maxima as a function of color. In these
diagrams, the interval between the dashed lines where the
absolute-magnitude PDF is extremely sharply peaked corresponds quite
well to the color interval for which, at that given metallicity,
He-burning stars are old enough (roughly $>1.5\Gyr$) to be in a
compact, ``classical'' RC; this feature naturally results from all
these He-burning stars having developed a degenerate core of similar
mass earlier in their evolution as RGB stars. To the blue of this
interval, the mean magnitude abruptly falls by $\approx0.4\magunit$,
and becomes far from constant at even bluer colors; this is
essentially the behavior expected from secondary-red-clump stars,
which have ignited He in non-degenerate conditions
\citep{Girardi99a}. The magnitude distribution at even bluer colors,
and at colors redder than the reddest dashed line, becomes
significantly more extended and featureless; those intervals no longer
reflect the behavior of He-burning stars, but rather those of
main-sequence, RGB, and asymptotic-giant-branch (AGB) stars.

We calculate the theoretical CMD in this manner for metallicities on a
grid with spacing $\Delta Z = 0.0005$ and determine the peak $M_{\ks}$
and FWHM as a function of color. The result is presented in
\figurename~\ref{fig:sig_jkz}, where the FWHM has been converted into
the equivalent Gaussian standard deviation $\sigma$. This figure
clearly shows the region in the $(J-\ks)_0$ and $Z$ plane where the
magnitude PDF has $\sigma \lesssim 0.1\magunit$, i.e., where distances
precise to $5\,\%$ can be determined spectro-photometrically. However,
it is clear that this locus is strongly dependent on metallicity. The
dashed lines approximately bound the low-$\sigma$ locus; they are
given by
\begin{align}\label{eq:jkzcuts}
Z & > 1.21\, [(J-\ks)_0-0.05]^9 + 0.0011\\
Z & < 2.58\,[(J-\ks)_0-0.40]^3 + 0.0034\,,
\label{eq:jkzcuts3}\end{align}
with additional bounds of
\begin{equation}\label{eq:jkzcuts2}
  Z \leq 0.06\,, \qquad \qquad (J-\ks)_0 \geq 0.5\,.
\end{equation}
The width of the magnitude distribution near the edges of this region
increases to $0.2\magunit$, which still yields $10\,\%$ precision in
the derived distances. As discussed in \sectionname~\ref{sec:apogee},
the contribution to the distance uncertainty from errors in the
extinction correction is typically $0.05\magunit$, smaller than the
intrinsic spread in the RC luminosity. \Eqnname
s~(\ref{eq:jkzcuts})--(\ref{eq:jkzcuts2}) are highly effective at
eliminating secondary-red-clump, RGB, and AGB stars from the sample.

\figurename~\ref{fig:sig_jkz} also demonstrates that the peak
magnitude of the PDF does not depend strongly on either color or
metallicity, with a minimum-to-maximum variation of about
$0.2\magunit$. We have also computed the surface in
\figurename~\ref{fig:sig_jkz} assuming an exponentially-declining
star-formation rate or a \citet{Kroupa03a} IMF. The difference with
respect to our fiducial assumptions is presented in
\figurename~\ref{fig:sig_jkz_diff}. This figure shows that there are
only minor differences within the region of our color and metallicity
cuts, especially for the peak of the magnitude PDF. Assumptions about
the SFH or IMF therefore do not introduce appreciable systematics,
with offsets typically less than $1\,\%$ in distance.

We have also computed the density near the \rc\ using the Padova
\citep{Girardi00a,Bonatto04a} and BaSTI \citep{Pietrinferni04a}
isochrone libraries using the same IMF and SFH assumptions. For the
latter, we combine the transformations of \citet{Carpenter01a} and
\citet{Bessel88a} to transform the BaSTI $K$ magnitudes to 2MASS
\ks\ magnitudes. The conditional density $p(M_{\ks} | [J-\ks]_0)$ for
four different metallicities for these alternative isochrone models is
also shown in \figurename~\ref{fig:mx_jkz}. It is clear that the
Padova densities are similar to those obtained from the PARSEC
isochrones, especially in the region where the color--magnitude locus
is narrow. A detailed comparison between the peak and FWHM of the
magnitude PDF shows that the Padova and PARSEC isochrones give the
same color--metallicity region over which the \rc\ is narrow; the
color-dependence of the peak magnitude is the same for the two models
(the Padova isochrones' maximum metallicity is $Z = 0.03$).

For the BaSTI isochrones we only use the values $Z =
[0.004,0.008,0.01,0.0198,0.03,0.04]$ provided by BaSTI and do not use
adjacent metallicity bins to increase the model sampling, as we do for
Padova and PARSEC models (the BaSTI models are sampled for a much
larger number of initial masses). \figurename~\ref{fig:mx_jkz} shows
that the \rc\ locus at high metallicity is similar to that obtained
from Padova and PARSEC models.

Giants on the RGB are affected by mass loss, the details of which are
not completely understood. Mass loss is a few times $0.01\msun$ for
giants with ages $\lesssim 5\Gyr$ and so is largely negligible, but
older giants can loose a few times $0.1\msun$
\citep[\eg,][]{Reimers75a,Schroeder05a}. However, mass loss
essentially only affects the mass in the stellar envelope and the core
mass that ignites He is unaffected, such that the effect on the RC
luminosity is small. The PARSEC models use the Reimers law for mass
loss with an efficiency parameter of $\eta=0.2$
\citep[\eg,][]{Miglio12a}, while the Padova and BaSTI models use
$\eta=0.4$. The good agreement among the different stellar isochrones
in the RC region therefore explicitly demonstrates that the details of
mass-loss prescriptions are unimportant. Additionally, we have
computed the peak magnitude of the RC for $\eta = 0.1$ and $\eta=0.3$
(roughly the range allowed by \citealt{Miglio12a}) and find that the
magnitude solely changes at the red end of the RC and only by
$<0.02\magunit$. The effect on the RC distances is therefore
negligible.

\begin{figure}[t!]
\includegraphics[width=0.48\textwidth,clip=]{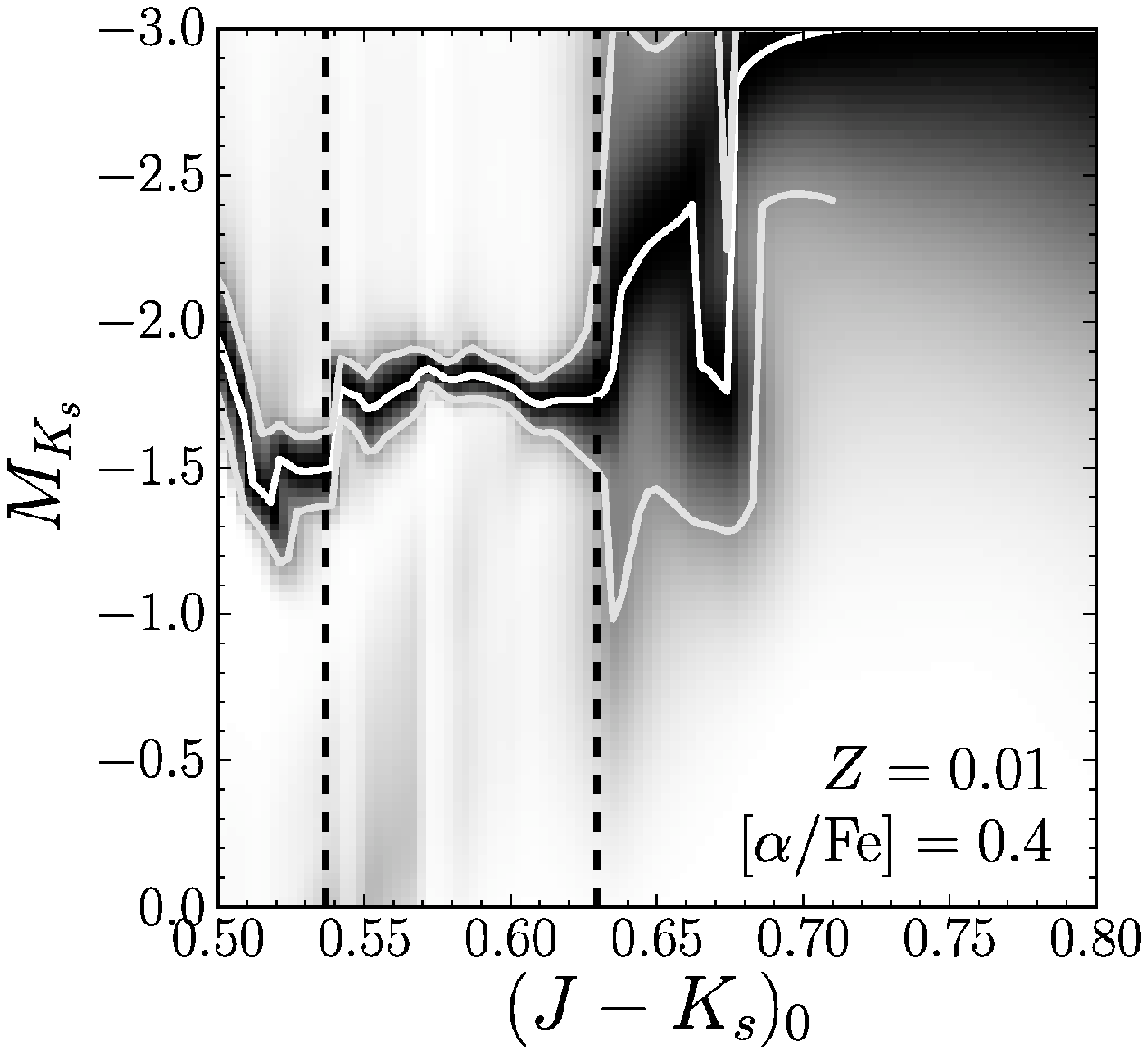}\\
\includegraphics[width=0.48\textwidth,clip=]{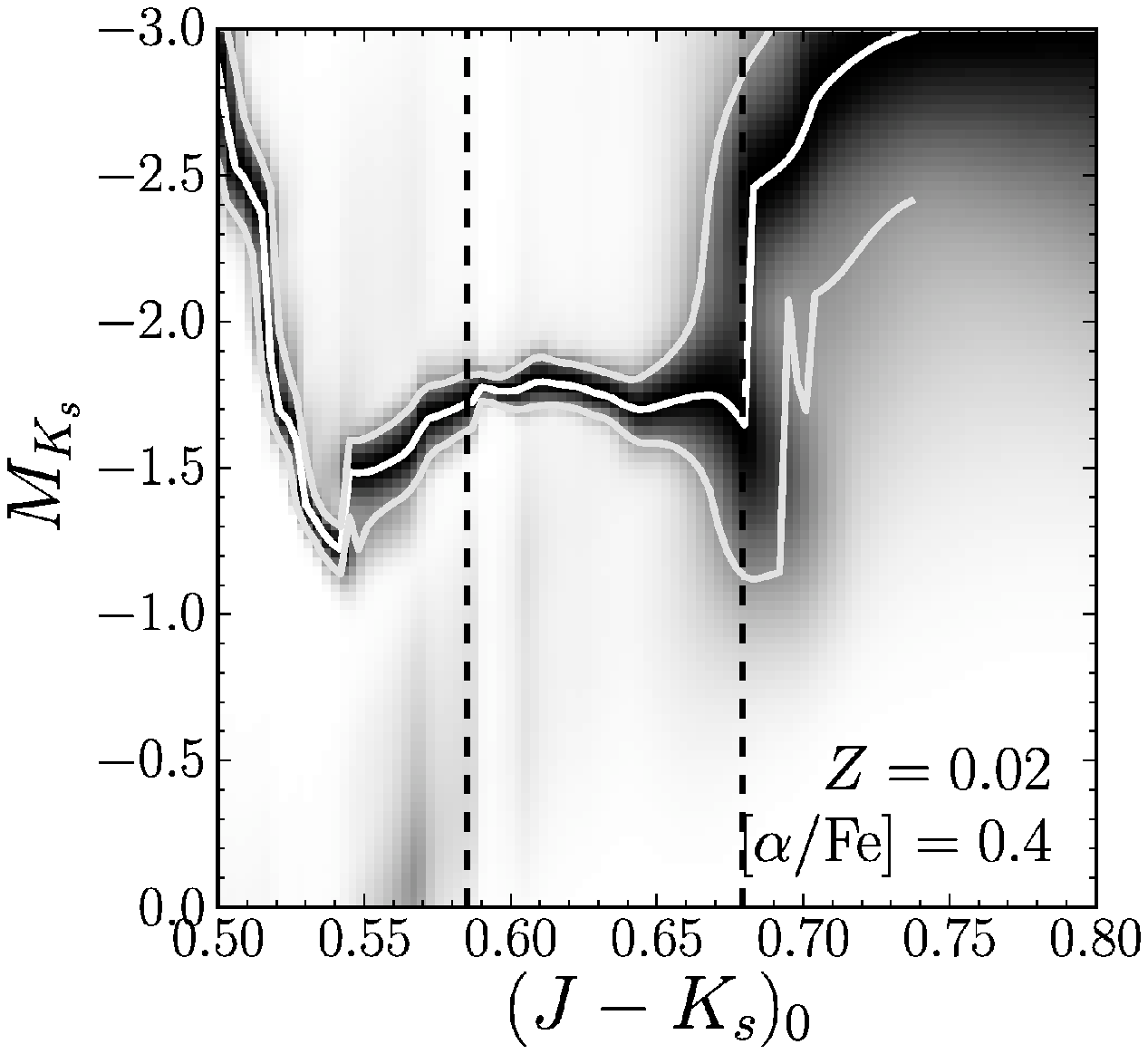}
\caption{Effect of $\alpha$-enhancement on the RC locus. This figure
  shows the same conditional PDF $p(M_{\ks}| [J-\ks]_0)$ for stars
  satisfying the cuts in \equationname s~(\ref{eq:loggteffcut}) and
  (\ref{eq:tefffeh}) for BaSTI models as in
  \figurename~\ref{fig:mx_jkz}, except that the models have $\afe =
  0.4$. The iron abundance \feh\ in these models is $\approx 0.35\dex$
  lower than for the equivalent non-$\alpha$-enhanced models. The
  dashed lines are the cuts in \equationname s~(\ref{eq:jkzcuts}) and
  (\ref{eq:jkzcuts3}) calculated based on the total metallicity
  $Z$. The RC locus is shifted approximately $0.025\magunit$ blueward
  for $\alpha$-enhanced models compared to solar-scaled models with
  the same overall metallicity.}\label{fig:afe}
\end{figure}

The models that we used in this section to characterize the RC locus
are all solar-scaled models. One of the main motivations of APOGEE is
to explore elemental-abundance ratios beyond the overall metallicity
and therefore we need to understand how the RC locus changes when
abundances are not solar scaled. In \figurename~\ref{fig:afe}, we
present the conditional PDF $p(M_{\ks}| [J-\ks]_0)$ for stars
satisfying the cuts in \equationname s~(\ref{eq:loggteffcut}) and
(\ref{eq:tefffeh}) for two models that are enhanced in $\alpha$
elements with respect to the solar ratios. We employ BaSTI models with
$\afe = 0.4$ \citep{Pietrinferni06a} with an overall metallicity of
$Z=0.01$ and $Z=0.02$; \feh\ for these models is about $0.35\dex$
lower than for the equivalent solar-scaled models. These
$\alpha$-enhanced models are moved blueward because of the decreasing
Fe opacity for $\alpha$-enhanced models at fixed overall
metallicity. In the relevant metallicity range for $\alpha$-enhanced
stars ($-1 \lesssim \feh \lesssim 0$), the blueward shift is
$\approx0.025\magunit$.

The metallicity presented in the APOGEE catalog that we use for the
basis of the RC selection is calibrated to \feh\ rather than the
overall metallicity. Therefore, when we calculate $Z(\feh)$ below for
applying the cuts in \equationname s~(\ref{eq:jkzcuts}) and
(\ref{eq:jkzcuts3}), we are underestimating the overall metallicity
for $\alpha$-enhanced stars. The bias in $\log Z$ is approximately
$0.35\dex \left(\afe/0.4\dex\right)$. Therefore, the
(color,metallicity) cut for a true RC stars at a given
$([J-\ks]_0,\log Z)$ will be evaluated at the wrong $([J-\ks]_0,\log
Z-0.35\dex)$ and we might spuriously remove some of the reddest RC
stars, because the RC cuts shift blueward for lower metallicity
stars. This is problematic as the red part of the RC is where the
oldest stars (ages $\gtrsim5\Gyr$) are located and these old stars are
the ones that could be $\alpha$-enhanced. However, the blueward shift
of the true RC locus with \afe\ means that we do not loose many
stars. The red cut in \equationname~(\ref{eq:jkzcuts}) has a slope in
$([J-\ks]_0,\log Z)$ of $\approx 6\dex\magunit^{-1}$. From inspecting
the isochrone models for old stars, we find that our cuts at low $Z$
are such that old stars are typically selected $\approx0.02\magunit$
bluer than the red cut of \equationname~(\ref{eq:jkzcuts}). Therefore
we do not remove true RC stars with $\afe \lesssim 0.25\dex$. Only
$8\,\%$ of stars with $\feh > -1$, $\teff > 4200\,\mathrm{K}$ and
$\afe > 0.05$ have $\afe > 0.25$ in the first two years of APOGEE
data. Therefore, \afe-bias in the RC selection at high \afe\ is small.

The different isochrone models do not entirely agree on the absolute
magnitude of the \rc, with PARSEC isochrones having a near-uniform
offset of $\sim\!0.1\magunit$ and $\sim\!0.05\magunit$ with respect to
the Padova and BaSTI models, respectively. This result shows the need
to externally calibrate the brightness of the \rc. However, the offset
is small and would only lead to systematics of order 5\,\% or less in
distance.

\subsection{Contamination by RGB stars}\label{sec:contam}

\begin{figure}[t!]
\includegraphics[width=0.48\textwidth,clip=]{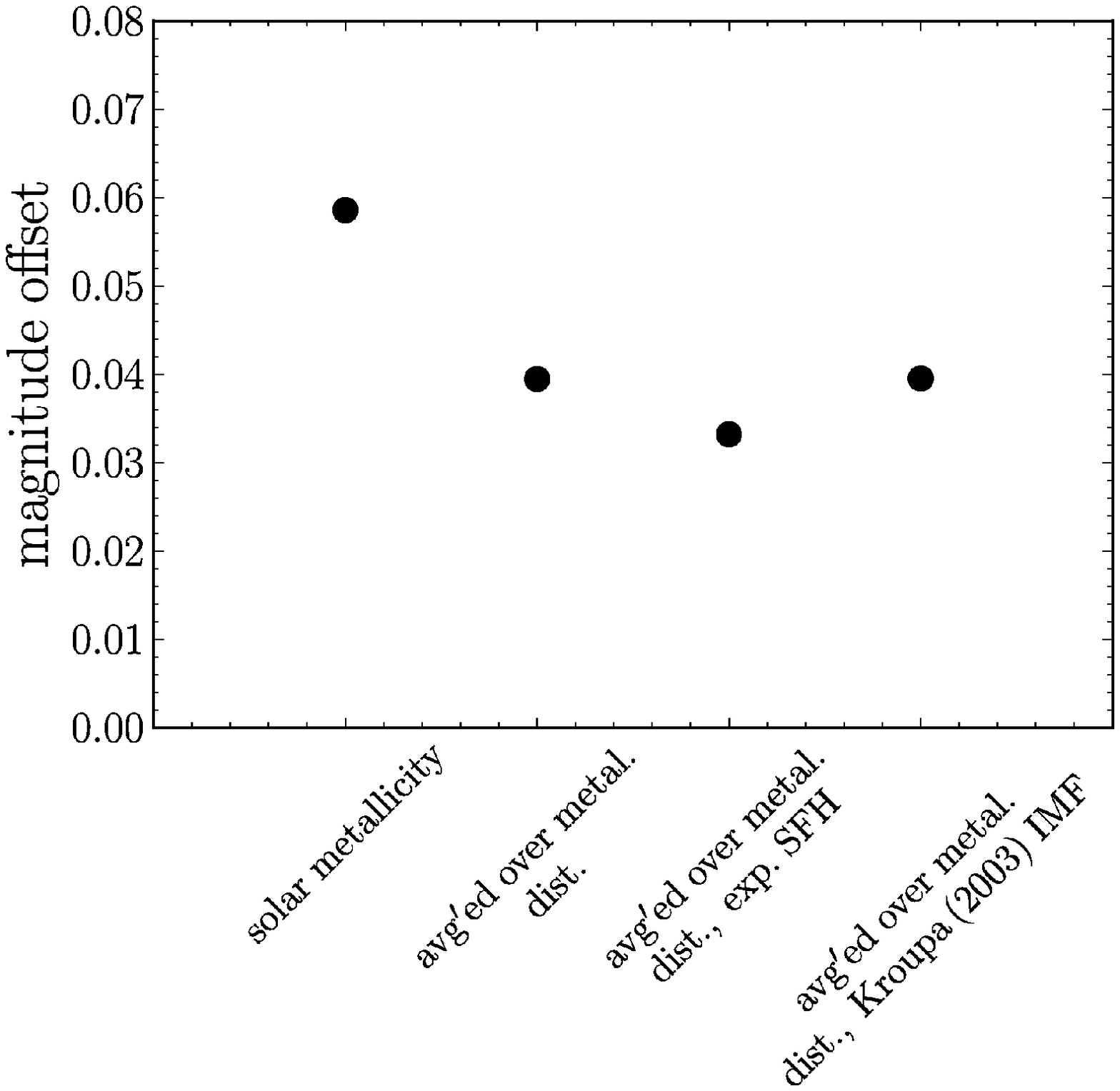}
\caption{External calibration of the absolute magnitude scale of the
  RC clump: This figure shows the difference between the average RC
  magnitude derived from \emph{Hipparcos} data \citep{Laney12a} and
  the mean RC magnitude calculated by averaging the peak magnitudes of
  the PARSEC RC over color and metallicity. The first point is the
  difference when using only the solar metallicity isochrones, while
  the other points are averaged over the metallicity distribution of
  local stars \citep{Casagrande11a}. The second point assumes a
  constant star-formation rate and a lognormal \citet{Chabrier01a}
  IMF, while the third and fourth points assume an
  exponentially-declining star-formation rate and a \citet{Kroupa03a}
  IMF, respectively. The assumptions about the star-formation history
  and IMF affect the magnitude calibration at a level that is below a
  percent.}\label{fig:magcorr}
\end{figure}

The RC selection technique presented in this section requires highly
accurate and precise measurements of \logg. Using the APOKASC data, we
can evaluate how well the selection technique works for various
precisions in \logg. We can estimate the contamination by applying the
selection to those stars in the APOKASC catalog for which the
evolutionary state was obtained by \citet{Stello13a} (which therefore
provides the ground truth of whether a star is in the RC or
not). Using the $\logg_{\mathrm{seismo}}$ measurements from
asteroseismology, the contamination of non-RC stars is
$4\pm2\,\%$. This is therefore the best that the presented technique
can do. Using more realistic errors of $0.1\dex$ and $0.2\dex$ in
$\logg_{\mathrm{spec}}$ from the analysis of high-resolution spectra,
the contamination rises to $9\pm2\,\%$ and $14\pm3\,\%$,
respectively. This contamination can be compared, for example, to the
selection technique used by \citet{Williams13a}, which consists of the
simpler cuts $0.55 \leq (J-K)_0 \leq 0.8$ and $1.8 \leq \logg \leq
3.0$. Applying these cuts to $\logg_{\mathrm{seismo}}$ yields a
contamination of $33\pm3\,\%$, eight times larger than the
contamination for the new technique presented here. Using RAVE
\citep{Steinmetz06a} data with spectroscopic \logg,
\citet{Williams13a} estimate their contamination to be
$\approx60\,\%$.

For the current spectroscopic pipeline for APOGEE, the errors in
\logg\ are approximately $0.2\,\dex$, but there are systematic offsets
in the measured $\logg_{\mathrm{spec}}$ for RC and RGB stars that are
such that the RC and RGB are closer together by $\approx0.2\dex$,
making it more difficult to separate them. From the \citet{Stello13a}
sample, the contamination by RGB stars using the current APOGEE
$\logg_{\mathrm{spec}}$ is $23\pm3\,\%$. Improved analysis of the
APOGEE spectra should in the future remove the relative bias in
\logg\ for RC and RGB stars.

\begin{figure}[t!]
\includegraphics[width=0.48\textwidth,clip=]{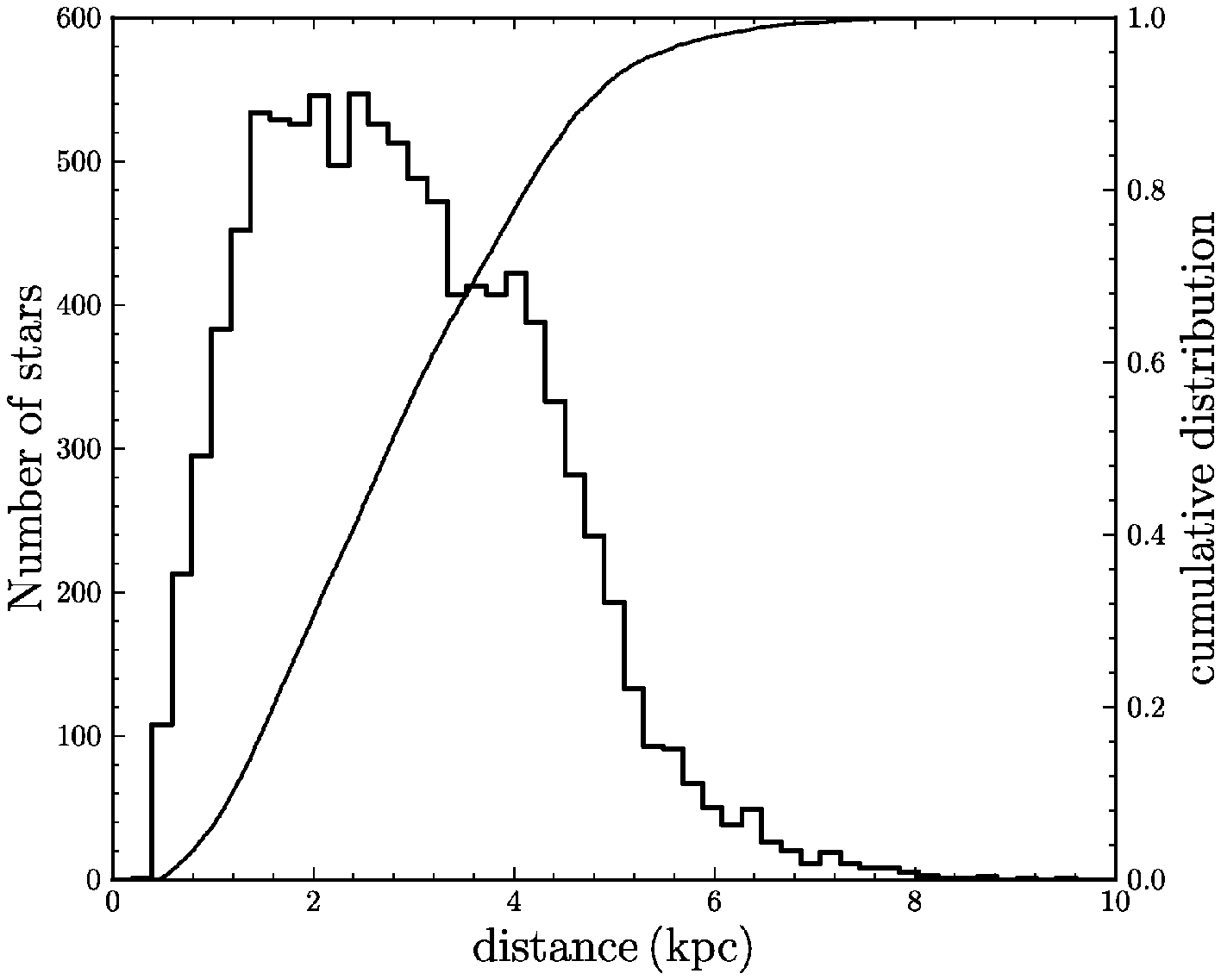}
\caption{Distance distribution of the \ncatalog\ stars in the DR11
  APOGEE-RC sample (histogram). The solid line shows the cumulative
  histogram. The median distance is $\approx2.75\kpc$ and 90\,\% of
  the stars lie between $1$ and $6\kpc$.}\label{fig:disthist}
\end{figure}

\begin{figure*}[t!]
\includegraphics[width=0.48\textwidth,clip=]{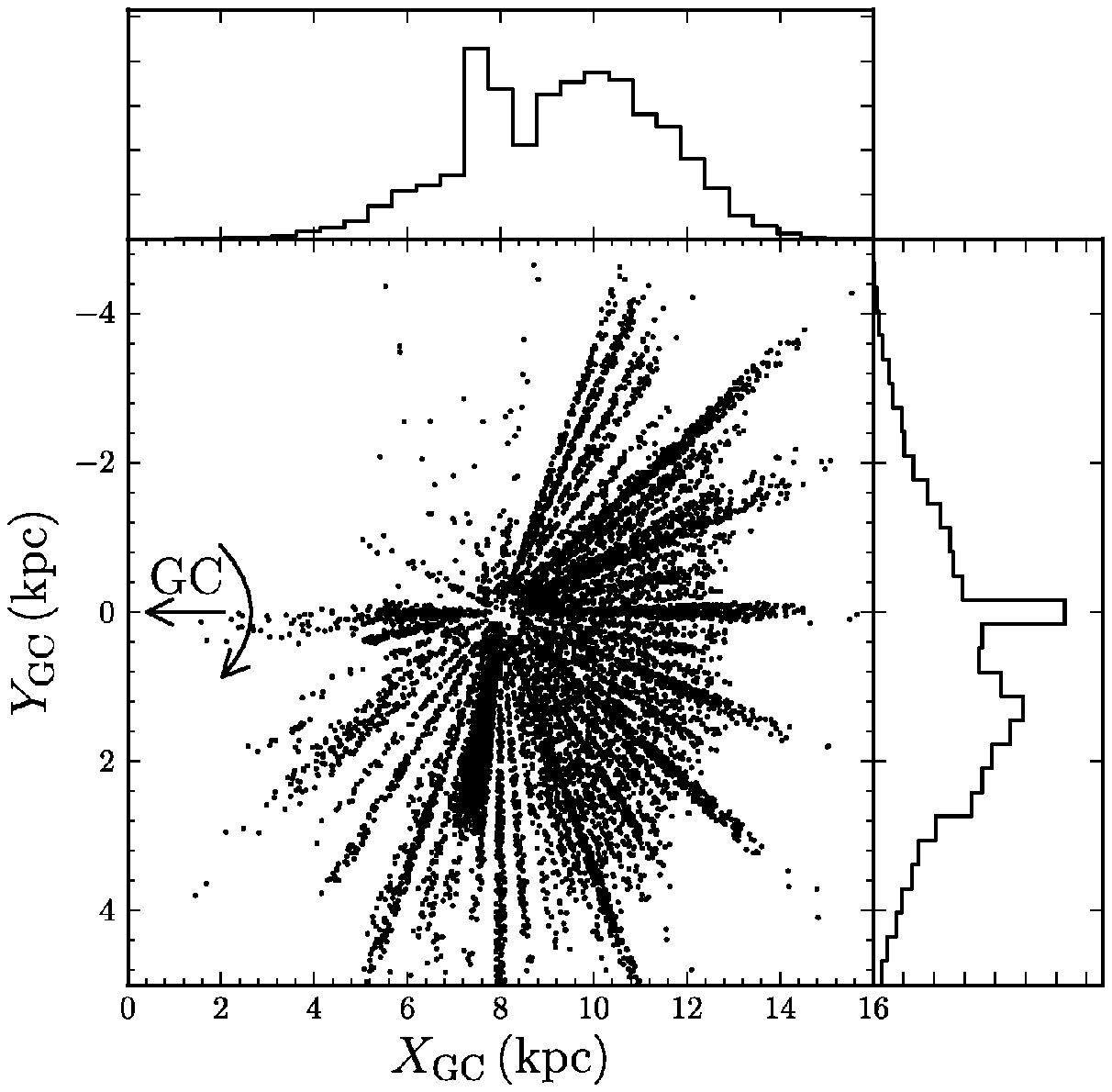}
\includegraphics[width=0.48\textwidth,clip=]{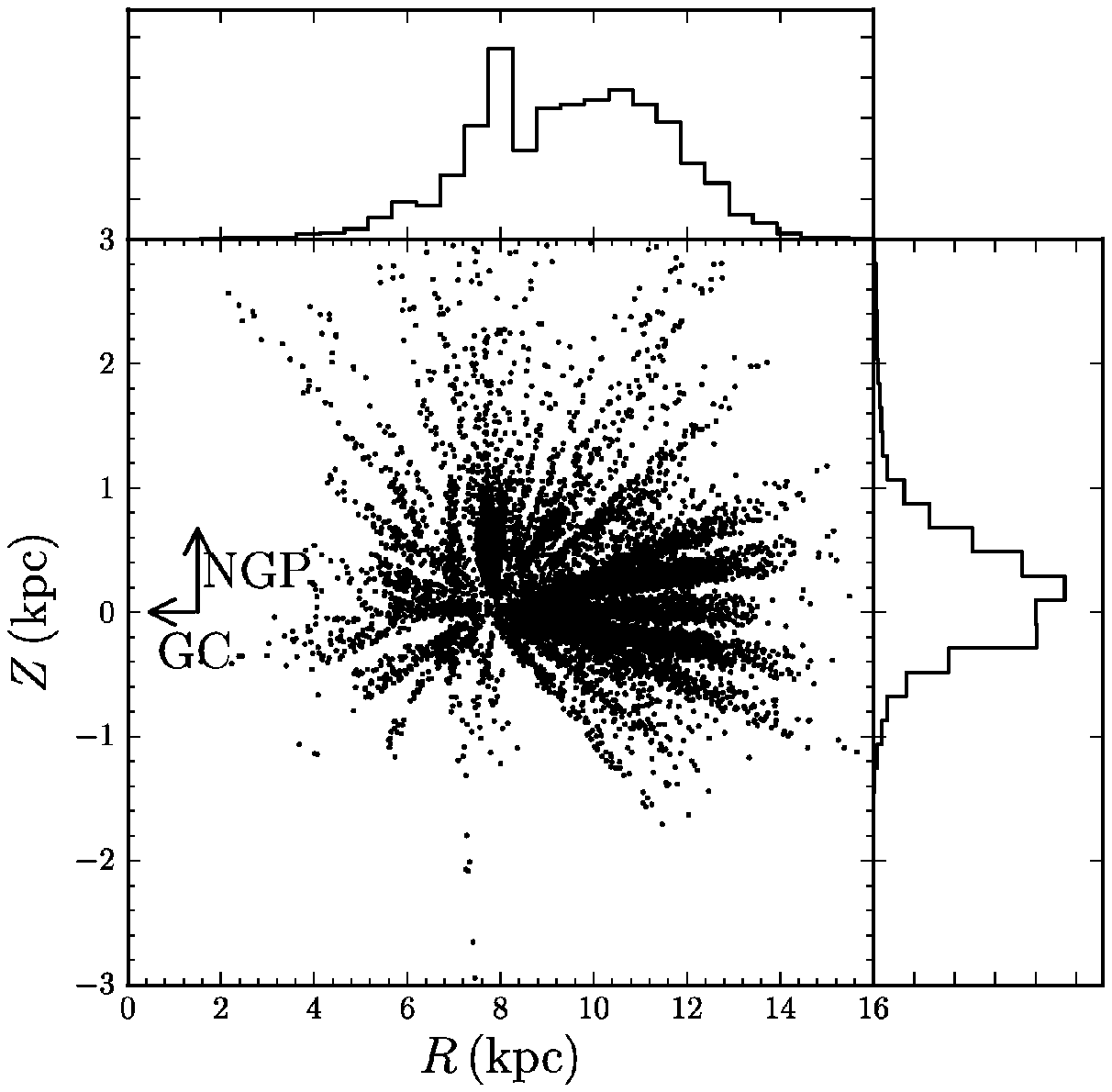}
\caption{Distribution of the \ncatalog\ stars in the DR11 APOGEE-RC
  sample in Galactocentric coordinates: distribution in the plane
  (\emph{left panel}) and in Galactocentric radius and distance from
  the plane (\emph{right panel}). The Sun is located at
  $(X_{\mathrm{GC}},Y_{\mathrm{GC}},Z) = (8,0,0.025)$ kpc. The
  overdensity of stars near $X_{\mathrm{GC}} = 8\kpc$ and $R = 8\kpc$
  is due to a large concentration of APOGEE targets in the
  \emph{Kepler} field.}\label{fig:distdist}
\end{figure*}

The current bias in the relative \logg\ measurements of RC and RGB
stars is such that \logg\ starts to be underestimated for RGB stars
below around $\teff=4900\,\mathrm{K}$ (see M.~Pinnsoneault et
al. 2014, in preparation). Practically, this means that we must change
the upper \logg\ limit used to define the RC sample to take into
account the biased \logg\ measurements for the RGB stars. This can be
done by specifying the following additional cut that follows the
\logg\ bias as a function of \teff\
\begin{equation}\label{eq:addllogg}
  \logg < 0.001\,\mathrm{\dex\,K}^{-1}\,(\teff-4800\,\mathrm{K})+2.75\,.
\end{equation}
In the current SDSS-III's Data Release 11 (DR11; years one and two)
sample (see below), this cut removes 2,261 out of 12,613 catalog
objects, or about 18\,\% of the sample. Using the APOKASC sample we
find that this additional constraint reduces the contamination to
$10\pm3\,\%$ compared to $23\,\%$ contamination without this extra
cut. We stress that this additional cut is a temporary solution to the
ASPCAP problems with the relative \logg\ of RC and RGB stars. Once
this bias is removed by pipeline improvements, the additional
\logg--\teff\ constraint will no longer be necessary. However, it can
still be applied to obtain a sample of RC stars with extremely high
purity: Using the APOKASC sample with evolutionary state measurements,
we estimate that applying the cut in \equationname~(\ref{eq:addllogg})
reduces the contamination from 14\,\% to less than 5\,\% and 10\,\% for
\logg\ uncertainties of $0.1\dex$ and $0.2\dex$, respectively.

We stress that the contamination fractions discussed in the previous
paragraphs are \emph{estimates}; the true contamination could be
different if the sample of APOKASC stars with evolutionary-state
measurements does not accurately represent the parent APOKASC sample,
or if the APOKASC sample does not reflect the RC and RGB populations
in the \emph{Kepler} field \citep{Koch10a} or in other parts of the
Galaxy\footnote{Additionally, the \citet{Stello13a} sample contains
  only a few stars with enhanced \afe\ with respect to the solar
  value. Therefore, we cannot currently test the RC selection for
  $\alpha$-enhanced stars. We will describe the outcome of tests with
  such stars in updates of the RC catalog, when seismological
  classifications become available for such stars.}. By comparing the
\citet{Stello13a} sample to the parent APOKASC sample we find that we
are likely slightly overestimating the contamination from RGB stars,
as RGB stars near the RC locus are disproportionally represented in
the sample for which the evolutionary state has been measured: 23\,\%
of the stars between the dashed lines in
\figurename~\ref{fig:logg-apokasc} versus 33\,\% of the stars below
the high-\logg\ line have measurements of their evolutionary state,
which implies that the contamination fraction in the full APOKASC
sample is likely $\approx\,40\%$ smaller at $7\pm2\,\%$ rather than
$10\,\%$. However, the APOKASC sample itself was constructed using a
complicated combination of cuts by \emph{Kepler} to create the
asteroseismic sample of giants and by the APOKASC collaboration to
provide spectroscopic follow-up for a subset of these (see
M.~Pinnsoneault et al. 2014, in preparation). Therefore, the relation
between the APOKASC catalog and the underlying Galactic populations is
hard to quantify. APOKASC preferentially selected first-ascent red
giants for spectroscopic follow-up, such that RGB stars are
overrepresented compared to RC stars in the asteroseismic sample and
the $7\,\%$ contamination fraction is probably still an overestimate.

To summarize our RC selection and characterization technique: We
select RC stars using the cuts in \equationname
s~(\ref{eq:loggteffcut}), (\ref{eq:tefffeh}), (\ref{eq:jkzcuts}),
(\ref{eq:jkzcuts3}), and (\ref{eq:jkzcuts2}). Because of the current
ASPCAP bias in the relative \logg\ measurements of RC and RGB stars,
we also apply the cut in \eqnname~(\ref{eq:addllogg}), although this
additional cut will no longer be necessary once the ASPCAP bias in
\logg\ is removed by improved modeling of the APOGEE spectra. For
stars satisfying these constraints, we calculate the absolute $\ks$
magnitude from the dereddened color $(J-\ks)_0$ and the metallicity
$Z$ using the two-dimensional surface in the top panel of
\figurename~\ref{fig:sig_jkz}. In \sectionname~\ref{sec:distance} we
discuss the uniform calibration offset that is applied to these
individual absolute magnitudes to place the distances on the
\emph{Hipparcos} scale.

\section{Distance calibration}\label{sec:distance}

We calibrate the absolute magnitude of the PARSEC stellar-isochrone
models for the \rc\ discussed in the previous section by using them to
calculate the average absolute magnitude for a RC sample of stars that
mimics that found close to the Sun. The average absolute
\ks\ magnitude of nearby RC stars is well-calibrated using
\emph{Hipparcos} parallaxes
\citep[\eg,][]{Alves00a,Groenewegen08a,Laney12a}. Our standard model
for a sample of stars that resembles local stars is a population that
has a distribution of metallicities similar to that found by
\citet{Casagrande11a} (with their $p(\feh)$ approximated here as a
mixture of two Gaussians with relative weights $0.8$ and $0.2$, means
of $0.016\dex$ and $-0.15\dex$, and dispersions of $0.15\dex$ and
$0.22\dex$, respectively) and a constant star-formation history with a
lognormal \citet{Chabrier01a} IMF. The average \ks\ absolute magnitude
of such a population as predicted by PARSEC is $-1.65$.

We choose the recent \citet{Laney12a} calibration of $M_{\ks} = -1.61$
of local RC stars, which is in very good agreement with the results of
\citet{Alves00a}, but is $0.07\magunit$ brighter than the calibration
of \citet{Groenewegen08a}. Therefore, there is a $3.5\,\%$ systematic
distance offset between distances derived from these two
calibrations. The calibration of \citet{Laney12a} is more accurate
than that of \citet{Groenewegen08a}, because they re-measure the
magnitudes of bright RC stars that are saturated in 2MASS ($K \lesssim
5$); their calibration is accurate to $2\,\%$.

Because the PARSEC models predict an average $\ks$ absolute magnitude
that is $0.04\magunit$ brighter than that found from local
observations, we apply a uniform $0.04\magunit$ correction to the
individual \ks\ magnitudes determined using the PARSEC models based on
a star's $([J-\ks]_0,Z)$ in \figurename~\ref{fig:sig_jkz} when
calculating distances for the sample of RC stars in \apogee. This
correction does not strongly depend on the assumptions made about the
age and metallicity composition of a local population of
stars. \figurename~\ref{fig:magcorr} shows the correction for
different assumptions about the age and metallicity distribution. For
a reasonable metallicity distribution, the systematic differences are
below $0.01\magunit$, or less than $0.5\,\%$ in distance.

We can further test the distance calibration using the APOKASC sample
by calculating ``direct seismic'' distances using spectroscopic \teff,
stellar radii determined using asteroseismic scaling relations, and
bolometric corrections \citep[\eg,][]{Marigo08a} (see T.~Rodrigues
\etal\ 2014, in preparation, for full details); these distances have
random and systematic uncertainties of $5\,\%$ each
\citep{Miglio13a}. Using $\logg_{\mathrm{seismo}}$ and the procedure
described in \sectionname~\ref{sec:sample} to select RC stars, we
compare the seismic with the RC distances (including the
\emph{Hipparcos} calibration) for 593 stars and find that the seismic
and RC distances agree to within better than $1\,\%$ using the median
difference, with a scatter of $\approx 7\,\%$. For the subset of
APOKASC stars with evolutionary-state measurements (see section
\ref{sec:contam}), the scatter is only $5\,\%$. Similarly, we compare
the RC distances with the seismic distances from the SAGA catalog
\citep{Casagrande14a}. For the 42 stars in common between the sample
of 593 APOKASC RC stars and the SAGA sample, the seismic distances
agree with the RC distances to within $3\pm1\,\%$, with a scatter of
$6\,\%$\footnote{T.~Rodrigues \etal\ 2014, in preparation, also show
  that their seismic distances agree to better than $1\,\%$ with the
  seismic distances in the SAGA catalog by considering all giants in
  common between the two samples.}. These values are consistent with
each method's random and systematic uncertainties and show that our
distance calibration is sound.

We therefore conclude that the distances that we assign to RC stars
are unbiased to within $\approx 2\,\%$ and have random uncertainties
of $\approx 5\,\%$.  The distance distribution of the \ncatalog\ RC
stars selected using the method described in
\sectionname~\ref{sec:sample} from DR11 of \apogee\ is presented in
\figurename~\ref{fig:disthist} (see \sectionname~\ref{sec:catalog} for
full details on this sample). Many of the RC stars in the sample are
only a few $\kpc$ from the Sun, with a median distance of $2.75\kpc$
and a $95\,\%$ interval of $0.8\kpc$ and $6\kpc$. The distribution of
this sample in Galactocentric coordinates is shown in
\figurename~\ref{fig:distdist}. This shows that the sample spans a
large area of the Galactic disk near the midplane.

\section{Sample selection function}\label{sec:ssf}

\begin{figure}[t!]
\includegraphics[width=0.49\textwidth,clip=]{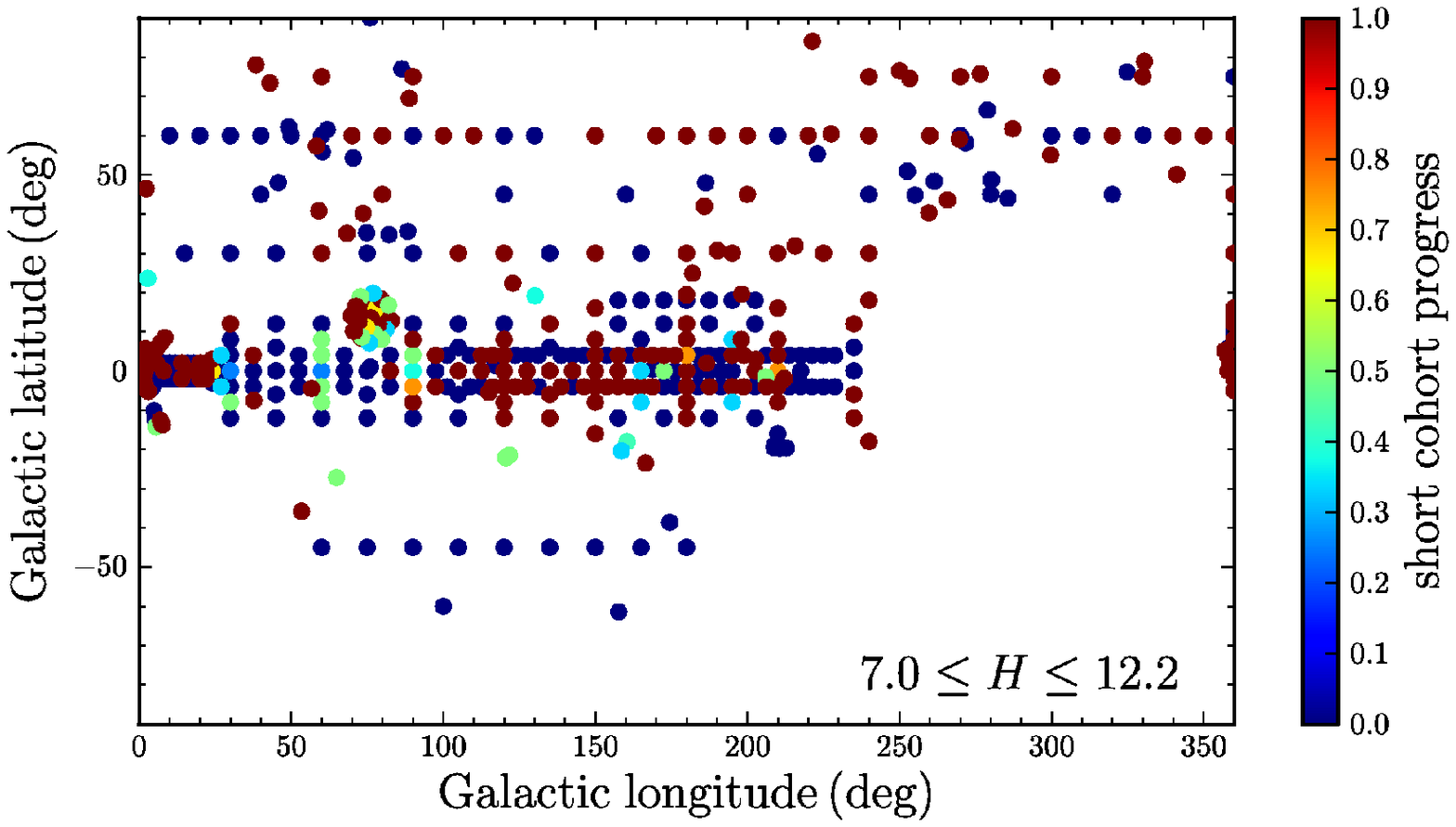}\\
\includegraphics[width=0.49\textwidth,clip=]{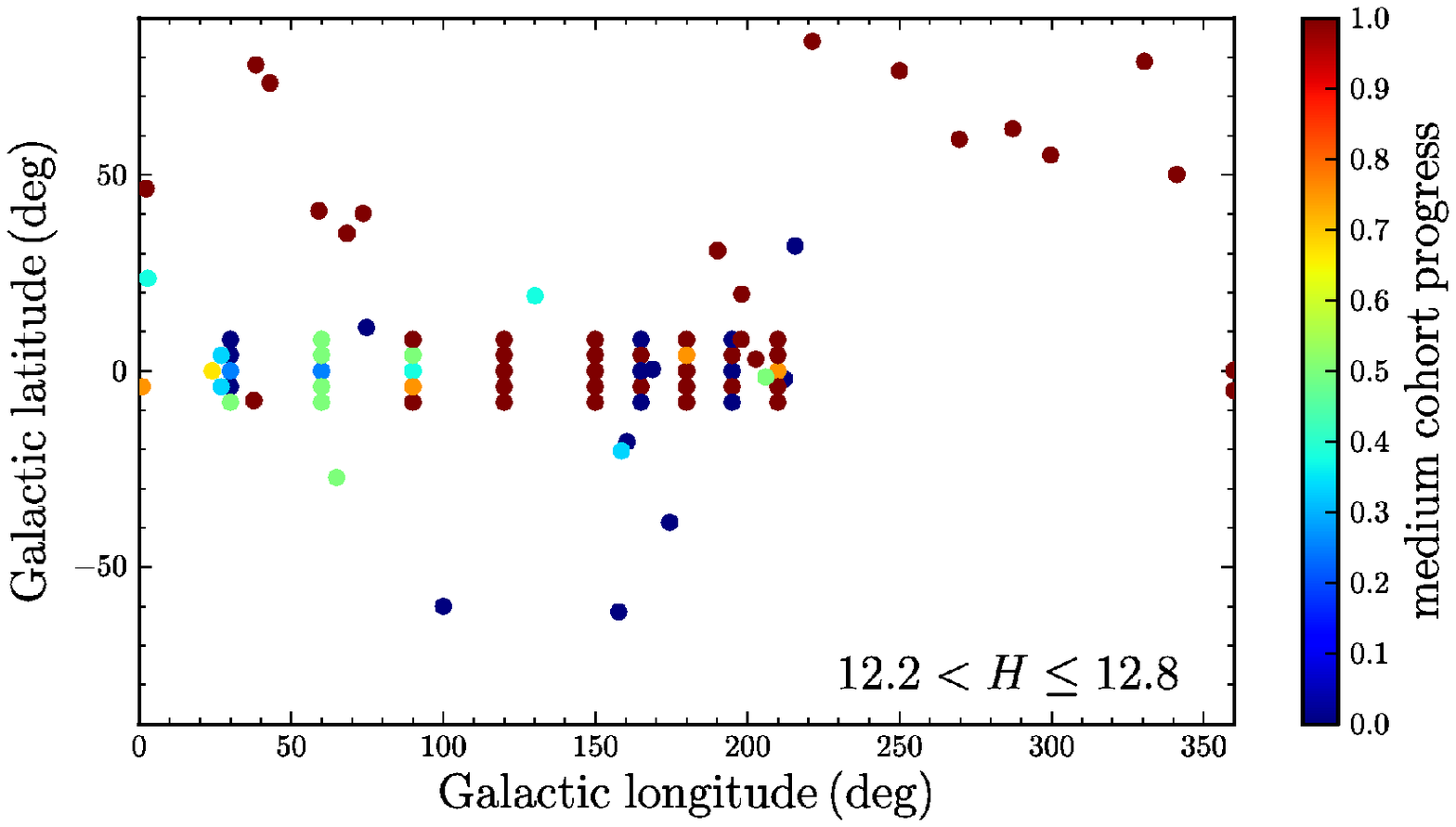}\\
\includegraphics[width=0.49\textwidth,clip=]{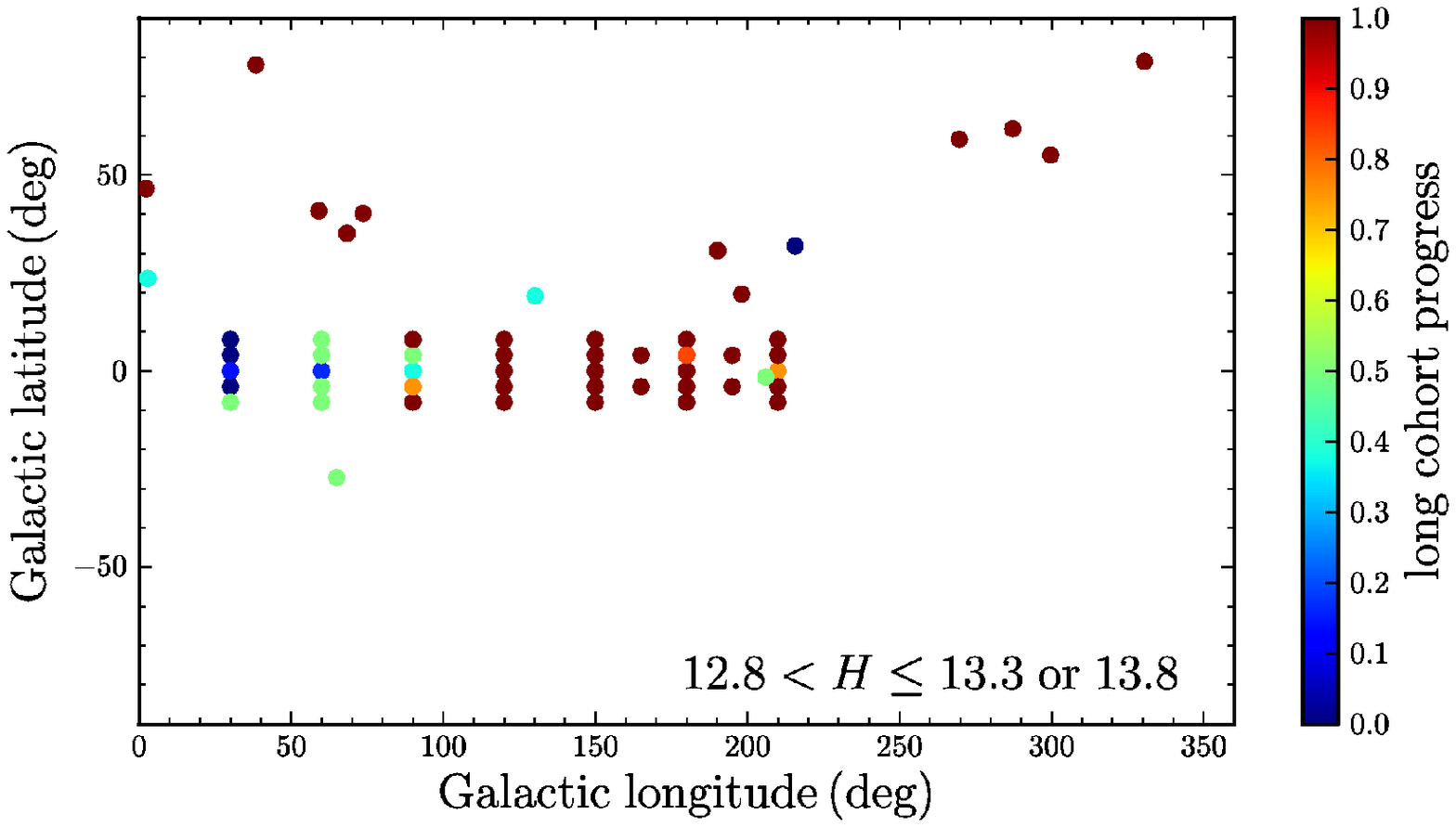}
\caption{\apogee's target completeness in different fields after the
  first two years. The different panels show the fractional
  completeness for different fields on the sky for ``short'' cohorts
  (\emph{top panel}), ``medium'' cohorts (\emph{middle panel}), and
  ``long'' cohorts (\emph{bottom panel}); see text for the meaning of
  the different cohorts. Overall, short cohorts are 50\,\% complete,
  medium cohorts 66\,\%, and long cohorts 78\,\%. The $H$-band limits
  do not apply to all fields; some fields have a short cohort with
  $7.0 \leq H \leq 11.0$ and some of those have a medium cohort with
  $11.0 \leq H \leq 12.2$ (this applies in particular to most bulge
  plates around $l = 0^\circ$ and all \emph{Kepler} fields around
  $(l,b) = (75^\circ,15^\circ))$.}\label{fig:obsprogress}
\end{figure}

As a spectroscopic survey observing a relatively small number of
objects in the photometric 2MASS catalog, \apogee\ requires decisions
as to what lines of sight to observe, what types of stars to target in
the chosen directions, and how to sample these stars as a function of
their photometric properties. These decisions affect how the
underlying photometric sample is reflected in the spectroscopic
data. To connect the high-dimensional distribution (or moments
thereof) of positions, velocities, and elemental abundances found in
the spectroscopic sample to the distribution of \emph{all} stars it is
necessary to correct for the effects of the selection (often referred
to as \emph{selection biases}), beyond the obvious angular selection
function encased in the APOGEE pointings (\eg,
\figurename~\ref{fig:obsprogress}). Generally, the relation between
the spectroscopic sampling on the underlying Galactic population can
be divided into three parts (see Sec.~4 of \citealt{Rix13a}): (a) the
procedure by which the survey selects stars from all potential
targets, (b) the relation between the kinds of stars observed and the
full underlying stellar population, and (c) the relation of the
observed spatial volume (typically a small number of lines of sight)
to the global volume. The last point requires one to extrapolate the
observed stellar distribution in the observed lines of sight to the
large volume between lines of sight; one must assume that spatial
gradients in the underlying distribution are well sampled by the
observed lines of sight, and one needs to estimate these gradients,
which can be done by modeling the spatial dependence of the stellar
distribution within the observed field pointings. However, as this
last step requires extensive modeling of the observed distributions,
we do not discuss it here further (see \citealt{BovyNoThickDisk} and
\citealt{BovyMAPstructure} for an example of this procedure).

We discuss the selection effects of (a) and (b) in this Section and
the next, respectively. The overall \apogee\ sample contains dozens of
sub-samples selected for a variety of reasons (\eg, observations of
members of stellar clusters or the Sgr dwarf galaxy, observations of
stars in the \emph{Kepler} field for which asteroseismology data
exist, etc.), most of which do not provide a fair sample in any sense
of the large-scale Galactic distribution of stars. However, most
\apogee\ targets are selected using a simple color cut, $(J-\ks)_0
\geq 0.5$, typically in three separate $H$-band magnitude ranges
\citep{Zasowski13a}, and this simple selection allows the fraction of
spectroscopically-observed objects to be determined as a function of
color $(J-\ks)_0$ and magnitude $H$. We will refer to this sample as
\apogee's \emph{main sample}, reserving the name \emph{statistical
  sample} for the part of the main sample for which observations are
complete (defined as the observations having a signal-to-noise ratio
of $100$ per half-resolution element).

\begin{figure}[t!]
\includegraphics[width=0.48\textwidth,clip=]{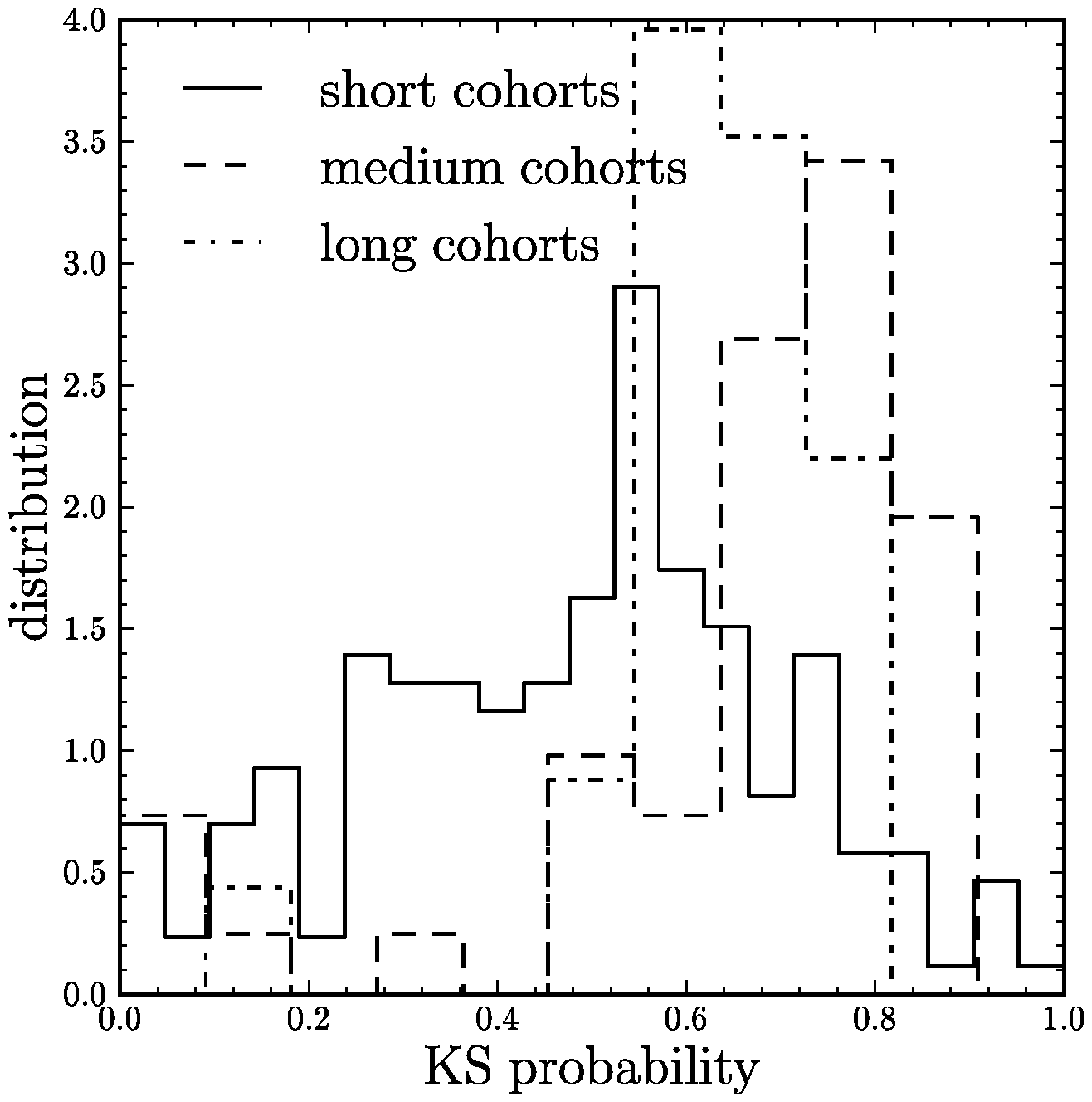}
\includegraphics[width=0.48\textwidth,clip=]{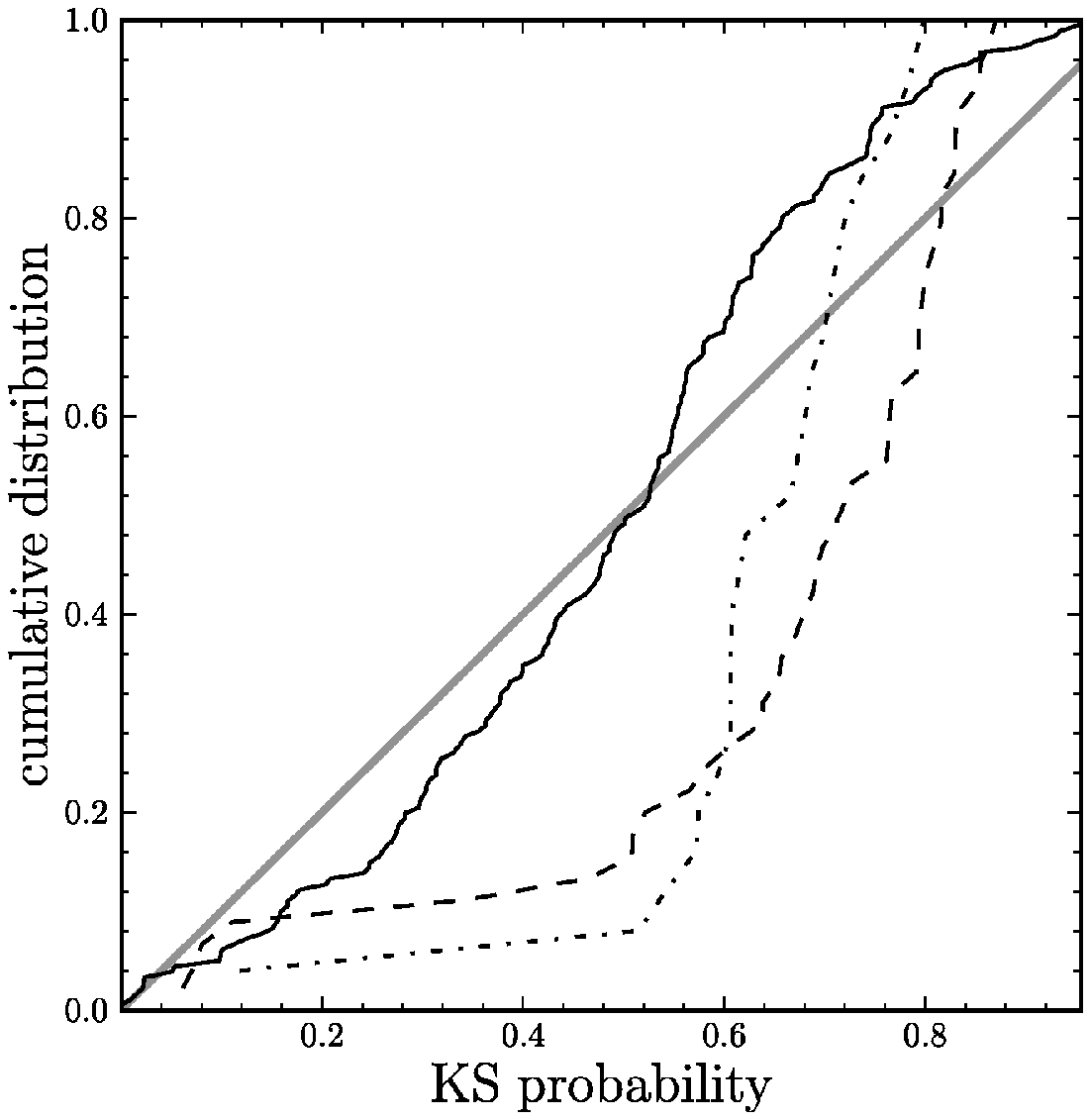}
\caption{Distribution of the KS probability that a field/cohort
  combination's statistical spectroscopic sample was drawn from the
  underlying photometric sample combined with the model selection
  function. This calculation is shown separately for short, medium,
  and long cohorts. The bottom panel shows the cumulative
  distribution. All field/cohort combinations have a large probability
  that their statistical spectroscopic sample was drawn from the
  underlying photometric sample, with the few field/cohort
  combinations with low probability consistent with statistical
  fluctuations (\ie, close to the one-to-one line in the cumulative
  distribution panel, shown in gray).}\label{fig:sfks}
\end{figure}

For the purposes of describing the details of the \apogee\ target
selection, it is necessary to give a brief description of the basic
data-collecting process (see also the \emph{Glossary} in Appendix A of
\citealt{Zasowski13a}). \apogee\ observes stars in a set of
\emph{fields}, which are circular locations on the sky with a radius
of $1^\circ\!\!.49$; see \figurename~\ref{fig:obsprogress} for the
location of all of these fields. Fields are observed in one ``visit''
sessions for up to 24 visit; each visit consists of a
$\approx64\,\mathrm{mins}$ observation. The vast majority of fields
are observed for three visits only, some fields are observed for only
1 visit (most fields in the \emph{Kepler} field, most fields in the
Galactic bulge, and fields targeting the core of the Sgr dwarf
galaxy), and others are observed for six, twelve, or twenty-four
visits. The main sample for fields observed for three visits or less
consists of a single magnitude \emph{cohort}. Other fields have either
two or three magnitude cohorts, corresponding to two or three
magnitude ranges. Typically, the brightest cohort, known as the
\emph{short cohort}, has $7.0 \leq H \leq 12.2$, the intermediate
\emph{medium cohort} has $12.2 < H \leq 12.8$, and the \emph{long
  cohort} has $12.8 < H \leq 13.3$ or $13.8$, depending on whether the
field is observed for 12 or 24 visits; we refer to this cohort
distribution as the \emph{standard cohort distribution}. The main
exceptions to this rule are fields with a single one-visit cohort, for
which the short cohort has $7.0 \leq H \leq 11$; some of the bulge
fields with a single-visit cohort have a medium cohort that has $11.0 <
H \leq 12.2$. To a good approximation, we can assume that within each
cohort's magnitude range, the color and magnitude are sampled
uniformly from the underlying population\footnote{In detail, the
  sampling is random in three bins, each containing 1/3 of the stars,
  in the short cohort; in the medium and long cohorts each $N$-th star
  is observed, with $N$ chosen such as to provide the desired total
  number of targets in each cohort. While these schemes are not
  entirely random, the large number of stars combined with the 2MASS
  magnitude errors and some stars not being observed because of
  practical observational reasons serve to make this sampling nearly
  indistinguishable from a random sampling (see
  \figurename~\ref{fig:sfks}).\label{foot:nth}}. For the purpose of
the discussion, we will assume the standard cohort distribution.

In fields observed for six visits or more, there are multiple
three-visit short cohorts (one for every three visits), and for fields
observed for 12 visits or more there are multiple medium cohorts (one
for every six visits); there is only ever a single long cohort per
field. A \emph{design} consists of a three-visit set of targets: this
contains a short cohort with a unique set of targets, a medium cohort
with targets shared among two designs, and a long cohort with targets
shared among all designs for a given field (see Figure 1 in
\citealt{Zasowski13a} for a visual representation of different designs
for a given field). A three-visit field has a single design; a
six-visit field has two designs, with a different short cohort on each
design and a shared medium cohort; and a 12- or 24-visit field has
four or eight designs, with four or eight unique short cohorts, two or
four medium cohorts, and a single long cohort. Each design is
implemented as a set of spectroscopic plates, which corresponds to a
physical plug-plate loaded onto the telescope. A design may be used on
a single plate or on multiple plates. Plates are observed in one-visit
increments. For the purpose of determining the selection function, we
only consider plates for which all planned observations were complete
by the deadline for DR11 (August 1,
2013). \figurename~\ref{fig:obsprogress} shows the fraction of
completed plates for short, medium, and long cohorts out of all
planned plates as part of \apogee's three-year run. Near the disk,
observations have high completeness, except for the region $30^\circ
\lesssim l \lesssim 100^\circ$.

Using the mapping between plates, designs, and cohorts, we can
determine for which fields and cohorts all of the observations for a
set of stars are complete. As an example, for a twelve-visit field, we
can determine which of its four different short-cohort sets of targets
have been observed to completion, the same for its two different
medium-cohorts sets of targets, and for its single long cohort. For
the purpose of building a \emph{statistical sample}, we then ignore
all spectroscopic targets for which observations have not reached the
full exposure time (such stars are present in the \apogee\ data
products). For each combination of a field and a cohort's magnitude
range \apogee\ provides a random sampling of the underlying
photometric sample. The selection function for a field and
magnitude-range combination is therefore simply given by the number of
stars in the statistical sample in this field and magnitude range,
divided by the number of stars in the photometric sample in the same
field and magnitude range. For example
\begin{equation}\label{eq:sf}
  \begin{split}
    &\mathrm{selection\ fraction}(\mathrm{field\ X},7.0 \leq H \leq 12.2) = \\
  & \frac{\mathrm{No.\ of\ stars\ in\ stat.\ sample\ in\ field\ X,\ with}\ 7.0 \leq H \leq 12.2}{\mathrm{No.\ of\ stars\ in\ phot.\ sample\ in\ field\ X,\ with}\ 7.0 \leq H \leq 12.2}\,
  \end{split}
\end{equation}
and similarly for the medium and long magnitude ranges for each
field. Thus, for each field the selection function is piecewise
constant with jumps at the magnitude limits of the (up to three)
cohorts of each field.

\begin{figure}[t!]
\includegraphics[width=0.48\textwidth,clip=]{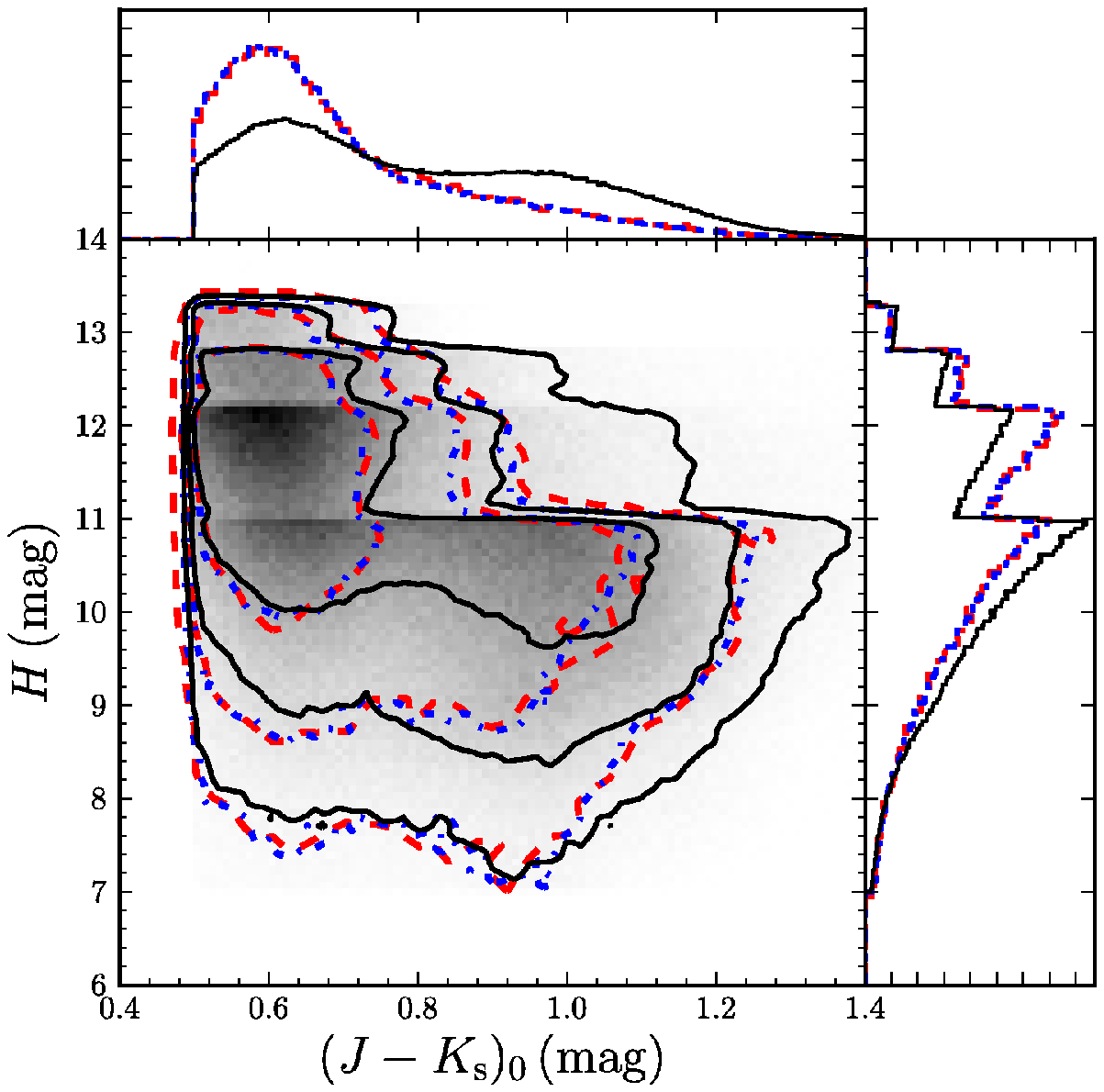}
\caption{Color--magnitude sampling of the statistical sample. The
  linear grayscale and the black lines show the density of potential
  targets in fields and cohorts contained in the statistical
  sample. The red dashed lines represent the distribution of stars in
  the statistical spectroscopic sample and the blue dash-dotted lines
  display the distribution of potential targets re-weighted by the
  selection function (see \equationname~[\ref{eq:sf}] and related
  discussion). The contours contain 68\,\%, 95\,\%, and 99\,\% of the
  distribution. The re-weighting of the photometric sample using the
  selection function perfectly reproduces the color--magnitude
  sampling provided by the statistical spectroscopic
  sample.}\label{fig:colormagSF}
\end{figure}

\begin{figure}[t!]
\includegraphics[width=0.48\textwidth,clip=]{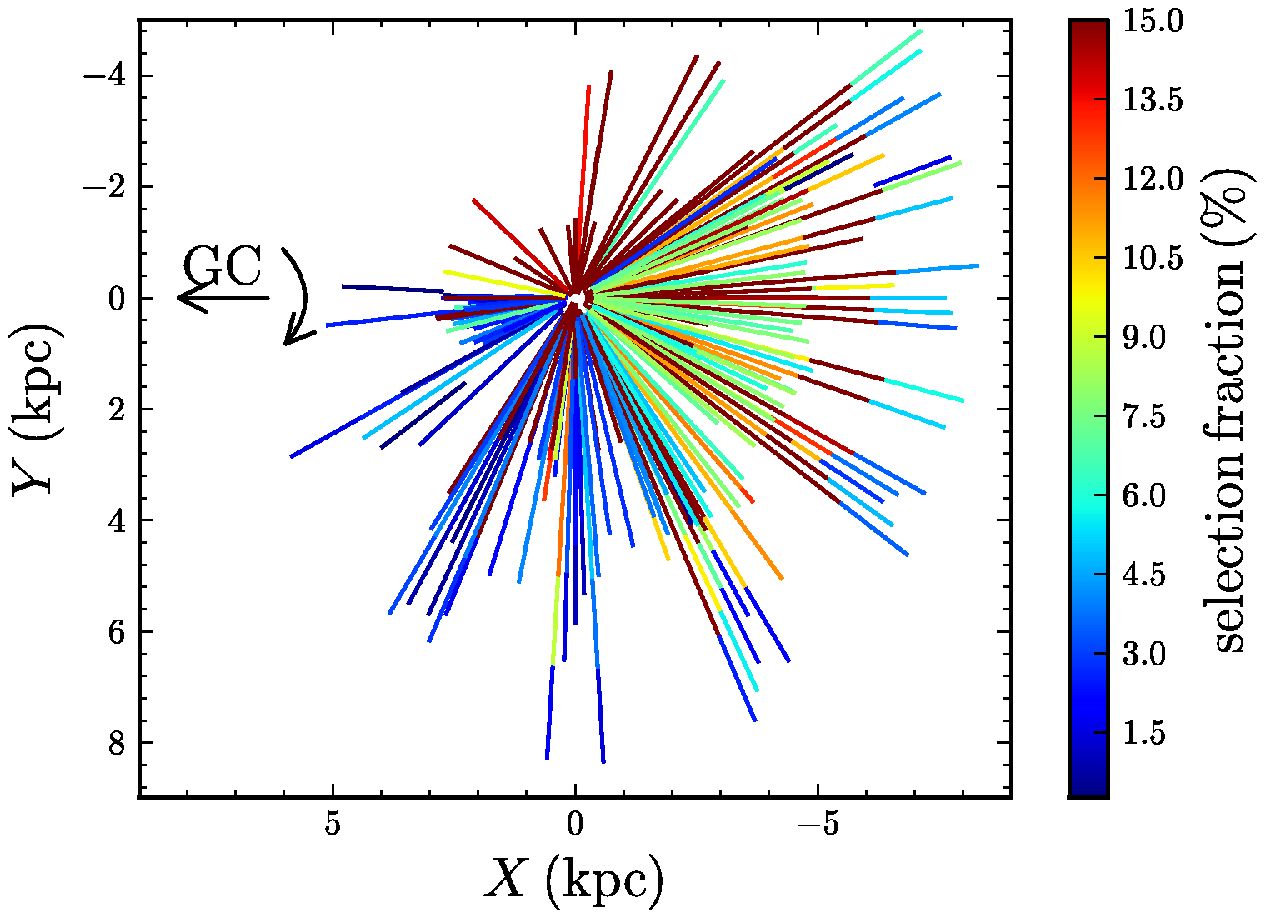}\\
\includegraphics[width=0.48\textwidth,clip=]{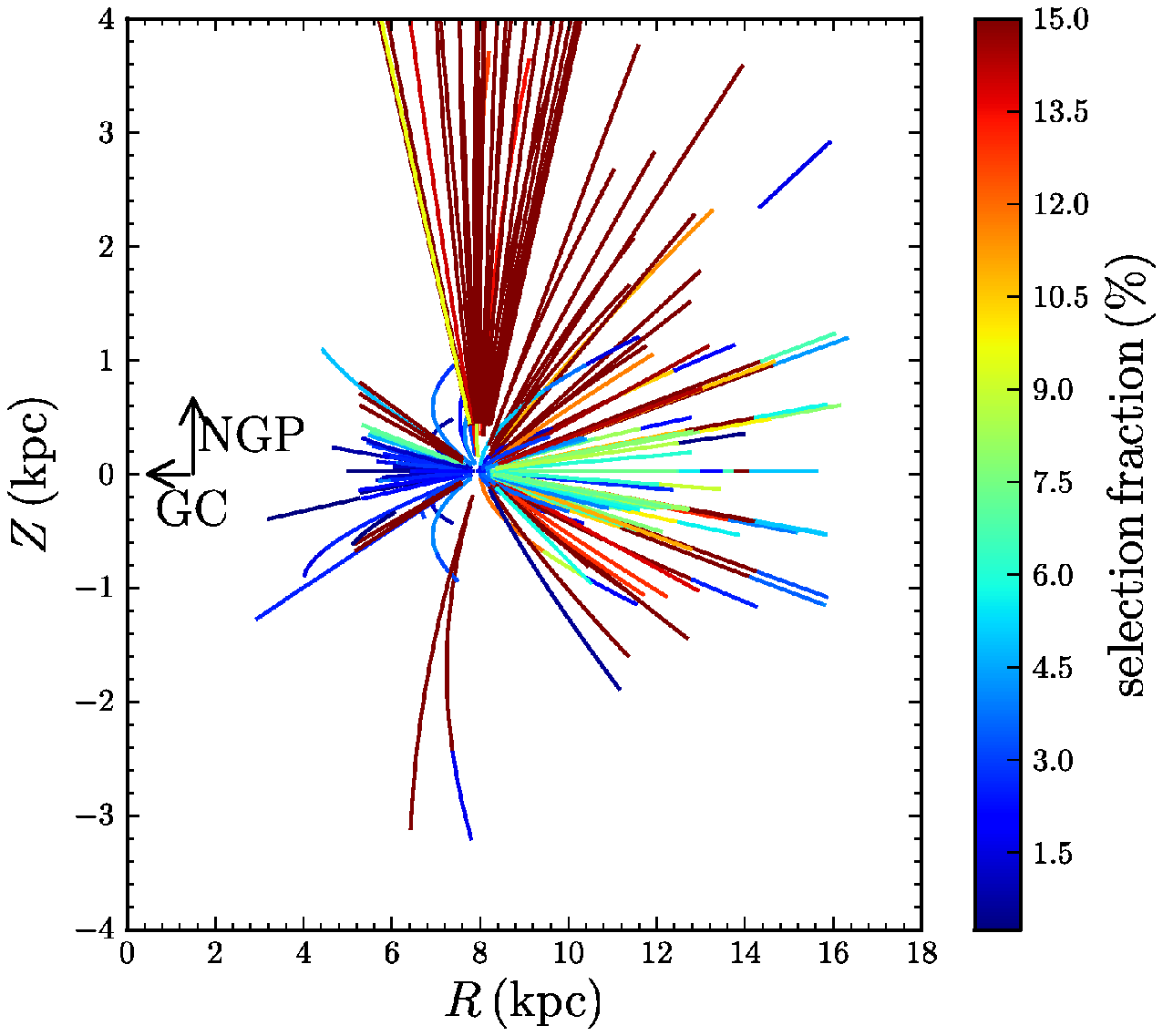}
\caption{The \apogee\ selection function, \ie, the fraction of all
  potential targets observed spectroscopically and present in the
  statistical sample. The selection function, which intrinsically is a
  function of $(l,b,H)$, is illustrated here as a function of Galactic
  coordinates $X$ and $Y$ (\emph{top panel}) and of Galactocentric
  radius $R$ and $Z$ (\emph{bottom panel}). $H$-band magnitudes are
  converted into distances using an absolute magnitude of $-1.49$
  \citep{Laney12a} and the median extinction in each field. Some
  high-latitude fields have selection fractions up to 95\,\% and are
  saturated in the color scale employed in this figure. Low latitude
  ($|b| < 10^\circ$) fields at the same Galactic longitude have been
  dithered by $2^\circ$ in the top panel for display purposes. The
  completeness is low near the Galactic midplane and toward the
  Galactic center.}\label{fig:sfxyrz}
\end{figure}

To determine whether the piecewise-constant model for the selection
function is a good approximation, we compare the magnitude
distribution of stars in the statistical spectroscopic sample to that
of stars in the photometric sample within the same field and magnitude
range. This comparison is done for the up to three different
cohort-magnitude-ranges of each field separately. In each field and
magnitude-range combination, the selection function model is that the
statistical sample is a fair sampling of the underlying photometric
distribution of targets, and this model is tested by computing the
Kolmogorov-Smirnov (KS) probability that the statistical sample
distribution of $H$-band magnitudes is the same as that of the
photometric sample. 

A histogram of these probabilities is shown in
\figurename~\ref{fig:sfks}. The vast majority of KS probabilities of
the short cohorts are large, such that the hypothesis that the
statistical sample was drawn from the photometric sample cannot be
rejected; the few fields whose short cohort has a small KS probability
are consistent with random noise (out of 102 short cohorts, one
expects $\approx5$ to have a KS probability $< 0.05$ and we find
$4$). The medium and long cohorts also mostly have large KS
probabilities, with the 1 out of 41 medium cohorts and 0 out of 25
long cohorts with $P_{\mathrm{KS}} < 0.1$, consistent with random
statistical fluctuations. We therefore conclude that the
piecewise-linear model of \eqnname~(\ref{eq:sf}) is a good model for
the selection function. We do caution that the KS probabilities of
medium and long cohorts are somewhat \emph{too consistent} with the
underlying distribution, which likely results from the non-random
`$N$'-th-star sampling used (see footnote \ref{foot:nth}) for these
cohorts. However, due to photometric magnitude errors, the
spectroscopic sampling is still an effectively-random sampling of the
intrinsic magnitude distribution (\ie, that which one would observe
without errors).

The distributions in color $(J-\ks)_0$ and magnitude $H$ of the
photometric sample in fields and magnitude ranges contained in the
statistical sample are shown as the density grayscale and black
contours and histograms in \figurename~\ref{fig:colormagSF}. This
distribution clearly shows the influence of the cohort magnitude
limits at $H = 11, 12.2, 12.8$, and $13.3$ (there are currently no
completed 24 visit fields that would reach $H = 13.8$). The fiber
allocation to different cohorts is such that stars in the fainter
magnitude bins are over-represented with respect to the brightest bin,
such that the magnitude distribution as well as the color distribution
of the statistical sample (shown as red dashed contours and
histograms) is weighted to fainter magnitudes and bluer colors. The
latter happens because, even though the color sampling is random over
the same color range for all magnitude ranges, the intrinsic
correlation between color and magnitude in the photometric sample
leads to an apparent bias in the color distribution of the
spectroscopic sample. Using the model for the selection function
described above, we can re-weight stars in the photometric sample such
that they trace the sampling of the spectroscopic sample, and this
distribution is shown as blue dash-dotted contours and histograms. It
is clear that correcting for the selection function leads to perfect
agreement between the re-weighted photometric sample and the
statistical spectroscopic sample.

The selection function is a function of location on the sky (field
location, or equivalently, Galactic longitude and latitude) and
$H$-band magnitude. To determine the spatial selection function
requires understanding how distances relate to dereddened $H_0$
magnitudes and what the extinction is as a function of distance for a
given line of sight. To get a sense of how the Milky Way spatial
volume is sampled by the RC sample of stars described in this paper,
we show the selection function as a function of position in
\figurename~\ref{fig:sfxyrz}, assuming a RC absolute $H$-band
magnitude of $-1.49$ \citep{Laney12a}, and using the median $A_H$ for
the stars in the statistical sample in each line of sight (in reality,
the extinction will be less closer to the Sun and greater at the faint
end, so the spatial sampling will cover a slightly smaller range than
shown here). This figure demonstrates that the statistical subsample
of the RC sample presented in this paper provides a large-scale
sampling of the Milky Way volume near the disk that is useful for
statistical analyses of the distribution of elemental abundances in
the Milky Way disk.

\section{Astrophysical sampling with the red clump}\label{sec:astrosf}

\begin{figure}[t!]
\includegraphics[width=0.48\textwidth,clip=]{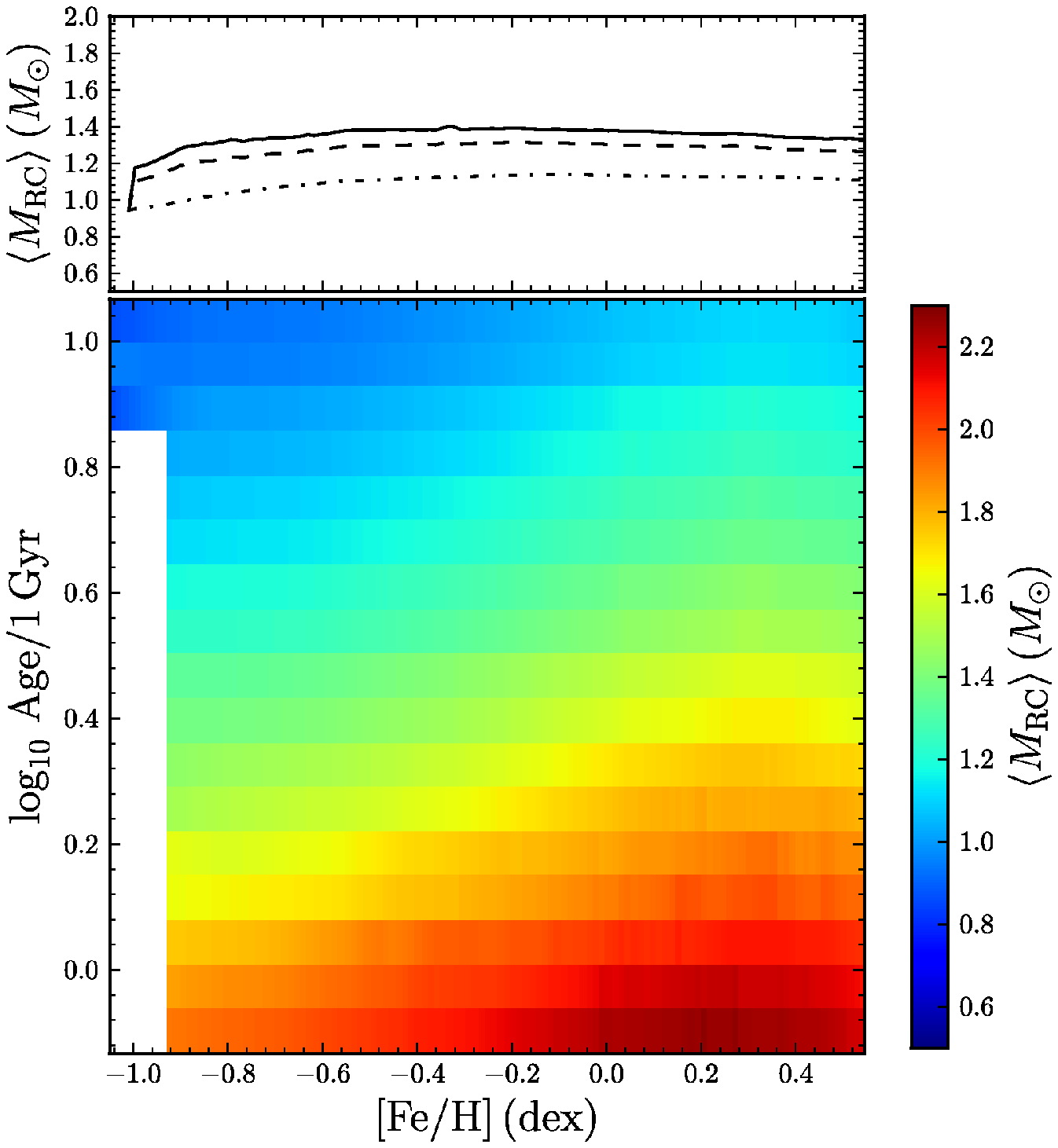}
\caption{The average initial mass of a star in the RC selection region
  defined in \sectionname~\ref{sec:sample} as a function of age and
  metallicity $Z$. The top panel shows the average mass integrated
  over a constant star formation rate (solid line) and
  exponentially-declining star-formation histories with $e$-folding
  times of $8\Gyr$ (dashed line) and $1\Gyr$ (dash-dotted
  line).}\label{fig:astrosf-mass}
\end{figure}

\begin{figure}[t!]
\includegraphics[width=0.48\textwidth,clip=]{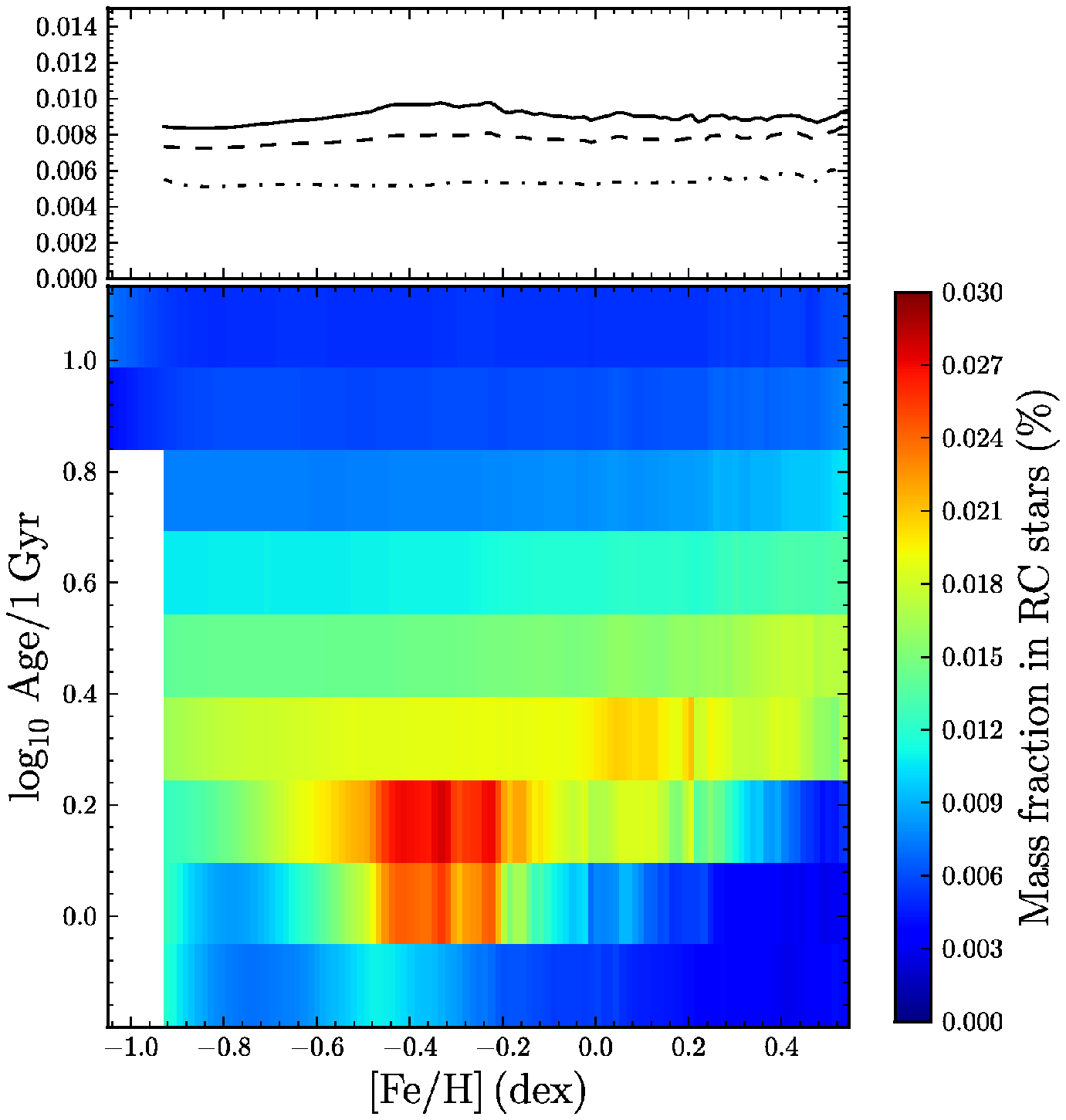}\\
\includegraphics[width=0.48\textwidth,clip=]{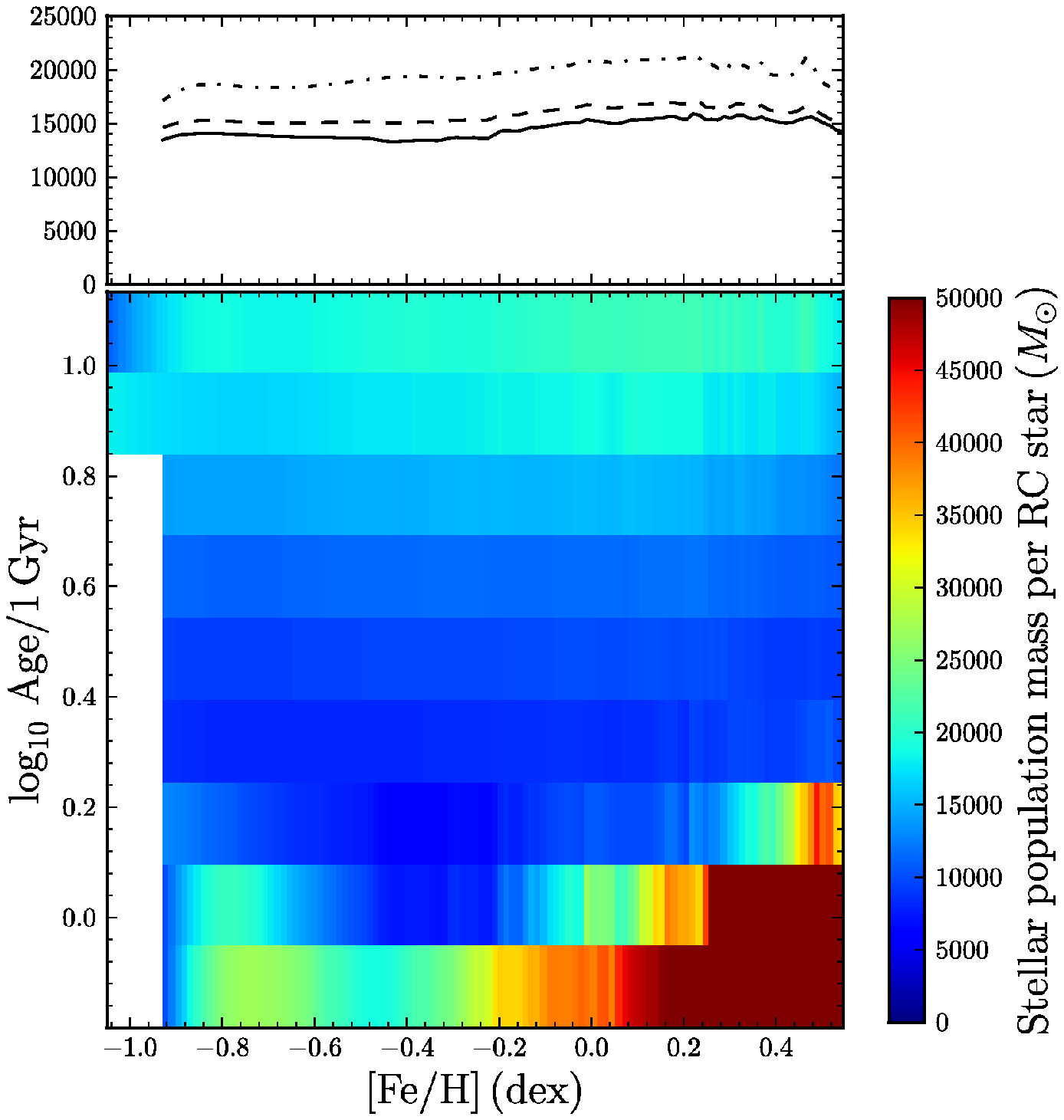}
\caption{Sampling of the underlying stellar population of the RC
  selection region of \sectionname~\ref{sec:sample}. The top figure
  displays the fraction of the mass of a population contained in the
  RC region and the bottom figure shows the amount of
  total-stellar-population mass each RC star represents. The top panel
  in each figure show these fractions and masses averaged over the
  different star-formation histories discussed in the caption of
  \figurename~\ref{fig:astrosf-mass}. Typically, the RC region
  contains about 0.01\,\% of the mass of a stellar population and each
  RC star represents about $15,000\msun$.}\label{fig:astrosf-frac}
\end{figure}

The second ingredient for relating the observed spectroscopic sample
of stars to the underlying Galactic stellar populations is to
determine how the spectroscopic tracer---RC stars in our
case---relates to the full population. The RC stage in the evolution
of stars with metallicities around solar is a relatively short phase
for stars with masses of approximately $1$ to $2\msun$ lasting for
about $100\Myr$. As such, the chance of finding stars in this stage of
stellar evolution is small compared to the total lifetime of the
stars, and only a small fraction of stars of a given age and
metallicity will at any given time find themselves within the RC
bounds given in \sectionname~\ref{sec:sample}. However, the RC is a
relatively \emph{long} phase compared to other evolved stages and in
that respect the RC is perhaps the best population tracer among
giants. In this Section, we use the PARSEC stellar-isochrone models to
determine the properties of stars in our RC selection region and how
they relate to the full stellar population.

Overall, the RC selection defined in this paper excludes stars with
(a) metallicities $\lesssim-1$ and (b) ages $\lesssim800\Myr$. These
are primarily due to the APOGEE color cut and the subsequent cuts in
\eqnname s~(\ref{eq:jkzcuts}), (\ref{eq:jkzcuts3}), and
(\ref{eq:jkzcuts2}). APOGEE's blue $(J-\ks)_0 \geq 0.5$ cut excludes
RC stars with low metallicities: \eqnname~(\ref{eq:jkzcuts}) evaluated
at $(J-\ks)_0 = 0.5$ gives a lower limit for the metallicity of
$\approx-0.9$. The additional blue cut in \eqnname
s~(\ref{eq:jkzcuts3}) eliminates secondary-red-clump stars with ages
$\lesssim1\Gyr$. Therefore, the RC selection in this paper selects
intermediate-age and old stars in the Milky Way disk.

\figurename~\ref{fig:astrosf-mass} shows the average initial mass of a
star in the RC selection region defined in
\sectionname~\ref{sec:sample} as a function of age and
metallicity. The typical mass is relatively constant as a function of
metallicity and is smaller for larger ages. For a constant SFH, the
average mass is $\approx 1.3\msun$ for near-solar metallicities and
declines to about $1.2\msun$ at $\feh \approx -1$. However, the
average mass is sensitive to the SFH: For an exponentially-declining
SFH with an $e$-folding time of $8\Gyr$, starting at $10\Gyr$, the
average mass is $\approx 0.1\msun$ lower than that for a constant SFH
for all metallicities. For an old population of stars, the average
mass drops to $\approx 1.0\msun$. The average mass can therefore
change by about $30\,\%$ depending on the SFH.

Because the RC stage (roughly the phase of core Helium burning, but
our considerations in this section are purely based on the selection
criteria from \sectionname~\ref{sec:sample}) is relatively short
($\approx 10^8\yr$), the fraction of stars of a single-burst stellar
population (assuming here a lognormal \citealt{Chabrier01a} IMF)
contained in the RC selection region is small. The relative fraction
of the mass of a single-burst stellar population contained in the RC,
as a function of age and metallicity, is shown in the top panel of
\figurename~\ref{fig:astrosf-frac}. Stars in the RC region are
typically between one and four $\Gyr$ old (see
\figurename~\ref{fig:astrosf-age} below) and the relative mass
fraction for ages above one \Gyr\ is about $10^{-4}$, a small fraction
of the stellar population's mass. Similarly, the bottom panel of
\figurename~\ref{fig:astrosf-frac} displays the fraction of the total
mass each star in the RC region represents (this is equal to
\figurename~\ref{fig:astrosf-mass} divided by the top panel). The
dependence on age and metallicity is similar to that of the relative
mass fraction. Integrating over a constant or exponentially-declining
SFH, these mass fractions are approximately constant with metallicity,
but depend at the $25\,\%$ to $50\,\%$ level on the SFH, as old
populations contain relatively less of a stellar-population's mass in
the RC region than younger populations.

\begin{figure}[t!]
\includegraphics[width=0.48\textwidth,clip=]{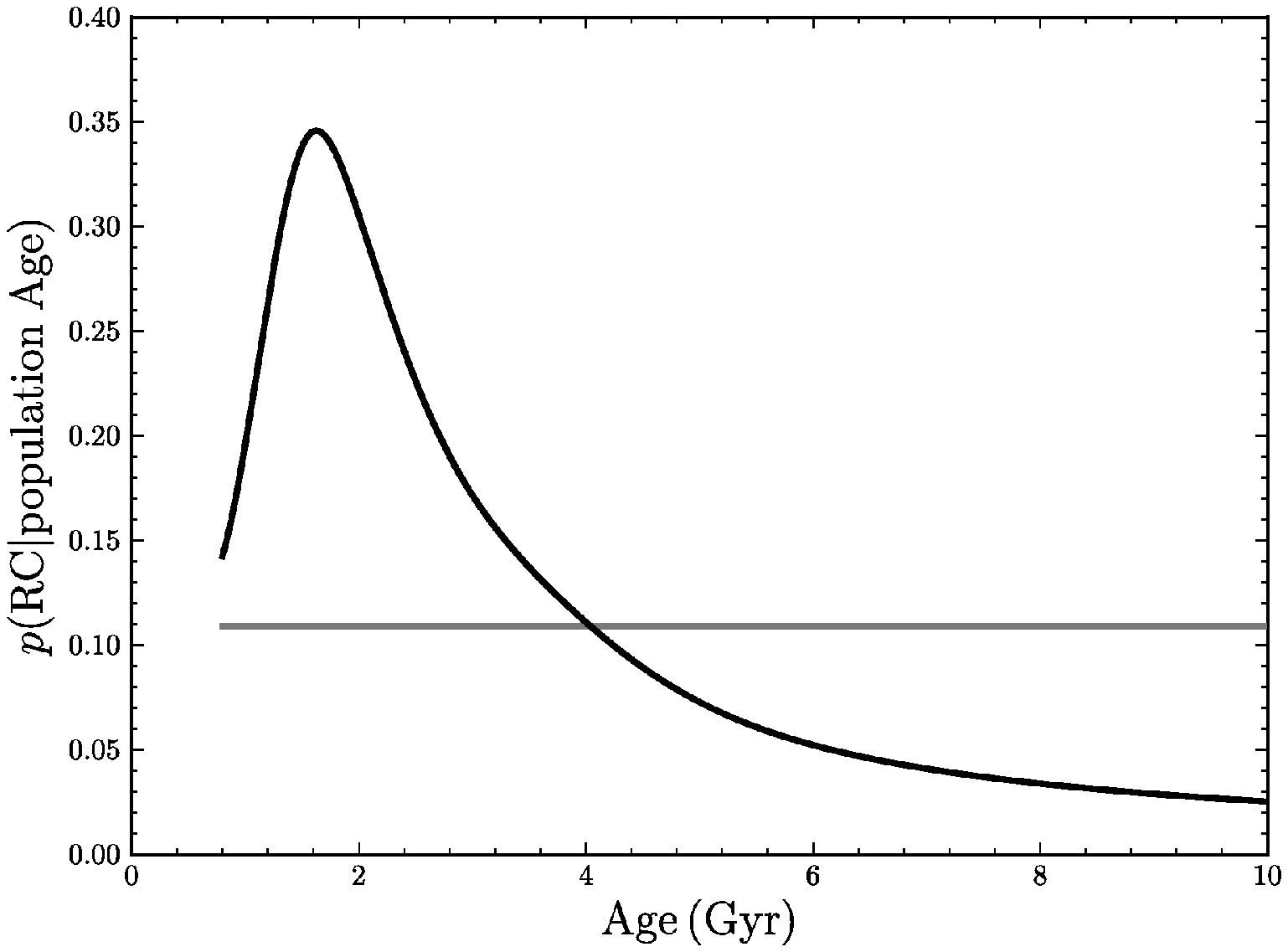}
\caption{Relative probability for stars in a stellar population to be
  present in the RC region as a function of age. For a flat SFH, this
  is equivalent to the age distribution for stars in the RC region;
  other SFHs need to be multiplied with this function to obtain the
  age distribution of the RC. We have assumed a solar-neighborhood
  metallicity distribution (\citealt{Casagrande11a}; see
  \sectionname~\ref{sec:distance}). An approximate functional form for
  the age distribution is given in
  \equationname~(\ref{eq:agedist}). Relative to a uniform age
  distribution (gray line) the RC is heavily weighted toward stars
  with ages between 1 and $4\Gyr$.}\label{fig:astrosf-age}
\end{figure}

The varying mass fractions in \figurename~\ref{fig:astrosf-frac} imply
that the RC as defined in this paper does not randomly sample the
underlying age distribution of stars, but is instead typically skewed
toward younger ages \citep[see
  also][]{Girardi01a}. \figurename~\ref{fig:astrosf-age} displays the
relative probability, as a function of age, for stars in a stellar
population that are present in the RC region. We calculate this
probability using a metallicity distribution similar to that in the
solar neighborhood (that of \citealt{Casagrande11a}, using the
representation described in \sectionname~\ref{sec:distance}) and it
can be approximated by the following functional form
\begin{align}\label{eq:agedist}
  \ln f(a) & = -1.6314+3.8230 a+2.2133 a^2-35.7414 a^3\nonumber\\
  & \qquad \qquad (-0.1 \leq a \leq 0.23)\,,\nonumber\\
  & = -1.0666+1.8664 a -9.0617 a^2 +4.5860 a^3\nonumber\\
  & \qquad \qquad (a > 0.23)\,,\nonumber\\
  & \qquad \mathrm{where}\ a = \log_{10} \left[\mathrm{Age} / 1\,\Gyr\right]
\end{align}
Relative to a constant age distribution, the RC selection biases the
sample to stars with ages between 1 and $4\Gyr$. We sample all ages in
the disk older than this, but the distribution is cut off below about
$800\Myr$. This effect is in large part due to our selection cuts: our
color--metallicity cuts in \sectionname~\ref{sec:sample} exclude the
bluest, youngest red-clump stars at a given metallicity, because their
luminosity distribution is not sufficiently narrow to allow precise
distances to be assigned for these stars. This is also the reason that
the red-clump age distribution derived by \citet{Girardi01a} peaks at
lower ages and extends to younger red clump stars; they considered the
full red clump, whereas we only consider the part of the RC that is
narrow in luminosity and which is typically older.

The discussion in this section is based on PARSEC stellar-evolution
models with mass loss on the giant branch modeled using the Reimers
law with an efficiency parameter of $\eta=0.2$ (see
\sectionname~\ref{sec:kde}). While mass loss affects a star's mass
during the RC phase, it neither influences the luminosity (see
\sectionname~\ref{sec:kde}) nor the lifetime of the RGB and RC
phases. Therefore, mass loss does not have any effect on how the RC
samples the underlying stellar population. We have explicitly checked
this by computing the total-stellar-population mass every RC star
represents (bottom panel of \figurename~\ref{fig:astrosf-frac}) and
the age sampling (\figurename~\ref{fig:astrosf-age}) for PARSEC models
with mass-loss efficiency parameters $\eta=0.1,0.2,0.3,$ and $0.4$. We
find small differences and in particular the age sampling of the RC is
robust with respect to changes in $\eta$.

From the discussion in this Section, it is clear that great care must
be taken when using the RC as a spectroscopic tracer of the Galactic
stellar distribution function, because the fraction of a stellar
population's mass contained in the RC region and the manner in which
other properties are sampled depends on the SFH---or, equivalently,
the age distribution. This is unavoidable when using evolved rather
than main-sequence stars as stellar-population tracers; as discussed
at the start of this section, the RC is a relatively long phase in the
evolution of post-main-sequence stars and is therefore among the best
population-tracers among giants stars. The fact that the relation
between the RC region and the full stellar population does not
strongly depend on metallicity for a given SFH, combined with the fact
that there is only a weak correlation observed in the solar
neighborhood between age and metallicity
\citep{Edvardsson93a,Nordstroem04a}, means that the RC sampling of the
metallicity distribution of stellar populations older than one
\Gyr\ should be relatively uniform. However, age is not sampled
uniformly (see \figurename~\ref{fig:astrosf-age}) and the detailed
distribution of elemental abundances likely depends strongly on age,
such that detailed studies need to take the likely age distribution of
the RC sample into account.

\section{The APOGEE-RC catalog}\label{sec:catalog}

In this Section we discuss the construction of the APOGEE-RC catalog
from the superset of data to be released as part of SDSS-III's DR11
and the contents of the catalog. We also briefly consider sources of
proper motions for the catalog stars. Because SDSS-III has a yearly
data-release schedule, we do not release the APOGEE-RC catalog with
this paper; it will instead be part of the SDSS-III DR11, currently
scheduled for Dec. 2014. At the time of writing this Dec. 2014 release
is planned to also contain the full three-year APOGEE data as part of
DR12. For public use, the DR11 APOGEE-RC catalog will therefore
immediately be made obsolete by the DR12 APOGEE-RC catalog. Full
details on how to obtain the APOGEE-RC catalog and what it contains
will be given in the DR11/DR12 documentation.

\subsection{Catalog creation}

The starting point is the catalog of all unique stars observed
spectroscopically, and to be released as part of DR11. For these
stars, we calculate extinction-corrected 2MASS $JH\ks$ magnitudes,
using the extinction corrections derived by the RJCE method
\citep{Majewski11a} for \apogee\ stars (we use the extinction
corrections calculated prior to target selection, for consistency). We
then employ the calibrated metallicities and surface gravities using
the corrections of \citet{Meszaros13a} and convert the metallicity
\feh\ to metal mass fraction $Z$ assuming $Z_\odot = 0.017$ and solar
abundance ratios. The dereddened color $(J-\ks)_0$, \logg, \teff\ and
metallicity $Z$ are then used to select RC stars using the cuts
described in \sectionname~\ref{sec:sample}. We also apply the
additional \logg--\teff\ cut of
\equationname~(\ref{eq:addllogg}). Stars selected in this way
constitute the full DR11 \apogee\ red-clump (\apogee-RC) catalog. This
sample consists of \ncatalog\ stars.

The absolute $\ks$-band magnitudes for all of the RC stars are
calculated using the model for $M_{\ks}$ shown in the top panel of
\figurename~\ref{fig:sig_jkz}, corrected for the calibration offset
determined in \sectionname~\ref{sec:distance}. Using the selection
function from \sectionname~\ref{sec:ssf}, we also determine which RC
stars are part of the statistical sample.

\subsection{Catalog contents}

\begin{figure}[t!]
\includegraphics[width=0.48\textwidth,clip=]{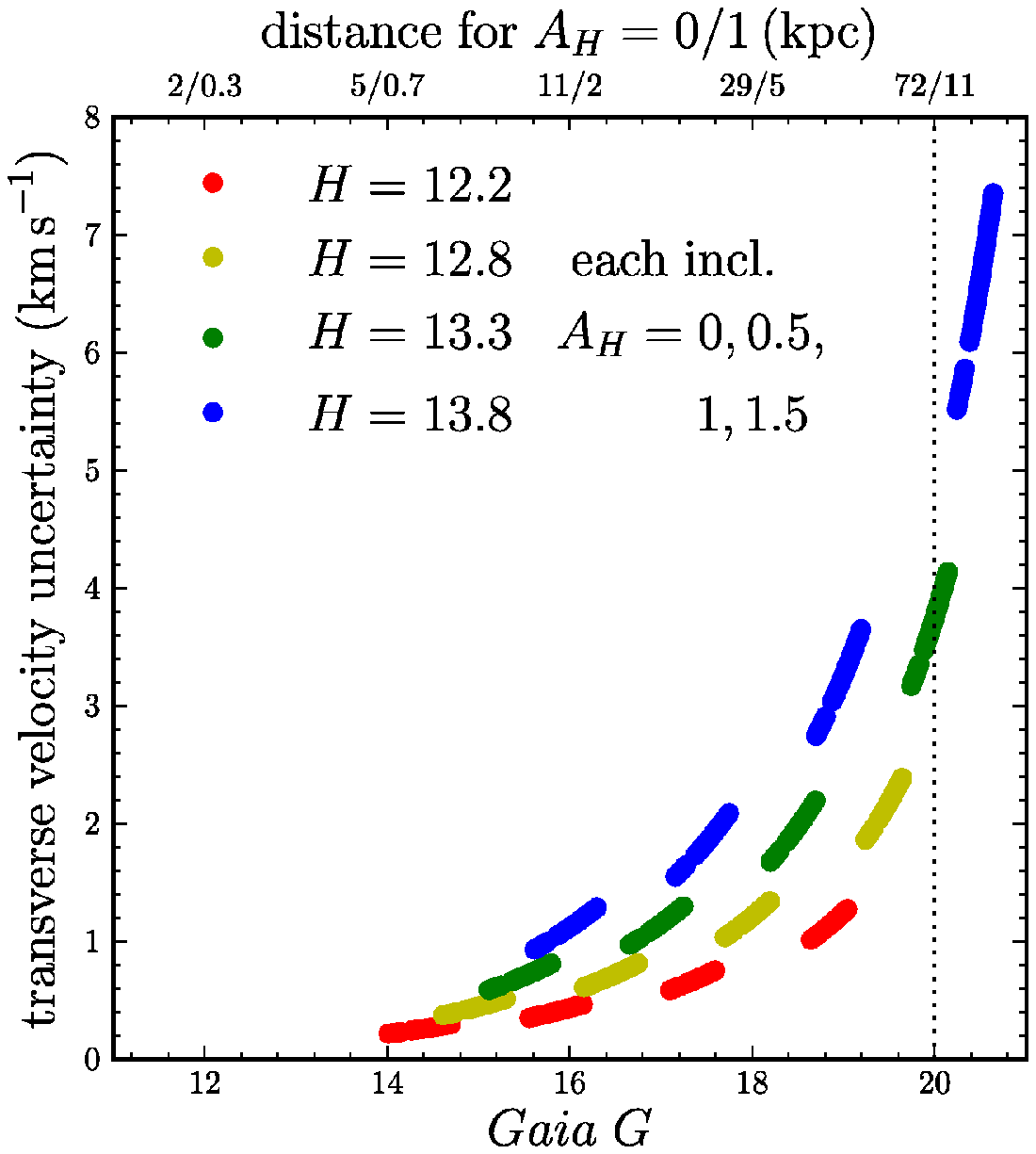}
\caption{Expected transverse-velocity precision of \emph{Gaia} for
  stars in the APOGEE-RC sample at different apparent magnitudes
  (colors) and different amounts of extinction (four bands for each
  color of increasing $G$ magnitude). The precisions only include the
  uncertainty arising from the proper motion, not that produced by the
  $5$ to $10\,\%$ RC distance uncertainty. These are end-of-mission
  precisions; the first \emph{Gaia} data release at launch+28 months
  with one third of the data will have proper-motion uncertainties
  that are $\approx3\sqrt{3}\approx5$ times larger. The top axis shows
  approximate distances computed assuming $M_G^{\mathrm{RC}} = 0.71$
  ($G-H \approx 2.2$ for the RC); $\approx90\,\%$ of stars in the DR11
  RC catalog have $A_H < 0.5$, with a median $A_H$ of $0.2$. The vast
  majority of stars in the APOGEE-RC sample will have precise proper
  motions out to $10\kpc$ from \emph{Gaia}, even in regions of high
  extinction.}\label{fig:gaiapm}
\end{figure}

The APOGEE-RC catalog contains all of the basic data derived from the
APOGEE spectra, such as the line-of-sight velocity, \teff, \logg,
\feh, as well as other abundances measured from the spectra. We also
include photometry from 2MASS, extinction corrections, APOGEE
targeting information, pipeline flags, and APOGEE identifiers to allow
the objects in the APOGEE-RC catalog to be matched to the full APOGEE
catalog. Full high-resolution spectra can be obtained from the
SDSS-III database by using the APOGEE identifiers. We include
distances derived from the absolute \ks\ magnitudes and, for
convenience, we add Galactocentric coordinates calculated assuming
$R_0 = 8\kpc$ and $Z_0 = 25\pc$.

The distance distribution of the \apogee-RC sample is presented in
\figurename~\ref{fig:disthist}. The distribution of the sample in the
volume around the Sun and in Galactocentric coordinates is shown in
\figurename~\ref{fig:distdist}. These distributions demonstrate that
the sample is dominated by disk stars at $|Z| \lesssim 1\kpc$, but the
sample also contains some stars in the stellar halo. Because the RC is
faint relative to more luminous red-giant \apogee\ targets and as
bulge fields are typically limited to $H \leq 11$ or $H \leq 12.2$,
there are relatively few bulge stars contained in this sample.

\subsection{Proper motions}\label{sec:pm}

We have added matches to the UCAC-4 \citep{Zacharias13a} and PPMXL
\citep{Roeser10a} proper-motion catalogs. However, the usefulness of
these catalogs, which have proper motions uncertainties
$\gtrsim2\mas\yr\inv$ (both statistical and systematic), for the
\apogee-RC sample is extremely limited. Stars in the \apogee-RC
catalog have typical distances of a few \kpc\ (see
\figurename~\ref{fig:disthist}). At these distances and farther,
Galactic rotation typically leads to proper motions of
$\approx2\mas\yr\inv$, while random motions---$\approx30\kms$ and
smaller at larger distances from the Galactic center---lead to proper
motions of about $6\mas\yr\inv\kpc$. Therefore, at a few \kpc\ the
proper motion uncertainties are similar or greater than the intrinsic
proper motions and the systematic uncertainties in these proper motion
catalogs make extracting proper-motions signals out of the noise
difficult.

\begin{figure}[t!]
\includegraphics[width=0.48\textwidth,clip=]{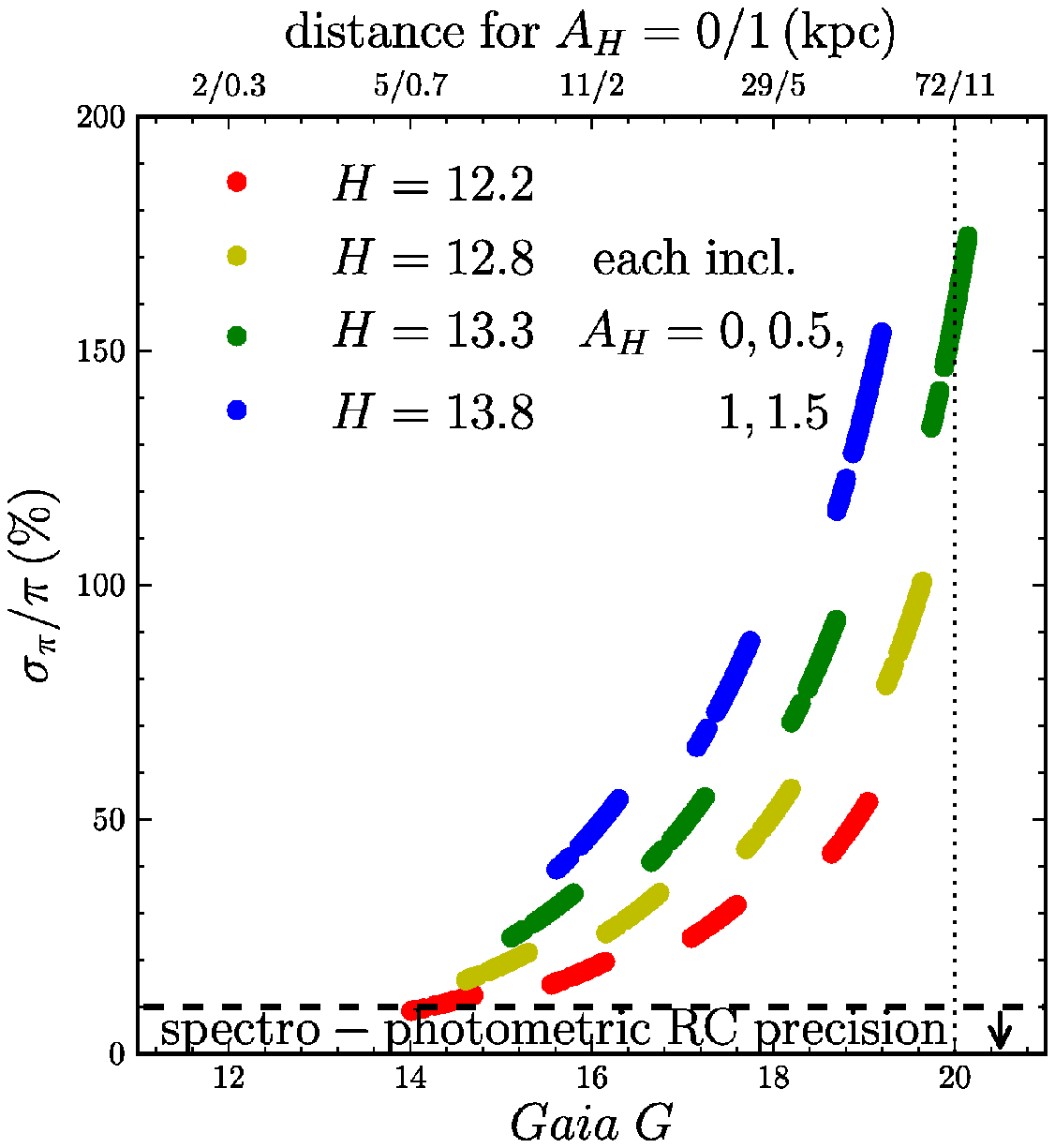}
\caption{Expected parallax precision of \emph{Gaia} for stars in the
  APOGEE-RC sample at different apparent magnitudes (colors) and
  different amounts of extinction (four bands for each color of
  increasing $G$ magnitude). These precisions are for the full five
  year mission; the first \emph{Gaia} data release at launch+28 months
  will have uncertainties that are $\approx\sqrt{3}\approx1.7$ times
  larger. The spectro-photometric RC distances are more precise than
  the \emph{Gaia} parallaxes for all but the brightest stars. At low
  extinction, the \emph{Gaia} parallaxes can identify RGB
  contamination in the RC sample.}\label{fig:gaiaplx}
\end{figure}

The \emph{Gaia} astrometric space mission, launched on Dec.~19 2013,
will measure proper motions with precisions better than
$200\muas\yr\inv$ for all stars with \emph{Gaia} $G\leq20$ over its
five-year mission. As $G$ is a white-light optical bandpass it is much
more strongly affected by dust extinction than the infrared
\apogee\ observations ($A_G / A_H \approx 4$). We estimate the
\emph{Gaia} $G$-band magnitudes of stars in the \apogee-RC sample by
using the PARSEC isochrones in the SDSS and 2MASS passbands in the RC
region defined in \sectionname~\ref{sec:sample} and using the
transformation between $g$, $g-z$ and $G$ of Table 5 of
\citet{Jordi10a}. We calculate $V-I_\mathrm{C}$ from $g-i$ as
\begin{align}
  V-I_\mathrm{C} & = 0.675 \,(g-i) + 0.364\ \ \mathrm{if} \ g-i \leq 2.1\\
  V-I_\mathrm{C} & = 1.110 \,(g-i) -0.520\ \ \mathrm{if} \ g-i > 2.1\,,
\end{align}
inspired by the data of \citet{Jordi10a}. For a given extinction $A_H$
we calculate the extinction in the \emph{Gaia} band using the highest
extinction entry in Table 8 of \citet{Jordi10a} and assuming $A_H /
A_V = 0.18307$. We then use the projected astrometric performance as a
function of $(G,V-I_\mathrm{C})$ from the \emph{Gaia} Performance
website\protect\footnote{Available at the following
  URL\\\protect{\scriptsize
    \url{http://www.rssd.esa.int/index.php?page=Science_Performance&project=GAIA}}~.}
to calculate the proper-motion precision. The resulting expected
precision in the transverse motion based on the \emph{Gaia} proper
motions at different apparent magnitudes and for different amounts of
extinction is shown in \figurename~\ref{fig:gaiapm}. These precisions
only include the effect of the proper motion uncertainty; additional
uncertainty comes from the $5$ to $10\,\%$ RC distance
uncertainty. The variation at a given magnitude in this figure is due
to the spread in $G$ and $V-I_\mathrm{C}$ in the RC region for a given
magnitude.

\begin{figure}[t!]
\includegraphics[width=0.48\textwidth,clip=]{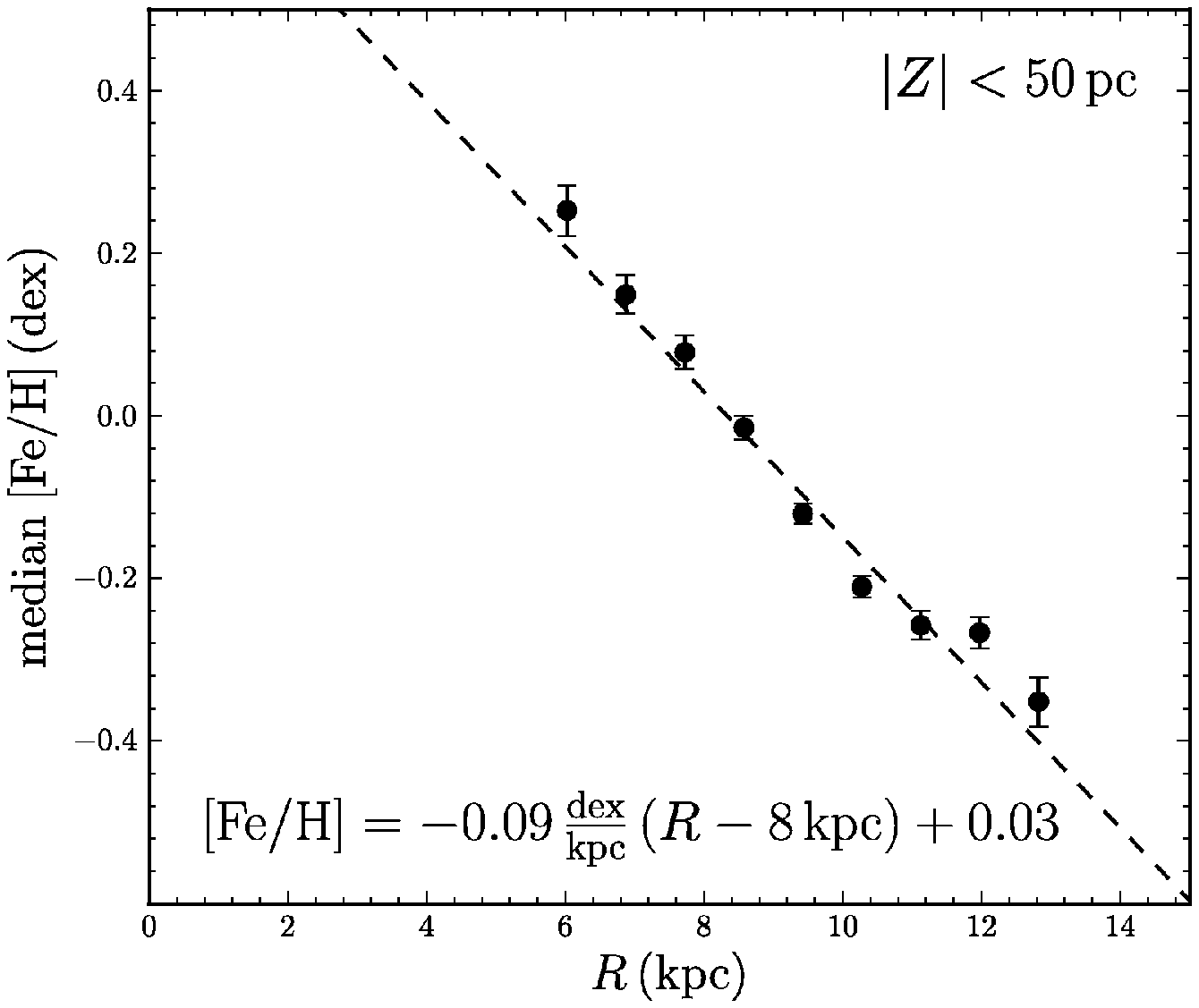}
\caption{Radial metallicity gradient in the midplane of the Milky Way
  (heights $<50$ pc) from 971 stars near the midplane in the DR11
  APOGEE-RC sample. The uncertainties in the slope and zero-point are
  both $0.01\dex$. The metallicity gradient of stars that are a few
  Gyrs old is quite steep in the midplane.}\label{fig:fehgrad}
\end{figure}

\figurename~\ref{fig:gaiapm} shows that, even at large extinctions,
most stars in the \apogee-RC catalog will be bright enough to have
\emph{Gaia} proper motions, with only the faintest and most heavily
extincted stars falling below the $G=20$ \emph{Gaia} magnitude
limit. Therefore, \emph{Gaia} will provide highly-precise proper
motions for the vast majority of our sample, leading to errors in the
transverse velocity $\lesssim 5\kms$. At the time of writing,
\emph{Gaia} is expected to release preliminary proper motions 28
months after launch using 20 months of data. The proper motion
uncertainties in that first data release are expected to be worse by a
factor of $\approx 5$ ($[20/60]^{-3/2}$). Even with these larger
uncertainties, most stars in the \apogee-RC sample will have
sub-$\mas\yr\inv$ proper motions and most will have uncertainties
below $250\muas\yr\inv$, or $10\kms$.

In \figurename~\ref{fig:gaiaplx} we display the expected \emph{Gaia}
relative parallax precision for stars in the APOGEE-RC catalog,
computed in a similar way as the proper motion uncertainties above
(the expected end-of-missio proper motion uncertainties are half the
expected parallax uncertainties). The parallaxes that will be
available in the launch+28 months release will have uncertainties that
are larger by a factor of $\approx 1.7$ ($[20/60]^{-1/2}$). The
spectro-photometric RC distances computed in Sections \ref{sec:sample}
and \ref{sec:distance} are more precise than the \emph{Gaia}
parallaxes for all but the brightest stars that do not suffer from
much extinction; these bright RC stars will be highly informative for
an improved calibration of the RC distances. However, for many RC
candidates, the \emph{Gaia} parallaxes could distinguish between true
RC stars or RGB interlopers by comparing the spectro-photometric
distance to the parallax.

\section{Azimuthal metallicity variations in the Milky Way midplane}\label{sec:science}

The RC catalog with high-resolution abundances and very accurate
distances will allow the distribution of abundances in the Milky Way's
disk to be mapped in unprecedented detail. As an illustration of this
capability, we discuss in this Section spatial variations in the
median metallicity close to the midplane of the Milky Way's
disk. Because there is a strong radial metallicity gradient for the
intermediate-age stars in the RC catalog (see below), the average
metallicity can be used as a tracer of the orbits of stars with
different mean orbital radii. For example, if closed orbits in the MW
disk are non-circular (\eg, elliptical as discussed below), then
the azimuthal dependence of the radius of a closed orbit maps into an
azimuthal dependence of the average \feh. Non-steady-state
non-axisymmetric flows, such as those induced by growth of the
Galactic bar or transient spiral structure, can also be traced by
azimuthal \feh\ variations, because the inhomogeneities in the
\feh\ distribution induced by these perturbations will remain visible
for multiple dynamical times. As a specific example, recent
simulations of the evolution of galactic disks \citep{DiMatteo13a}
have shown that for galaxies with a radial metallicity gradient, the
presence (or absence) of azimuthal metallicity variations in the stars
can constrain recent radial mixing induced by the bar. Mixing induced
by transient spiral structure \citep{Sellwood02a} would lead to
smaller, but similar effects. Such mixing would induce azimuthal
\feh\ variations of comparable magnitude to the radial variations that
last for $\gtrsim1\Gyr$.

We start by showing the radial midplane metallicity gradient in
\figurename~\ref{fig:fehgrad} , specifically the median metallicity as
a function of Galactocentric radius for all stars in the RC catalog
contained within $50\pc$ from the midplane. The data can be well fit
by a single linear relation (except for a discrepant point at
$R\approx12\kpc$) with a slope of $-0.09\pm0.01\dex\kpc^{-1}$ and a
local normalization at $R_0 = 8\kpc$ of $0.03\pm0.01\dex$. Thus, the
median metallicity in the solar neighborhood is approximately solar
and the metallicity gradient is quite steep for stars that are
typically between $1$ and $5\Gyr$ old (see
\sectionname~\ref{sec:astrosf}). This measurement agrees with previous
determinations of the metallicity gradient of intermediate-age stars
(see \citealt{Nordstroem04a,Hayden14a}, and references therein) and
open clusters older than $\approx1\Gyr$
\citep[\eg,][]{Yong12a,Frinchaboy13a}.

\begin{figure}[t!]
\includegraphics[width=0.48\textwidth,clip=]{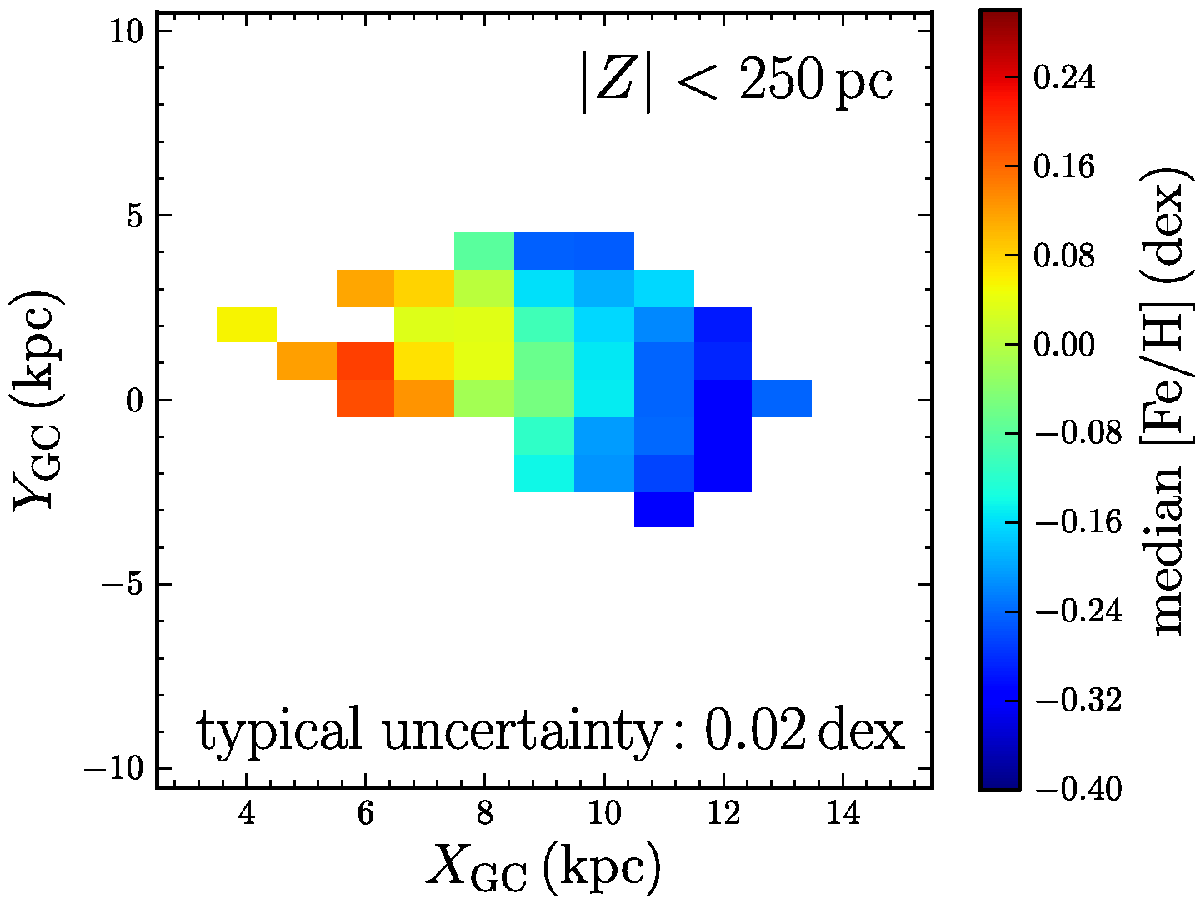}\\
\caption{Two-dimensional distribution of the median metallicity near the
  midplane (heights $<250$ pc) from 4,330 stars in the DR11 APOGEE-RC
  sample. This figure shows a map of the median metallicity in
  rectangular coordinates.}\label{fig:fehaz}
\end{figure}

\begin{figure}[t!]
\includegraphics[width=0.48\textwidth,clip=]{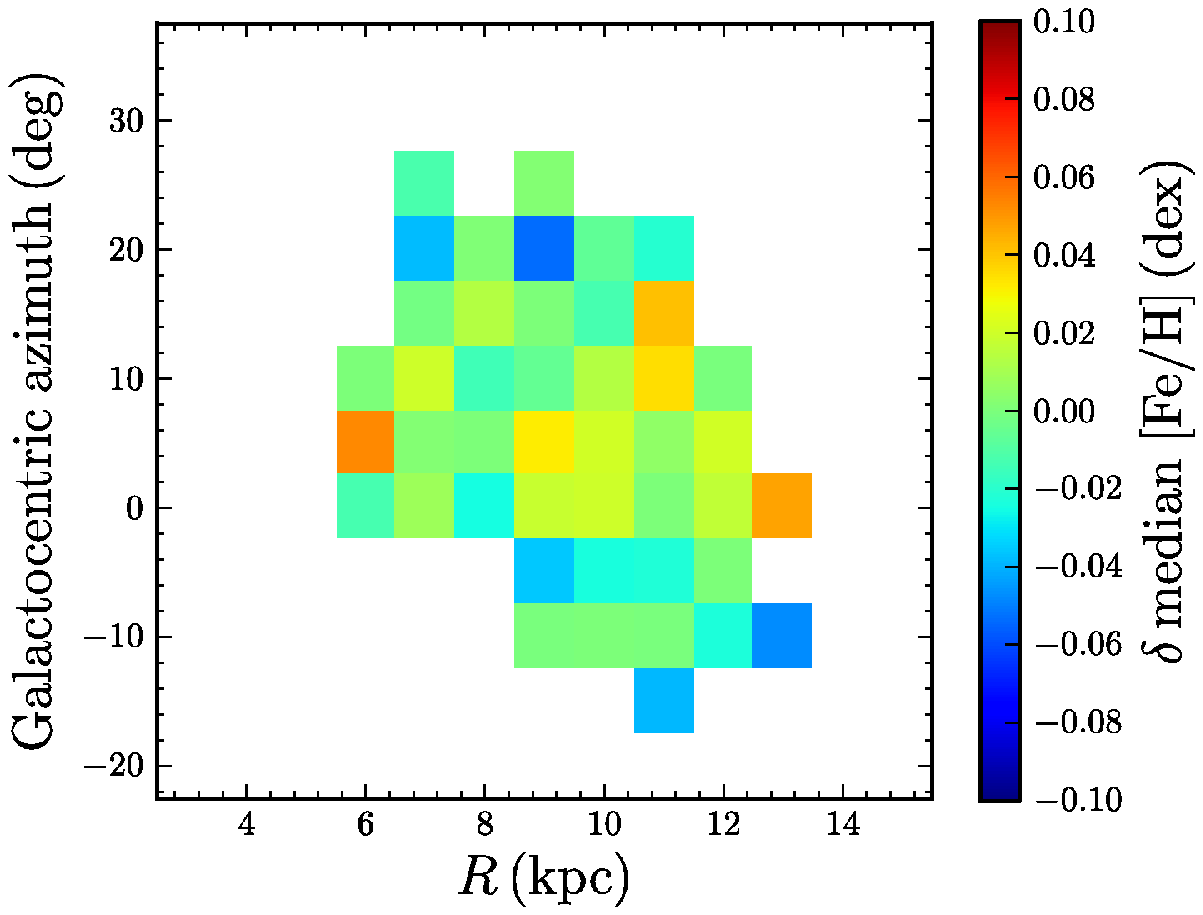}
\caption{Absence of azimuthal metallicity variations in the Milky Way
  disk. This figure shows the azimuthal variation of the median
  metallicity (with respect to the median for each radius, see
  \figurename~\ref{fig:fehgrad}) in cylindrical coordinates. There are
  no azimuthal variations in the $\approx45^\circ$ region near the Sun
  to within the uncertainties of $\approx0.02\dex$.}\label{fig:fehaz2}
\end{figure}

The RC catalog covers a region around the Sun that spans about
$45^\circ$ in Galactocentric azimuth $\phi$, allowing the mapping of
the median metallicity in both $R$ and $\phi$. We extend the range in
vertical heights to include all stars in the RC catalog within
$250\pc$ from the midplane such that we can use small pixels in $R$
and $\phi$, as allowed by our precise
distances. \figurename~\ref{fig:fehaz} presents the two-dimensional
behavior of the median metallicity in rectangular
coordinates. \figurename~\ref{fig:fehaz2} shows the azimuthal
metallicity residuals with respect to the median metallicity at each
radius in cylindrical coordinates. The typical uncertainty in the
median metallicity of a given two-dimensional spatial pixel is about
$0.02\dex$. The residual map in \figurename~\ref{fig:fehaz2} clearly
demonstrates that there are no significant azimuthal variations within
the observed octant above the uncertainty of $0.02\dex$; deviations
that are present have a standard deviation of $0.02\dex$. Remarkably,
this constraint is more than an order of magnitude tighter than the
radial variation of the metallicity
(\figurename~\ref{fig:fehgrad}). This is the first time that azimuthal
variations in the metallicity distribution of intermediate-age stars
have been constrained. Previous investigations of azimuthal gradients
in the abundance distribution of young tracers of the disk (ages
$\lesssim 1\Gyr$) have obtained mixed results, with some studies
seeing variations \citep{Davies09a}, while others do not
\citep{Luck11a}. Because young tracers have ages that are not much
larger than the rotational period ($T_\phi\approx250\Myr$; see below),
their azimuthal distribution likely more strongly reflects their birth
properties than dynamical stellar mixing. The opposite is the case for
the RC stars that we use, which are all older than $1\Gyr \approx
4\,T_\phi$ and are typically $2\Gyr=8\,T_\phi$
(\figurename~\ref{fig:astrosf-age}).

The absence of significant variations in the median metallicity with
Galactocentric azimuth limits the extent to which closed orbits in the
Galactic disk are non-circular. As an example we consider that the MW
disk is elliptical due to a $m=2$ perturbation $\Delta \Phi =
\varepsilon_\Psi V_c^2 / 2 \cos 2\,(\phi-\phi_b)$ to the potential
$\Phi_0$ where $\varepsilon_\Psi$ is the dimensionless amplitude of
the perturbation, $V_c$ is the circular velocity, and $\phi_b$ is a
position angle \citep{Kuijken94a}; we assume that the unperturbed
circular velocity curve is flat, \ie, $\Phi_0 \propto \ln R$. The
ellipticity of the equipotential surfaces and of the closed orbits is
$\approx 1-\varepsilon_\Psi$. Such ellipticity in the MW is only
mildly constrained, especially if the Sun lies near the minor or major
axis of the perturbation \citep{Kuijken91a,Kuijken94a}, but
ellipticities of $\approx5\,\%$ are common in external MW-like
galaxies \citep{Rix95a}. If $\varepsilon_\Psi \approx 0.05$ in the MW,
then there will be coherent non-circular flows with an amplitude of
$\approx10\kms$ and estimates of the circular velocity could be
impacted by similar amounts \citep{Kuijken94a}.

\begin{figure}[t!]
\includegraphics[width=0.48\textwidth,clip=]{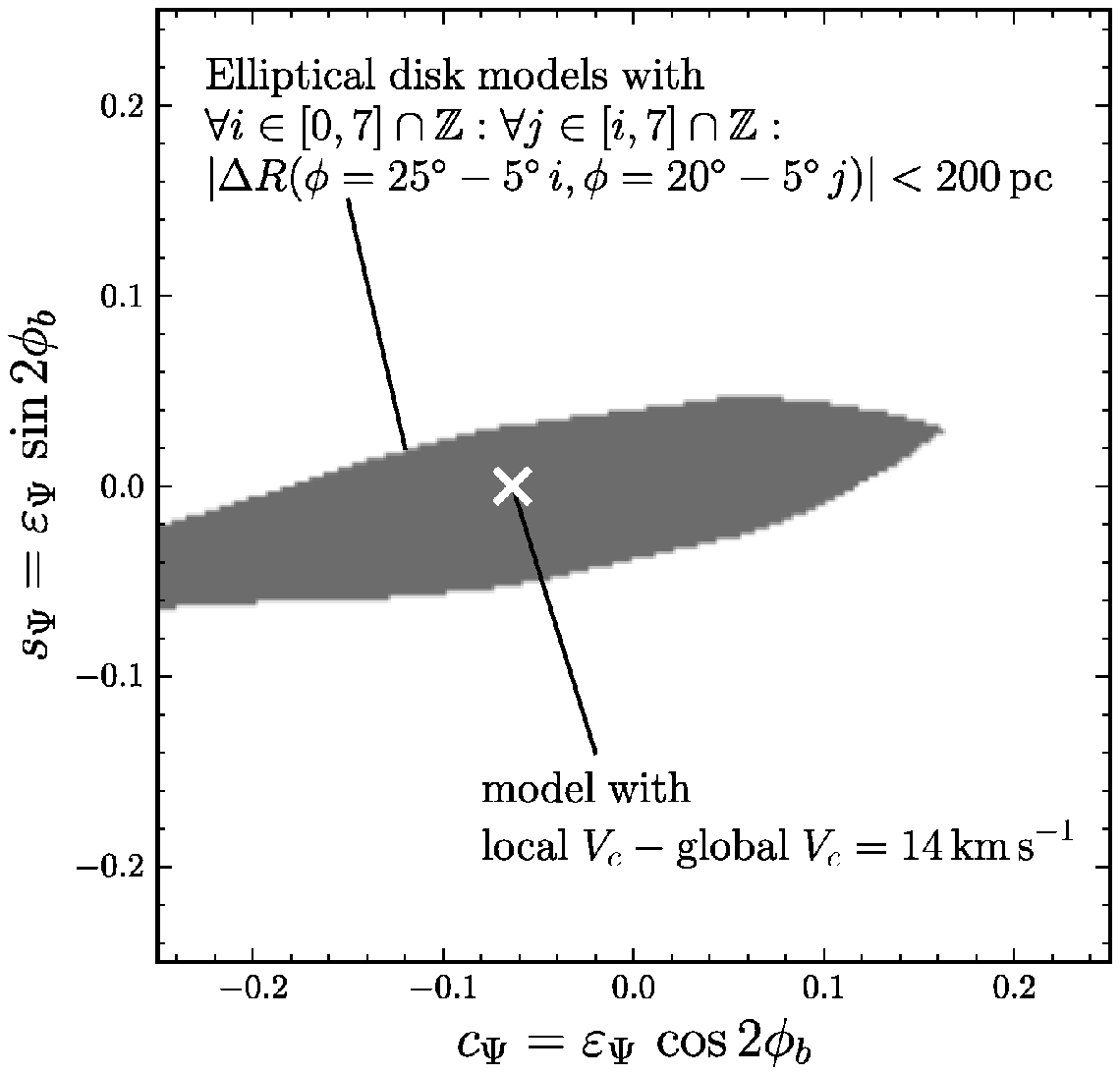}
\caption{Constraints on the ellipticity of the MW's disk from the
  absence of azimuthal metallicity variations within a
  $\approx45^\circ$ region near the Sun. The ellipticity is induced by
  a static $\cos 2\phi$ perturbation to an axisymmetric,
  flat-rotation-curve background potential with equipotential surfaces
  with ellipticity $1-\varepsilon_\Psi$ and a position angle
  $\phi_b$. The constraints are shown in terms of parameters $c_\Psi$
  and $s_\Psi$ that describe perturbations that are symmetric and
  asymmetric, respectively, with respect to the Sun--Galactic-center
  line. Ellipticity that leads to radial variations larger than
  $200\pc$ for closed orbits within the observed azimuthal range would
  induce azimuthal \feh\ variations $\gtrsim0.02\dex$ and are
  therefore ruled out. Asymmetric ($|s_\Psi| \gtrsim 0.05$) and
  positive, symmetric ($c_\Psi \gtrsim0.05$) are strongly
  constrained. Models in which the Local Standard of Rest (LSR), \ie,
  the rotational velocity of the local closed orbit, is ahead of the
  axisymmetric circular velocity without an accompanying mean radial
  motion of the LSR ($c_\Psi < 0, s_\Psi \approx 0$) are allowed. An
  example of such a model is indicated by the
  $\times$.}\label{fig:ellip}
\end{figure}

Elliptical motions that lead to radial variations larger than $200\pc$
along closed orbits within the $\approx45^\circ$ azimuthal region near
the Sun lead to $\gtrsim0.02\dex$ azimuthal variations in the average
metallicity. As we do not observe such variations, these closed orbits
are disfavored by the data. Specifically, we consider any elliptical
closed orbit that leads to $|\Delta R| > 200\pc$ for any of the
$|\Delta \phi|$ between $25^\circ$ and $-15^\circ$ to be ruled out. We
present these constraints in \figurename~\ref{fig:ellip}. This figure
uses an alternative parameterization of the elliptical-disk potential
where the perturbation is split into a component $c_\Psi =
\varepsilon_\Psi\,\cos 2\phi_b$ that is symmetric with respect to the
Sun--Galactic-center line and a component $s_\Psi =
\varepsilon_\Psi\,\sin 2\phi_b$ that is asymmetric. Elliptical models
where the Sun lies on the major or minor axis ($s_\Psi \approx 0$ and
$c_\Psi > 0$ and $< 0$, respectively) are much less constrained than
models where the Sun is in between the major and minor axis ($c_\Psi
\approx 0$). This is because for the latter models $R(\phi)$ is almost
entirely monotonic with $\phi$ within the observed region, such that
large $|\Delta R|$ occur; if the Sun lies near the major or minor
axis, much smaller $|\Delta R|$ exist. Models where the Sun lies near
the minor axis are the least constrained, with large $c_\Psi < 0$
allowed by the data. In such models the closed orbit at the Sun's
position has a rotational velocity that is larger than the average,
axisymmetric $V_c$ by $|c_\Psi|\,V_c$; offsets up to $40\kms$ are
allowed by these data. In particular, a model where the closed orbit
near the Sun is ahead of the average $V_c$ by $14\kms$ as proposed by
\citet{BovyVc} to explain the APOGEE stellar kinematics in the disk,
is allowed; this model is indicated in \figurename~\ref{fig:ellip} by
a white cross. RC data from the final year of APOGEE combined with
data from APOGEE-2 will extend the mapping of azimuthal
\feh\ variations over at least $\phi=45^\circ$ to
$\phi=-45^\circ$. This mapping will allow limits on $c_\Psi$ and
$s_\Psi$ that are $\lesssim 0.05$.

If the Milky Way's bar or spiral structure now or in the recent past
had induced significant radial mixing of stars, this generically leads
to azimuthal variations over similar spatial scales. The fact that we
observe azimuthal metallicity variations to be smaller than the radial
metallicity variation over $\approx200\pc$ (the metallicity gradient
is $\approx0.1\dex\kpc^{-1}$ or $0.02\dex (200\pc)^{-1}$), indicates
that no significant radial mixing over these spatial scales has
recently occurred (see \citealt{DiMatteo13a} for a detailed discussion
of this in the context of mixing induced by a bar; see also the
azimuthal variations in the density of migrating stars in Figure 10 in
\citealt{Roskar12a}); specifically, $\lesssim10\,\%$ of stars can have
migrated $\gtrsim2\kpc$ with $\Delta \phi \approx 45^\circ$ near the
Sun. Any azimuthal variation will smooth out after the mixing
ends. Because the Milky Way's rotation curve is approximately flat in
the region where we map the median metallicity
\citep[\eg,][]{GKT79,BovyVc}, $|\dd \Omega / \dd R| \approx \Omega/R$,
such fluctuations over a scale $\Delta R$ across an azimuthal range
$\Delta \phi$ near the Sun would disappear in approximately
$\mathrm{few} \times T_\phi \times (\Delta \phi/ 360^\circ) \times
\left(R_0 / \max(\Delta R,\sigma_{R_g})\right)$, where $T_\phi$ is the
rotational period ($\approx250\Myr$) and $\sigma_{R_g}$ is the spread
in guiding-star radii ($\approx 0.1\,R$ near $R_0$ or about $1\kpc$
for the intermediate-age disk; \citealt{BovyVc}). For fluctuations
below $1\kpc$ over the $45^\circ$ range probed by the DR11 APOGEE-RC
catalog, this timescale is approximately $\mathrm{a\ few}\times
250\Myr$ near the Sun and about $50\,\%$ larger in the outer disk
regions shown in \figurename~\ref{fig:fehaz2}.  Therefore, our
constraints on mixing on kpc scales apply to approximately the last
$\Gyr$ of the MW's evolution. However, quantitative constraints on
realistic migration scenarios will require more detailed modeling of
the azimuthal signature of radial migration, its dissolution over
time, and its observability within the APOGEE survey volume.

The absence of azimuthal variations in the median metallicity can also
be used to constrain distance systematics. Any systematic distance
offsets in the RC sample would induce azimuthal gradients in
metallicity: such offsets would distort rings of stars at constant $R$
onto a locus spanning a range of apparent $\tilde{R}$, with a strong
correlation between $\tilde{R}-R$ and azimuth $\phi$. The metallicity
distribution of stars at a constant apparent $\tilde{R}$ would
therefore show apparent azimuthal variations that correspond to the
real radial variations. Changing all of the distances in the RC
catalog by $50\,\%$ upward or downward induces clear azimuthal
variations at large and small $R$, respectively. \emph{Assuming that
  there are no intrinsic azimuthal variations}, this result excludes
such large distance systematics. Smaller distance systematics induce
smaller azimuthal gradients, and azimuthal variations over the volume
probed by the RC catalog are minimized for distance systematics within
a few percent of the distances that we assigned in
\sectionname~\ref{sec:distance} (using the maximum of the median
absolute azimuthal deviations $\delta \feh$ in $\Delta R = 1\kpc$ bins
to quantify the accuracy of the distances). Thus, we conclude from
this test that the distances in the RC catalog are accurate to a few
percent, in agreement with the considerations in
\sectionname~\ref{sec:distance}. Because contamination by RGB stars at
significantly different distances (typical differences of
$\sim\!70\,\%$) would lead to distance systematics, this test also
indicates that the overall RGB contamination must be $\lesssim5\,\%$,
in accordance with the direct contamination estimates using
asteroseismic evolutionary-state classifications in
\sectionname~\ref{sec:sample}.

\section{Conclusion}\label{sec:conclusion}

We have presented a new method for selecting individual likely RC
stars from spectro-photometric data. This approach was used to produce
a sample of RC stars with high purity from the APOGEE data set. The
selection technique is a combination of simple cuts in
$(\logg,\teff,\feh,[J-\ks]_0)$ motivated using stellar isochrone
models and calibrated using high-quality seismic \logg\ data from a
subsample of APOGEE stars with measured oscillation frequencies from
\emph{Kepler}. These cuts in particular select RC stars for which the
intrinsic absolute-magnitude distribution is very narrow ($\sigma
\lesssim 0.1\magunit$), such that highly precise distances for these
stars can be obtained. Tests using \emph{Kepler} stars for which the
evolutionary state was measured from their photometric variability
show that the purity of the sample is $\gtrsim 93\,\%$. Using stellar
isochrone models we calculate the small ($\lesssim 5\,\%$) deviations
from a standard candle RC magnitude and we calibrate the distances
against RC stars in \emph{Hipparcos}. This procedure results in
distances that are unbiased to within $\approx 2\,\%$ and have random
uncertainties of $\approx 5\,\%$.

We produce the APOGEE-RC catalog using the new RC selection method and
distance calculation. This catalog is currently based on SDSS DR11,
and contains \ncatalog\ stars with accurate and precise distances,
stellar parameters, and elemental abundances. The catalog will be
released publicly as part of the public combined DR11/DR12 data
release (currently scheduled for Dec. 2014); full details on the
contents of the catalog will be given in the documentation
accompanying this data release. Future data releases of SDSS-IV will
expand the current RC catalog and the data it contains.

The APOGEE target selection and the subsequent RC selection introduces
biases in the way the underlying stellar populations are represented
in the sample. These biases are important for many detailed
investigations of the abundance structure of the disk and need to be
corrected for. As discussed in \sectionname~\ref{sec:ssf}, we have
determined the manner in which stars in the RC catalog were selected
from the underlying 2MASS photometry for a subsample of stars for
which this is possible. Selection weights for this statistical sample
are included in the catalog and code that evaluates the selection
function will be made publicly available with the release of the
catalog. Beyond this first bias are the astrophysical selection
effects due to the fact that RC stars as any other kind of giants are
an age- and metallicity-biased tracer of the underlying
populations. These effects are discussed in detail in
\sectionname~\ref{sec:astrosf}. The main conclusion from this analysis
are that the biases are relatively constant with metallicity for a
given star-formation history, but they depend at the $25\,\%$ level on
the star-formation history.

The RC catalog will be useful for exploring the full stellar
distribution function of spatial location, kinematics, and abundances
in the Milky Way disk, because of the large volume covered by the
data, the precise distances permitting study of the smooth structure
and any substructure to be resolved, and the availability of a large
number of elemental abundances from the high-resolution APOGEE
spectra. In \sectionname~\ref{sec:pm} we showed that Gaia will soon
provide highly precise proper motions for the vast majority of stars
in the RC catalog, giving transverse velocities precise to $< 5\kms$
over the volume covered by the RC catalog, while the photometric RC
distances will remain more precise than the Gaia parallaxes. This
information, combined with the excellent APOGEE line-of-sight
velocities with uncertainties $\lesssim0.1\kms$, will allow
explorations of the kinematics in the Milky Way disk that are not
limited by velocity uncertainties. Currently, stellar parameters
measured from the APOGEE spectra are limited to \teff, \logg, \feh,
and \afe, but these already lead to qualitatively new constraints on
the abundance distribution in the Milky Way's disk. In
\sectionname~\ref{sec:science} we used the current RC catalog to make
the first map of the azimuthal metallicity distribution of stars of a
few Gyrs old and we limit azimuthal variations of the median
metallicity to be $\lesssim0.02\dex$. This result limits the overall
ellipticity of the MW disk and the amount of large-scale stellar
radial mixing within the last Gyr. In D.~Nidever \etal\ (2014, in
preparation) we explore the distribution of (\feh,\afe) over $5\kpc <
R < 12\kpc$ and $|Z| < 2\kpc$ with the large data set of RC stars,
which allows strong, novel constraints on the chemical evolution of
the Milky Way's disk.


\acknowledgements It is a pleasure to thank Friedrich Anders, Victor
Debattista, Rok Ro{\v s}kar, Martin Smith, and Scott Tremaine for
helpful comments and discussions. J.B. was supported by NASA through
Hubble Fellowship grant HST-HF-51285.01 from the Space Telescope
Science Institute, which is operated by the Association of
Universities for Research in Astronomy, Incorporated, under NASA
contract NAS5-26555. J.B. also gratefully acknowledges the hospitality
of the Asia Pacific Center for Theoretical Physics (APCTP) during the
7th Korean Astrophysics Workshop where parts of this research were
performed. D.L.N. was supported by a McLaughlin Fellowship at the
University of Michigan. H.W.R. has been supported by the European
Research Council under the European Union's Seventh Framework
Programme (FP 7) ERC Grant Agreement n. [321035]. P.M.F. was supported
by a NSF AST-1311835 grant. T.S.R. has been supported by CNPq-Brazil
scholarship. S.B. acknowledges partial support from NSF grant
AST-1105930 and NASA grant NNX13AE70G. T.C.B. acknowledges partial
support from grant PHY 08-22648: Physics Frontier Center / Joint
Institute for Nuclear Astrophysics (JINA), awarded by the
U.S. National Science Foundation. W.J.C. and Y.E. acknowledge the
support of the UK Science and Technology Facilities Council
(STFC). Funding for the Stellar Astrophysics Centre is provided by The
Danish National Research Foundation (grant agreement No.:
DNRF106). D.A.G.H. and O.Z. acknowledge support provided by the
Spanish Ministry of Economy and Competitiveness under grant
AYA-2011-27754. The research leading to the presented results has
received funding from the European Research Council under the European
Community's Seventh Framework Programme (FP7/2007-2013) / ERC grant
agreement no 338251 (StellarAges) and from the Deutsche
Forschungsgemeinschaft (DFG) under grant SFB 963/1 ``Astrophysical
flow instabilities and turbulence''. The research is supported by the
ASTERISK project (ASTERoseismic Investigations with SONG and Kepler)
funded by the European Research Council (Grant agreement no.:
267864). A.S. is partially supported by the MICINN grant
AYA2011-24704.

Funding for SDSS-III has been provided by the Alfred P. Sloan
Foundation, the Participating Institutions, the National Science
Foundation, and the U.S. Department of Energy Office of Science. The
SDSS-III web site is http://www.sdss3.org/.

SDSS-III is managed by the Astrophysical Research Consortium for the
Participating Institutions of the SDSS-III Collaboration including the
University of Arizona, the Brazilian Participation Group, Brookhaven
National Laboratory, Carnegie Mellon University, University of
Florida, the French Participation Group, the German Participation
Group, Harvard University, the Instituto de Astrofisica de Canarias,
the Michigan State/Notre Dame/JINA Participation Group, Johns Hopkins
University, Lawrence Berkeley National Laboratory, Max Planck
Institute for Astrophysics, Max Planck Institute for Extraterrestrial
Physics, New Mexico State University, New York University, Ohio State
University, Pennsylvania State University, University of Portsmouth,
Princeton University, the Spanish Participation Group, University of
Tokyo, University of Utah, Vanderbilt University, University of
Virginia, University of Washington, and Yale University.



\begin{thebibliography}{}

\bibitem[Alvarez \& Plez(1998)]{Alvarez98a}
  Alvarez,~R. \& Plez,~B. 1998, \aap, 330, 1109
\bibitem[Alves(2000)]{Alves00a}
  Alves, D.~R. 2000, \apj, 539, 732
\bibitem[Anders \etal(2014)]{Anders14a}
  Anders,~F., Chiappini,~C., Santiago,~B.~X., \etal\ 2014, \aap, 564, A115
\bibitem[Bedding \etal(2011)]{Bedding11a}
  Bedding,~T.~R., Mosser,~B., Huber,~D., \etal\ 2011, \nat, 471, 608
\bibitem[Bessell \& Brett(1988)]{Bessel88a}
  Bessell,~M.~S. \& Brett,~J.~M., 1988, \pasp, 100, 1134
\bibitem[Binney \etal(2013)]{Binney13a}
  Binney,~J., Burnett,~B., Kordopatis,~G., \etal\ 2014, \mnras, 437, 351
\bibitem[Bonatto \etal(2004)]{Bonatto04a}
  Bonatto,~C., Bica,~E., \& Girardi,~L.\ 2004, \aap, 415, 571
\bibitem[Bovy \etal(2009)]{Bovy09b}
  Bovy,~J., Hogg,~D.~W., \& Rix, H.-W. 2009, \apj, 704, 1704
\bibitem[Bovy \etal(2012a)]{BovyVc}
  Bovy,~J., Allende~Prieto,~C., Beers,~T.~C., \etal\ 2012a, \apj, 759, 131
\bibitem[Bovy et al.(2012b)]{BovyNoThickDisk}
  Bovy, J., Rix, H.-W., \& Hogg, D.~W.\ 2012b, \apj, 751, 131
\bibitem[Bovy et al.(2012c)]{BovyMAPkinematics}
  Bovy, J., Rix, H.-W., Hogg, D.~W., et al.\ 2012c, \apj, 755, 115
\bibitem[Bovy et al.(2012d)]{BovyMAPstructure}
  Bovy, J., Rix, H.-W., Liu, C., et al.\ 2012d, \apj, 753, 148
\bibitem[Bovy \& Rix(2013)]{Bovy13a}
  Bovy,~J. \& Rix,~H.-W. 2013, \apj, 779, 115
\bibitem[Bressan \etal(2012)]{Bressan12a}
  Bressan,~A., Marigo,~P., Girardi,~L., \etal\ 2012, \mnras, 427, 127
\bibitem[Brown \etal(2011)]{Brown11a}
  Brown,~T.~M., Latham,~D.~W., Everett,~M.~E., \& Esquerdo,~G.~A. 2011, \aj, 142, 112
\bibitem[Carpenter(2001)]{Carpenter01a}
  Carpenter,~J.~M. 2001, \aj, 121, 2851
\bibitem[Casagrande \etal(2011)]{Casagrande11a}
  Casagrande,~L., Sch\"{o}nrich,~R., Asplund,~M., \etal\ 2011, \aap, 530, 138
\bibitem[Casagrande \etal(2014)]{Casagrande14a}
  Casagrande,~L., Silva Aguirre,~V., Stello,~D., \etal\ 2014, \apj, 787, 110
\bibitem[Chabrier(2001)]{Chabrier01a}
  Chabrier,~G.\ 2001, \apj, 554, 1274
\bibitem[Chen \etal(2001)]{Chen01a}
  Chen,~B., Stoughton,~C., Smith,~J.~A., \etal\ 2001, \apj, 553, 184
\bibitem[Churchwell \etal(2009)]{Churchwell09a}
  Churchwell,~E., Babler,~B.~L., Meade,~M.~R., \etal\ 2009, \pasp, 121, 213
\bibitem[Davies \etal(2009)]{Davies09a}
  Davies,~B., Origlia,~L., Kudritzki,~R.-P., \etal\ 2009, \apj, 696, 2014
\bibitem[de Bruijne(2012)]{deBruijne2012} 
  de Bruijne, J.~H.~J.\ 2012, \apss, 341, 31
\bibitem[Di Matteo \etal(2013)]{DiMatteo13a}
  Di Matteo,~P., Haywood,~M., Combes,~F., Semelin,~B., Snaith,~O.~N. 2013, \aap, 553, A102
\bibitem[Edvardsson \etal(1993)]{Edvardsson93a}
  Edvardsson,~B., Andersen,~J., Gustafsson,~B., \etal\ 1993, \aap, 275, 101
\bibitem[Eisenstein \etal(2011)]{Eisenstein11a}
  Eisenstein,~D., Weinberg,~D.~H., Agol,~E. \etal\ 2011, \aj, 142, 72
\bibitem[ESA (1997)]{ESA97a}
  ESA 1997, The \emph{Hipparcos} and Tycho Catalogues (Noordwijk: ESA: ESA
  SP-1200)
\bibitem[Frinchaboy \etal(2013)]{Frinchaboy13a}
  Frinchaboy,~P.~M., Thompson,~B., Jackson,~K.~M. 2013, \apj, 777, L1
\bibitem[Ghez \etal(2008)]{Ghez08a}
  Ghez,~A.~M., Salim,~S., Weinberg,~N.~N., \etal\ 2008, \apj, 689, 1044
\bibitem[Gillessen \etal(2009)]{Gillessen09a}
  Gillessen,~S., Eisenhauer,~F., Trippe,~S., Alexander,~T., Genzel,~R., Martins,~F., \& Ott, T. 2009, \apj, 692, 1075
\bibitem[Girardi(1999)]{Girardi99a}
  Girardi,~L. 1999, \mnras, 308, 818
\bibitem[Girardi \etal(2000)]{Girardi00a}
  Girardi,~L., Bressan,~A., Bertelli,~G., \& Chiosi,~C. 2000, \aaps, 141, 371
\bibitem[Girardi \& Salaris(2001)]{Girardi01a}
  Girardi,~L. \& Salaris,~M. 2001, \mnras, 323, 109
\bibitem[Groenewegen(2008)]{Groenewegen08a}
  Groenewegen,~M.~A.~T. 2008, \aap, 488, 935
\bibitem[Gunn et al.(1979)]{GKT79} 
  Gunn, J.~E., Knapp, G.~R., \& Tremaine, S.~D.\ 1979, \aj, 84, 1181
\bibitem[Gunn \etal(2006)]{Gunn06a}
  Gunn,~J.~E., Siegmund,~W.~A., Mannery,~E.~J. \etal\ 2006, \aj, 131, 2332
\bibitem[Gustafsson \etal(2008)]{Gustafsson08a}
  Gustafsson,~B., Edvardsson,~B., Eriksson,~K., J{\o}rgensen,~U.~G., Nordlund,~\AA, \& Plez,~B. 2008, \aap, 486, 951
\bibitem[Hayden \etal(2014)]{Hayden14a}
  Hayden,~M.~R., Holtzman,~J.~A., Bovy,~J., \etal\ 2014, \aj, 147, 116
\bibitem[Hekker \etal(2013)]{Hekker13a}
  Hekker,~S., Elsworth,~Y., Mosser,~B., Kallinger,~T., Basu,~S., Chaplin,~W.~J., \& Stello,~D. 2013, \aap, 556, A59
\bibitem[Jordi \etal(2010)]{Jordi10a}
  Jordi,~C., Gebran,~M., Carrasco,~J.~M., \etal\ 2010, \aap, 523, A48
\bibitem[Juri\'{c} \etal(2008)]{Juric08a}
  Juri\'{c},~M., Ivezi\'{c},~\v{Z}., Brooks,~A., \etal\ 2008, \apj, 673, 864
\bibitem[Kaiser \etal(2002)]{Kaiser02}
  Kaiser,~N., Aussel,~H., Burke,~B.~E., \etal\ 2002, \procspie, 4836, 154
\bibitem[Kallinger \etal(2010)]{Kallinger10a}
  Kallinger,~T., Mosser,~B., Hekker,~S., \etal\ 2010, \aap, 522, 1
\bibitem[Keller \etal(2007)]{Keller07a}
  Keller,~S.~C., Schmidt,~B.~P., Bessell,~M.~S., \etal\ 2007, \pasp, 24, 1
\bibitem[Koch \etal(2010)]{Koch10a}
  Koch,~D.~G., Borucki,~W.~J., Basri,~G., \etal\ 2010, \apj, 713, 79
\bibitem[Koesterke(2009)]{Koesterke09a}
  Koesterke,~L. 2009, American Institute of Physics Conference Series, 1171, 73
\bibitem[Koesterke \etal(2008)]{Koesterke08a}
  Koesterke,~L., Allende Prieto,~C., \& Lambert,~D.~L. 2008, \apj, 680, 764
\bibitem[Kroupa(2003)]{Kroupa03a}
  Kroupa,~P. 2003, \mnras, 322, 231
\bibitem[Kuijken \& Tremaine(1991)]{Kuijken91a}
  Kuijken~K. \& Tremaine,~S.\ 1991, in Dynamics of Disk Galaxies, ed. B.~Sundelius (G\"{o}teborg: G\"{o}teborg Univ. Press), 257
\bibitem[Kuijken \& Tremaine(1994)]{Kuijken94a}
  Kuijken~K. \& Tremaine,~S.\ 1994, \apj, 421, 178
\bibitem[Kurucz(1979)]{Kurucz79a}
  Kurucz,~R.~L.\ 1979, \apjs, 40, 1
\bibitem[Laney \etal(2012)]{Laney12a}
  Laney,~C.~D., Joner,~M.~D., \& Pietrzy{\'n}ski,~G. 2012, \mnras, 419, L1637
\bibitem[Levine \etal(2008)]{Levine08a}
  Levine,~E.~S., Heiles,~C., \& Blitz,~L.\ 2008, \apj, 679, 1288
\bibitem[Luck \& Lambert(2011)]{Luck11a}
  Luck,~R.~E. \& Lambert,~D.~L. 2011, \aj, 142, 136
\bibitem[Majewski \etal(2011)]{Majewski11a}
  Majewski,~S.~R., Zasowski,~G., Nidever,~D.~L.\ 2011, \apj, 739, 25
\bibitem[Marigo \etal(2008)]{Marigo08a}
  Marigo,~P., Girardi,~L., Bressan,~A., Groenewegen,~M.~A.~T., Silva,~L., \& Granato,~G.~L.\ 2008, \aap, 482, 883
\bibitem[McWilliam \& Zoccali(2010)]{McWilliam10a}
  McWilliam,~A. \& Zoccali,~M. 2010, \apj, 724, 1491
\bibitem[Miglio \etal(2012)]{Miglio12a}
  Miglio,~A., Brogaard,~K., Stello,~D., \etal\ 2012, \mnras, 419, 2077
\bibitem[Miglio \etal(2013)]{Miglio13a}
  Miglio,~A., Chiappini,~C., Morel,~T., \etal\ 2013, \mnras, 429, 423
\bibitem[M\'{e}sz\'{a}ros \etal(2013)]{Meszaros13a}
  M\'{e}sz\'{a}ros,~Sz., Holtzman,~J., Garc\'{\i}a~P\'{e}rez,~A.~E., \etal\ 2013, \aj, 146, 133
\bibitem[Mosser \etal(2011)]{Mosser11a}
  Mosser,~B., Barban,~C., Montalb\'{a}n, J., \etal\ 2011, \aap, 532, 86
\bibitem[Nataf \etal(2010)]{Nataf10a}
  Nataf,~D.~M., Udalski,~A., Gould,~A., Fouqu\'{e},~P., \& Stanek,~K.~Z., \apj, 721, L28
\bibitem[Nidever \etal(2012)]{Nidever12a}
  Nidever,~D.~L., Zasowski,~G., Majewski,~S.~R. 2012, \apj, 755, L25
\bibitem[Nordstr\"{o}m \etal(2004)]{Nordstroem04a}
  Nordstr\"{o}m,~B., Mayor,~M., Andersen,~J., \etal\ 2004, \aap, 418, 989
\bibitem[Paczynski \& Stanek(1998)]{Paczynski98a}
  Paczynski,~B. \& Stanek,~K.~Z. 1998,\apjl, 494, L219
\bibitem[Pietrinferni \etal(2004)]{Pietrinferni04a}
  Pietrinferni,~A., Cassisi,~S., Salaris,~M., \& Castelli,~F. 2004, \apj, 612, 168
\bibitem[Pietrinferni \etal(2006)]{Pietrinferni06a}
  Pietrinferni,~A., Cassisi,~S., Salaris,~M., \& Castelli,~F. 2006, \apj, 642, 797
\bibitem[Plez(2012)]{Plez12a}
  Plez,~B. 2012, Turbospectrum: Code for spectral synthesis, Astrophysics Source Code Library, record ascl:1205.004
\bibitem[Reimers(1975)]{Reimers75a}
  Reimers, D. 1975,  Mem. Soc. R. Sci. Liege, 6, 369
\bibitem[Rix \& Zaritsky(1995)]{Rix95a}
  Rix,~H.-W. \& Zaritsky,~D. 1995, \apj, 447, 82
\bibitem[Rix \& Bovy(2013)]{Rix13a}
  Rix,~H.-W. \& Bovy,~J. 2013, \aapr, 21, 61
\bibitem[Roeser \etal(2010)]{Roeser10a}
  Roeser,~S., Demleitner,~M., Schilbach,~E. 2010, \aj, 139, 2440
\bibitem[Ro{\v s}kar \etal(2012)]{Roskar12a}
  Ro{\v s}kar,~R., Debattista,~V.~P., Quinn,~T.~R., \& Wadsley,~J. 2012, \mnras, 426, 2089
\bibitem[Santiago \etal(2014)]{Santiago14a}
  Santiago,~B.~X., Brauer,~D.~E., Anders,~F., \etal\ 2014, \aap, submitted
\bibitem[Schr\"{o}der \& Cuntz(2005)]{Schroeder05a}
  Schr\"{o}der,~K.-P. \& Cuntz,~M. 2005, \apj, 630, L73
\bibitem[Scott(1992)]{Scott92a}
  Scott,~D.~W. 1992, Multivariate Density Estimation: Theory, Practice, and Visualization (New York, Chicester: John Wiley \& Sons)
\bibitem[Sellwood \& Binney(2002)]{Sellwood02a}
  Sellwood,~J.~A. \& Binney,~J.~J. 2002, \mnras, 336, 785
\bibitem[Siebert \etal(2011)]{Siebert11a}
  Siebert,~A., Famaey,~B., Minchev,~I., \etal\ 2011, \mnras, 412, 2026
\bibitem[Silverman(1986)]{Silverman86a}
  Silverman,~B.~W. 1986, Density Estimation for Statistics and Data Analysis
(London: Chapman and Hall)
\bibitem[Skrutskie \etal(2006)]{Skrutskie06a}
  Skrutskie,~M.~F., Cutri,~R.~M., Stiening,~R., \etal\ 2006, \aj, 131, 1163
\bibitem[Sobeck \etal(2014)]{Sobeck14a}
  Sobeck,~J., Majewski,~S., Hearty,~F., \etal\ 2014, in American Astronomical Society Meeting Abstracts, 223, \#440.06
\bibitem[Stanek \& Garnavich(1998)]{Stanek98a}
  Stanek,~K.~Z. \& Garnavich,~P.~ M. 1998, \apjl, 503, L131
\bibitem[Stanek \etal(1998)]{Stanek98b}
  Stanek,~K.~Z., Zaritsky,~D., Harris,~J. 1998, \apjl, 500, L141
\bibitem[Steinmetz \etal(2006)]{Steinmetz06a}
  Steinmetz,~M., Zwitter,~T., Siebert,~A., \etal\ 2006, \aj, 132, 1645
\bibitem[Stello \etal(2013)]{Stello13a}
  Stello,~D., Huber,~D., Bedding,~T.~R. \etal\ 2013, \apjl, 765, L41
\bibitem[Udalski(1998)]{Udalski98a}
  Udalski,~A., Szymanski,~M., Kubiak,~M., Pietrzynski,~G., Wozniak,~P., \& Zebrun,~K. 1998, Acta~Astron., 48, 1
\bibitem[Valentini \& Munari(2010)]{Valentini10a}
  Valentini,~M., \& Munari, U. 2010, \aap, 522, A79
\bibitem[Williams \etal(2013)]{Williams13a}
  Williams,~M.~E.~K., Steinmetz,~M., Binney,~J., \etal\ 2013, \mnras, 436, 101
\bibitem[Wilson \etal(2010)]{Wilson10a}
  Wilson,~J.~C., Hearty,~F., Skrutskie,~M.~F., \etal\ 2010, Proc.~SPIE, 7735, 46
\bibitem[Wright \etal(2010)]{Wright10a}
  Wright,~E.~L., Eisenhardt,~P.~R.~M., Mainzer,~A.~K., \etal\ 2010, \aj, 140, 1868
\bibitem[Yong \etal(2012)]{Yong12a}
  Yong,~D., Carney,~B.~W., \& Friel,~E.~D. 2012, \aj, 144, 95
\bibitem[Zacharias \etal(2013)]{Zacharias13a}
  Zacharias,~N., Finch,~C.~T., Girard,~T.~M., \etal\ 2013, \aj, 145, 44
\bibitem[Zasowski \etal(2009)]{Zasowski09a}
  Zasowski,~G., Majewski,~S.~R., Indebetouw,~R., \etal\ 2009, \apj, 707, 510
\bibitem[Zasowski \etal(2013)]{Zasowski13a}
  Zasowski,~G., Johnson,~J.~A., Frinchaboy,~P.~M., \etal\ 2013, \aj, 146, 81
\end{thebibliography}
\end{document}